\documentclass[a4paper, twoside,
twocolumn,aps,superscriptaddress,showpacs]{revtex4}

\usepackage{amsmath}
\usepackage{amssymb}
\usepackage{amsfonts}
\usepackage{graphicx}
\usepackage{epstopdf}
\usepackage{bm}
\usepackage{graphicx}
\usepackage{dcolumn}

\newcommand{\beq}{\begin{equation}}
\newcommand{\eeq}{\end{equation}}
\newcommand{\bea}{\begin{eqnarray}}
\newcommand{\eea}{\end{eqnarray}}

\newcommand{\showlabel}[1]{} 
\newcommand{\internalcom}[1]{}

\def\bea{\begin{eqnarray}}
\def\eea{\end{eqnarray}}
\def\be{\begin{equation}}
\def\ee{\end{equation}}
\def\etal{{\it et al.}}

\newcommand{\pib}{\mbox{\boldmath$\pi$}}
\newcommand{\alb}{\mbox{\boldmath$\alpha$}}

\newcommand{\pbar}{$\bar{p}$~}
\newcommand{\Hbar}{$\bar{\rm H}$~}

\newcommand{\pos}{e$^+$~}

\begin{document}
\title{Cold and Ultracold Rydberg Atoms in Strong Magnetic Fields}

\author{T. Pohl}
\affiliation{Max-Planck-Institute for the Physics of Complex Systems, N\"otnitzer 38, D-01187, Dresden, Germany}
\affiliation{ITAMP, Harvard Smithsonian Center for Astrophysics, 60 Garden Street, MS 14, Cambridge, MA 02138, USA}
\author{H. R. Sadeghpour}
\affiliation{ITAMP, Harvard Smithsonian Center for Astrophysics, 60 Garden Street, MS 14, Cambridge, MA 02138, USA}
\author{P. Schmelcher}
\affiliation{Physikalisches Institut, Universit\"at Heidelberg, Philosophenweg 12, 69120 Heidelberg, Germany}
\affiliation{Theoretische Chemie, Institut f\"ur Physikalische Chemie, Universit\"at Heidelberg, Im Neuenheimer Feld 229,
D-69120 Heidelberg, Germany}



\date{\today}

\begin{abstract}
Cold Rydberg atoms exposed to strong magnetic fields possess unique properties
which open the pathway for an intriguing many-body dynamics taking place in Rydberg
gases consisting of either matter or anti-matter systems. We review both the
foundations and recent developments of the field in the cold and ultracold regime
where trapping and cooling of Rydberg atoms have become possible. Exotic states
of moving Rydberg atoms such as giant dipole states are discussed in detail, 
including their formation mechanisms in a strongly magnetized cold plasma. Inhomogeneous
field configurations influence the electronic structure of Rydberg atoms, and we describe the utility of 
corresponding effects for achieving tightly trapped ultracold Rydberg atoms. 
We review recent work on large, extended cold Rydberg gases in magnetic fields and their 
formation in strongly magnetized ultracold plasmas through collisional recombination. Implications 
of these results for current antihydrogen production experiments are pointed out, and techniques for 
trapping and cooling of such atoms are investigated.
\end{abstract}

\maketitle

\setcounter{tocdepth}{100}
\tableofcontents

\clearpage


\section{Introduction}\label{sec:introduction}
Rydberg atoms, by virtue of their special properties, are of fascination in atomic
and optical physics \cite{GallagherBook}. These atoms whose size scales as the square of the Rydberg
principal quantum number $n$, are ideally suited for fundamental quantum interrogations as well as detailed 
classical analyses which test the correspondence principle. 

Owing to the large excursions of the bound Rydberg electron, highly excited states have extremely long lifetimes, scaling as $n^3$.
Their exaggerated dimensions also imply large geometrical cross sections, such that atomic processes like 
Penning ionization or inelastic electron collisions (ionization, recombination and (de)-excitation) occur with 
high rates in an ultracold plasma environment. Moreover, the large polarizabilities of Rydberg atoms makes 
them highly susceptible to external electric and magnetic fields $-$ the latter being the primary subject of the 
present article.

The largest discovered Rydberg atoms present in a natural environment were carbon atoms in interstellar space
with a principal quantum number of $n\sim 1009$ \cite{stepkin} which have been detected due to their radio recombination lines.
Rydberg atoms in cool tenuous interstellar media offer a valuable opportunity for measuring magnetic fields. 
Measurements of H$n\alpha$ lines ($\Delta n=1$ radio recombination transitions) allow for the 
determination of the Zeeman splitting in hydrogen, from which typically small magnetic field strengths in the range 
of $\mu$G can be inferred \cite{crucher,troland}. 
Although small in strength, such magnetic fields in tenuous environments, where
neutral densities are less than $10^4$ cm$^{-3}$, can play an important role for the interstellar gas dynamics \cite{crucher}.

The laboratory creation of cold and dilute highly-excited media has become possible with the development of 
techniques to laser-cool and trap neutral atoms \cite{cooling1,cooling2,cooling3,wpwRMP99,Metcalf99}. Laser-excitation 
of cold atomic gases leads to an intriguing many-body system with very strong long-range interactions. Together with their 
extremely long lifetimes the strong interactions between Rydberg atoms make for promising candidates in quantum 
information processing \cite{qi1,Lukin01,Hyafil04,qi3,qi2}. 
A range of questions, arising from the collective excitation dynamics in cold Rydberg 
gases \cite{uc1,uc2,uc3,uc4,uc5,uc6,uc7,uc8,uc9,uc10} and the formation and non-equilibrium evolution of 
ultracold neutral plasmas \cite{uc1,uc2,uc3,uc4,uc5,uc6,uc7,uc8,uc9,uc10} have been studied in experiment and theory. 
Since outlining these emerging topics is certainly beyond the scope of the present article, we refer to some recent review 
articles \cite{kppr,ckll,gapi} for more comprehensive discussions.

A very different class of Rydberg atoms yet appears in strong magnetic fields, as realized through recombination 
processes in cold antimatter plasmas at CERN \cite{athena,atrap1}, in ion storage rings \cite{gwinner}, and in cold 
 atom traps, subject to magnetic fields of several Tesla \cite{ChoiRaithel2005}. Due to their fragility, the effects of 
 such laboratory magnetic fields on the electronic structure of Rydberg atoms are similarly pronounced as those of 
 very strong, extraterrestrial fields ($\sim 10^4$Tesla) \cite{highmag} on ground state atoms. Highly excited Rydberg 
 states, hence, provide a well-accessible venue for combined theoretical and experimental studies of strong 
 magnetic field effects on atomic systems.
 From a different perspective, highly inhomogeneous magnetic fields may offer a promising possibility for a controlled 
 manipulation of Rydberg atoms, which are otherwise difficult to trap by standard optical techniques, developed for ground 
 state atoms.

Here, we will focus on magnetic field induced static and dynamical effects on Rydberg atoms, i.e. on their 
internal and translation degrees of freedom, their formation in cold plasmas as well as their evolution in external traps.
This review article is organized as follows. Section \ref{sec:maryga} is devoted to existing experiments in magnetized 
Rydberg gases and plasmas, with emphasis on the atomic excitation from laser-cooled gases of ground state atoms.
Antihydrogen production in Penning traps through recombination in magnetized plasmas is briefly outlined here. 
Section \ref{sec:atprohomag} first focuses on the fundamental properties of atomic systems in homogeneous external magnetic 
fields. Following a discussion of the symmetries and constants of motion, the latter are exploited
to perform a gauge-independent pseudoseparation of the center of mass motion of neutral atoms in
magnetic fields or combined perpendicular electric and magnetic fields. Effects due to the 
magnetic field-induced coupling of the center of mass motion to the internal motion are reviewed 
on a classical-mechanics level. The main focus of this section are the giant
dipole states whose preparation, properties as well as applications to matter-antimatter systems
are studied in detail. We conclude with a discussion of the concept of the guiding center approximation to describe strongly magnetized atoms. In section \ref{sec:atproinhomag} the structure and quantum dynamics of
atomic systems in inhomogeneous magnetic fields is addressed. We first describe their properties assuming 
immobile nuclei, followed by a systematic approach to 
trapping of ultracold Rydberg atoms in 3D quadrupole and Ioffe-Pritchard traps. An analysis of the lifetime
of trapped Rydberg atoms and their long-time dynamics in strong field traps complement this section.
Section \ref{sec:ryatffp} is dedicated to Rydberg atom formation and subsequent collisions in cold magnetized 
plasmas, with particular reference to antihydrogen experiments.
We conclude with an outlook in section \ref{outlook} pointing out some perspectives for future studies.

\section{Magnetized Rydberg gases and plasmas}\label{sec:maryga}
Over the last decade, experimental efforts in photoexcitation and photoionization
of laser-cooled atoms have led to creation of cold highly-excited Rydberg gases \cite{uc12,uc13,uc1,uc2,uc3,uc4,uc5} and
ultracold neutral plasmas \cite{unp1,unp2,unp3,unp4,unp5,unp6,unp7,fzr07}. At strong magnetic fields $B$ and sufficiently 
low thermal velocities $v_{\rm th}=\sqrt{k_{\rm B}T/m}$ a situation can be realized, for which the cyclotron radius 
$r_{\rm c}=v_{\rm th}/\omega_{\rm c}$ becomes much smaller than 
the distance of closest approach $b=e^2/k_{\rm B}T$, and the cyclotron frequency  $\omega_{\rm c}=eB/mc$ provides the 
shortest dynamical times scale of charges with mass $m$.
Such strongly magnetized conditions are naturally realized in non-neutral plasmas of either electrons or laser-cooled 
ions, confined in the strong magnetic fields of a Penning trap \cite{dun99}. Their intense study over the last few decades has 
led, for example, to a deeper understanding of strongly correlated plasmas and various transport processes.
Neutrality, as realized in cold Rydberg gases, adds another degree of complexity, by providing a collisionally rich environment 
in which phenomena such as an avalanche-like plasma formation 
\cite{RobinsonGallagher2000}, angular momentum mixing collisions \cite{DuttaRaithel2001}, Penning ionization \cite{penn04} and
Rydberg atom formation through three-body recombination \cite{fzr07} have been observed.

\subsection{Atomic excitation in strong magnetic fields}\label{dimag}
Rydberg atoms are generally produced by collision (charge particle excitation and electron exchange) or optical 
excitation. The production of cold Rydberg atoms is typically achieved through optical excitation. Here laser cooled gases 
in a magneto-optical \cite{uc1,uc2,uc3,uc5} or a dipole \cite{uc4} trap are used as a starting point, which subsequently are excited to Rydberg states 
either by UV excitation \cite{uc1} or by two-photon excitation \cite{uc2,uc3,uc4,uc5}. Typically two-photon excitation of Rubidium atoms proceeds via excitation of the 
 $5S_{1/2}$ ground state to the first excited $5P_{3/2}$, which is coupled by a second laser to either $nS_{1/2}$ or $nD_{3/2,5/2}$ Rydberg states. 
The maximum excitation efficiency achieved with such laser-pulses is, however, typically limited by a high sensitivity to spatial laser-pulse inhomogeneities and fluctuations in laser power and frequency.
Potentially much higher efficiency can be achieved using the method of stimulated Raman adiabatic passage (STIRAP), in which
the upper transition pulse is applied ahead of the lower one, in an counterintuitive pulse sequence \cite{BergmannRMP1998}. 
The method utilizes the existence of a dark state in a three-level system, which is composed only of the ground and excited Rydberg state. 
By adiabatically following this dark state during the evolution of the laser pulse, in principle $100\%$ excitation efficiency can be achieved under perfect 
adiabatic conditions. In cold Rydberg gases this approach was first demonstrated  by Cubel et al. \cite{uc11}, who reported an excitation efficiency of $\sim70\%$.

The effects of strong magnetic fields, i.e. of the diamagnetic coupling, on the electronic structure of Rydberg atoms were originally studied by 
Jenkins and Segre \cite{Jenkins39} through absorption spectroscopy of sodium and potassium atoms for principal 
quantum numbers of $n\le35$. A later refinement and extension to Rydberg states of up to $n=75$ by Garton and 
Tomkins \cite{Garton69} revealed important additional features in the absorption spectrum of the Barium I series.
Unexpectedly, they found a series of sharp resonances, at frequency intervals of $3\omega_{\rm c}/2$ and extending well into
the zero-field ionization continuum. In contrast regular Landau levels are spaced by the electron cyclotron 
frequency $\omega_{\rm c}=eB/mc$. The continuum features in the spectra of magnetized atoms $-$ later 
termed quasi-Landau resonances $-$ have been studied in a number of subsequent experiments \cite{Econ78,Fonck80,Castro80}. 
The diamagnetic Rydberg atom in strong magnetic fields has, moreover, received considerable attention as a paradigm system for quantum chaos \cite{Friedrich89}, permitting detailed and precise comparisons between theory and experiment \cite{rydchaos2}.
For further details on these aspects of Rydberg atoms in strong magnetic fields we refer the reader
to the corresponding reviews \cite{Friedrich89,Friedrich97,Schmelcher98a,Cederbaum97} and cited references therein.

\subsection{Cold gases in strong magnetic fields}
Magnetic trapping of ground state atoms, as an important prerequisite to forming magnetically trapped Rydberg gases, has been 
demonstrated in the group of Raithel \cite{GuestRaithelPRL2005,ckll}. Several millions of rubidium atoms were accumulated 
and laser cooled in a superconducting Ioffe-Pritchard trap operating at large bias fields of up to 2.9 Tesla.

Strongly inhomogeneous magnetic fields whose magnitude varies over length scales comparable to the size of a Rydberg atom 
drastically alter its electronic structure and lead to intricate coupling between the atom's internal and center-of-mass dynamics.
As will be worked out in detail in section \ref{sec:atproinhomag}, such configurations provide very tight Rydberg atom confinement with
laboratory magnetic fields.

Ref. \cite{ChoiRaithel2005} reported Rydberg atom confinement in magnetic traps with weaker field gradients,
based on the trapping configuration described above \cite{GuestRaithelPRL2005}. The experiment could monitor
collective atom cloud oscillations in the trap, which was used to extract average magnetic moments of the trapped
Rydberg atoms. Importantly, the experiment found extremely long Rydberg gas lifetimes of up to $200$ ms. This observation 
can be traced back to a drastically suppressed spontaneous decay rate due to the external magnetic field, which will be
discussed in detail in section \ref{lifetime}. The combined measurement of the lifetimes and magnetic moments provided strong experimental 
evidence for the production of so-called guiding center atoms. Such atoms are characterized by large regular electron orbits transverse to the magnetic field lines, superimposed by tightly confined cyclotron oscillations . The basic properties of nearly circular $-$ and, hence, long-lived $-$ guiding center atoms and their response to additional electric will be worked out in 
section \ref{gca}. While the experimental two-photon excitation scheme, described in the previous section, exclusively produces
Rydberg states with low angular momenta, the observed high-angular momentum, circular guiding center are formed through 
$l$-mixing collisions with free plasma electrons. There are several processes that lead to the formation of free electrons in a cold 
Rydberg gas, whose relative importance is currently studied intensively by several groups in zero-field Rydberg gases 
\cite{RobinsonGallagher2000,unp11,unp8b,ioni3,ioni4,ioni2,ioni1}

An even more collisionally rich and strongly magnetized environment has been realized in \cite{ckz05} by trapping an ultracold neutral
rubidium plasma in the strong magnetic fields of a, so-called nested Penning trap (see next section). Beyond providing a well suited
platform to study interesting collective and collisional dynamics in the strongly magnetized regime, such experiments can also permit important
case studies for ongoing antihydrogen experiments, that produce antimatter hydrogen atoms through collisional recombination
in ultracold two-component positron-antiptoton plasmas \cite{athena,atrap1}.

\subsection{Antihydrogen production from cold, magnetized plasmas}\label{HbarExp}
Several collaborations (ATRAP \cite{atrap1}, ATHENA \cite{athena}, ALPHA \cite{alpha07}, ASACUSA \cite{asacusa}, AEGIS \cite{aegis}), operating at the Antiproton Decelerator at CERN, are working on the production of cold antihydrogen atoms, to provide an important testing ground for fundamental physics. One produced, cold antihydrogen atoms could be employed for comparative high precession spectroscopy with the hydrogen atom for accurate tests of CPT (charge, parity and time) invariance \cite{cpt2,cpt1,grav2,GabRev05} and could be used to measure gravitational forces between neutral matter and antimatter, i.e. to test the weak equivalence principle \cite{cpt2,grav1,grav2,aegis}. 

\begin{figure} [b!]
\centerline{\includegraphics[width=3.2in]{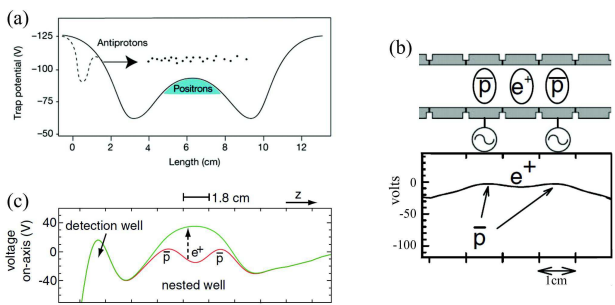}}
\caption{\label{NestedPenning}
A nested Penning trap. The \pbar accumulate in the potential valleys, while the \pos are deposited at 4K at the saddle
of the Penning potential. The magnetic field axis is shown.
}
\end{figure}

The basic ingredient for producing cold antihydrogen (\Hbar) are cold antiprotons (\pbar) and positrons (\pos). 
The techniques for cooling the initially highly energetic antiprotons, produced at CERN, have been developed 
over the two decades by Gabrielse and co-workers (see e.g. \cite{GabRev05}). In brief, initially, fast \pbar ions 
are matter degraded and captured in a Penning trap and sympathetically cooled by simultaneously confined electrons, 
which dissipate energy through synchrotron radiation \cite{Gab86}. 
Positrons are typically obtained from a sodium source \cite{Greaves94} and later cooled in subsequent steps 
(see e.g. \cite{poscool3,poscool1,poscool2} for more details).

In order to simultaneously confine the cold positron and antiproton plasma in the same spatial region Gabrielse 
{\it et al.} developed a specially designed magnetic trap, termed a nested Penning trap \cite{Gab88}. The basic principle is schematically
shown in Fig.\ref{NestedPenning}. As in a standard Penning trap radial confinement is provided by a homogenous 
magnetic field oriented along the z-direction \cite{dun99}. Axial confinement is achieved through a sequence of ring
electrodes that realize an effective double well potential. Since antiprotons and positrons have opposite charge, 
antiprotons are accumulated in the side-wells while the positrons are trapped in the central maximum of the double well 
potential. 

Initiating the \Hbar formation requires a spatial overlap between the two plasma components, which has been realized 
experimentally  using several different approaches \cite{athena,atrap1,atrap2,atrap4}. In the first ATHENA and ATRAP 
experiments \cite{athena,atrap1}, antiprotons were launched into the trapping region with energies above the central well
such that they would oscillate throughout the trap \cite{athena,atrap1} (see Fig.\ref{NestedPenning}a). During each passage 
through the positron cloud antiprotons are cooled via collisions and upon reaching sufficiently low velocities start to form highly 
excited antihydrogen atoms, predominantly via three-body recombination (see section \ref{sec:ryatffp}). In this approach the 
velocities of formed atoms critically depends on the characteristic time of recombination compared to the cooling timescale, 
and was later found to be rather high \cite{athena3,robicheaux}. 

In order to achieve lower atomic velocities, a driven production scheme was developed in \cite{atrap2}. 
It starts from antiprotons, which are initially cooled into the side-wells of the trap, and subsequently heated by rf-fields such that they slowly come into overlap with the central positron cloud 
(see Fig.\ \ref{NestedPenning}b). In section \ref{sect_chx}, we give a further discussion of this method and a theoretical 
interpretation of the observed \Hbar velocities \cite{AtrapVelocity,pohl1}. Finally, in a recent experiment the ATRAP collaboration used a third 
approach, based on a gradual increase of the central well voltage, which slowly merges both plasma components \cite{atrap4}.
The various detection techniques used to probe the properties of produced \Hbar atoms are reviewed for example
in \cite{GabRev01,athena4,GabRev05}

To be applicable for the aforementioned spectroscopy and gravitation measurements the produced, \Hbar atoms necessarily have 
to be translationally cold and in their ground state. Since atom formation in cold plasmas predominantly proceeds via three-body 
capture into highly excited Rydberg states, the production efficiency of trappable ground state antihydrogen and their resulting 
velocities is crucially determined by the properties of highly excited Rydberg atoms and their recombination dynamics in strong 
magnetic fields (see section \ref{lifetime}, \ref{selfcooling} and \ref{sec:ryatffp}). For a comprehensive overview over the multitude of atomic 
 processes occurring in antihydrogen experiments we also refer to a recent review by Robicheaux \cite{robrev}.

\section{Atomic properties in homogeneous magnetic fields}\label{sec:atprohomag}
The present chapter aims at reviewing the fundamental properties of moving atomic systems
in homogeneous magnetic and/or crossed homogeneous electric and magnetic fields.
They are of immediate relevance to the dynamics and processes taking place in matter as well
as antimatter Rydberg gases and plasmas in a strong magnetic field and determine many of their properties.
Firstly, i.e. in section \ref{hamsymcon}, we will introduce the Hamiltonian for both neutral
and charged atomic systems and discuss its symmetries and resulting constants of motion.
Interpretations of the conserved pseudomomentum are provided. Section \ref{pseudsep}
is devoted to the pseudoseparation of the center of mass motion with a focus on neutral
systems and an outline on ions. The gauge-independent pseudoseparation provides us with
a generalized potential whose properties are discussed firstly here. Dynamical effects due to the
coupling of the center of mass and internal (electronic) motion in a homogeneous magnetic
field are studied in section \ref{comc} for the underlying classical dynamics. The characteristics for both
the case of a vanishing and a nonvanishing pseudomomentum are explored.
The giant dipole states emerging due to the double-well structure of the generalized potential
for sufficiently strong motional and/or external electric fields are investigated in section \ref{gidista}
first at hand of the hydrogen atom. A possible pathway for their systematic experimental preparation 
is outlined. A substantial part of this section analyzes the giant dipole states i.e. resonances of 
multiply-excited Rydberg atoms in crossed fields. Section \ref{matantimat} contains an application
of the giant dipole physics to matter-antimatter systems specifically the positronium atom.
As a major outcome it is shown that giant dipole states of positronium represent extremely long-lived
and therefore quasistable states of matter-antimatter systems \cite{Ackermann97}.

The fundamental properties of moving atomic systems
in homogeneous magnetic and/or crossed homogeneous electric and magnetic fields
are of immediate relevance to the dynamics of matter as well
as antimatter Rydberg gases and plasmas in  strong magnetic fields.
The coupling of the center of mass (CM) and internal (electronic) motion in a homogeneous B field has 
consequences for the electronic structure of the atomic system. The emergence of giant dipole states (GDS) due to 
the double-well structure of the generalized potential is one such consequence being a result of the gauge-independent
pseudoseparation of the CM motion.

\subsection{The Hamiltonian - symmetries and constants of motion} \label{hamsymcon}

We consider a finite number of interacting charged
particles exposed to an external homogeneous and static magnetic field employing
the non-relativistic Coulomb interaction. Relativistic effects induced by the
external magnetic field become relevant only \cite{Lindgren79,Virtamo79,Chen99} for field strengths well-above the 
regime considered here, i.e. particularly for field strengths occuring in the photospheres of
neutron stars ($B \gg 10^{5}$ T). The starting-point of our analysis is the nonrelativistic Hamiltonian in
Cartesian coordinates in the laboratory frame

\begin{equation}\label{hamlab}
{\mathcal{H}}_L = \sum \limits_{i} \frac{1}{2 m_i} {\pib}_{i}^2 + \sum \limits_{i > j}{V_c} 
\left( |{\bf{r}}_i - {\bf{r}}_j | \right)
\end{equation}

where we have omitted the interaction of the spins with the magnetic field. The latter is indispensable
to obtain the right ordering of the energy levels in the astrophysically strong field regime
\cite{Ivanov98,Ivanov99,Ivanov00,Ivanov01a,Ivanov01b}
but plays no role for the following considerations focusing on the spatial symmetries and constants
of motion and the resulting transformations to the Hamiltonian.
We remark that the following analysis holds not only for the case of ${V_c}$ representing the Coulomb potential
but for any translation- and rotation invariant potential.
${\pib}_i$ is the kinetic (mechanical) momentum for the i-th particle of mass $m_i$, charge $q_i$, position  
${\bf{r}}_i$ with the canonical momentum ${\bf{p}}_i$, and the vector potential ${\bf{A}}_i$ 

\begin{equation}\label{kinmom}
{\pib}_i = {\bf{p}}_i - q_i {\bf{A}}_i
\end{equation}
The Schr\"odinger equation ${\mathcal {H}}_L \Psi_L = E \Psi_L$ belonging to the 
Hamiltonian in Eq. (\ref{hamlab})
is invariant with respect to the gauge transformations

\begin{eqnarray}\label{gaugetrans}
\Psi_L & \longmapsto & \exp\left( + i \sum \limits_{i} q_i \Lambda \left ( {\bf{r}}_i \right) \right) \Psi_L\\ \nonumber
{\bf{A}}_i & \longmapsto &  {\bf{A}}_i + \nabla_i \Lambda \left ( {\bf{r}}_i \right)
\end{eqnarray}

In this and the following section we will, as far as possible, use a gauge-independent formalism.
In those cases where it is necessary to introduce an explicite gauge, we make use of the symmetric Coulomb gauge,
${\bf{A}}_i = \frac{1}{2} \left( {\bf{B}} \times {\bf{r}}_i \right)$, where ${\bf{B}}$ is the
magnetic field vector. The kinetic mechanical momentum (\ref{kinmom}) obeys the commutation relation

\begin{equation}\label{commkinmom}
\left[ \pib_{i\alpha} , \pib_{j\beta} \right] = i q_i \epsilon_{\alpha \beta \gamma}
B_{\gamma} \delta_{ij}
\end{equation}

where the Greek indices denote the components of the corresponding vectors. $\epsilon_{\alpha \beta \gamma}$
is the completely antisymmetric tensor of rank three. 
The so-called pseudomomentum ${\bf{k}}_i$ \cite{Lamb59,Avron78,Johnson83}
\begin{equation}\label{pseudmom}
{\bf{k}}_i = \pib_i + q_i \left[ {\bf{B}} \times {\bf{r}}_i \right]
\end{equation}
obeys the following commutation relations
\begin{eqnarray}\label{commpseud}
\left[ {\bf{k}}_{i\alpha} , \pib_{j\beta} \right] &=& 0 \\ \nonumber
\left[ {\bf{k}}_{i\alpha} , {\bf{k}}_{j\beta} \right] &=& - i q_i \epsilon_{\alpha \beta \gamma}
B_{\gamma} \delta_{ij}
\end{eqnarray}

The mathematical and physical interpretation of the pseudomomentum \cite{Avron78,Johnson83}
depends on the situation under consideration and is closely connected to the phase space symmetries and conserved
quantities of the underlying Hamiltonian. In the symmetric gauge,
the kinetic and pseudomomenta differ only by the sign of their field-dependent terms. 
The classical Hamiltonian corresponding to the Eq. (\ref{hamlab}) 
is then invariant under the phase space translation

\begin{equation}\label{pttgroup}
({\bf{r}}_i,{\bf{p}}_i) \mapsto ({\bf{r}}_i + \alb,{\bf{p}}_i + (\frac{q_i}{2}) {\bf{B}} \times \alb)
\end{equation}

which is a combined translation and gauge transformation. In the absence of the external magnetic field,
the phase space translation group reduces to a translation in coordinate space forming an Abelian group
for both  underlying classical and quantum systems.
The coordinate translation group is generated by the conserved total canonical momentum ${\bf{P}}=\sum_i {\bf{p}}_i$
being equal to the total kinetic momentum. The phase space translation group in the presence of the field
for a quantum system is non-Abelian, i.e. phase space translations do not commute
in general. The generators of the phase space translation group are the components of the total
pseudomomentum
\begin{equation}\label{totpseudlab}
{\bf{K}}_L=\sum_i {\bf{k}}_i 
\end{equation}

which represent constants of motion in the presence of the field

\begin{equation}\label{conspseud}
\left[{\mathcal{H}}_L,{\bf{K}}_L \right] = 0
\end{equation}

As indicated above, the components of the pseudomomentum ${\bf{K}}_L$ can, in general, not be made sharp
simultaneously,

\begin{equation}
\left[ {\bf{K}}_{L\alpha}, {\bf{K}}_{L\beta} \right] = - i Q \epsilon_{\alpha \beta \gamma} B_{\gamma}
\end{equation}

where $Q = \sum_i q_i$ is the total charge of the system. Only in the case of a neutral particle system,
i.e. for $Q=0$, the components of ${\bf{K}}_L$ commute altogether and can be exploited equally as constants of motion
in the context of a pseudoseparation of the CM motion. 
We remark that since $[K_{L\|},K_{L\perp}]=0$, the CM motion along the B field is exactly decoupled from the relative motion.

For a single charged particle in a homogeneous magnetic field, the coordinate vector of the center of the classical Landau orbit
is provided by the so-called guiding center, see a discussion of the guiding center approximation later in Sec. \ref{gca},
\begin{equation} \label{guidingcenter}
{\bf{r}}_c = {\bf{r}} + \frac{1}{q B^2} \left( {\pib} \times {\bf{B}} \right) 
\end{equation}
The guiding center vector is closely related to the pseudomomentum vector 

\begin{equation}
{\bf{k}}_{\perp} = q {\bf{r}}_c \times {\bf{B}};~~~ {\bf{r}}_{c \perp} = \frac{1}{q B^2} \left( {\bf{B}} \times {\bf{k}} \right)
\end{equation}

This extends naturally to a system of noninteracting particles where the total pseudomomentum is interpreted as the 
superposition vector of the guiding centers of the individual particles. 
For a neutral two-body system, we have

\begin{equation}
{\bf{K}}_L = q \left( {\bf{r}}_{c1} - {\bf{r}}_{c2} \right) \times {\bf{B}} \label{neutguid}
\end{equation}

i.e. ${\bf{K}}_L$ is proportional to the distance vector of the guiding centers of the two oppositely charged particles.
The dual interpretation
of the pseudomomentum becomes evident if one takes into account that in case of the interaction with radiation \cite{Avron78}, the
sum of the photon momentum and the pseudomomentum is conserved. 
In the absence of the homogeneous field, but for
interacting particles, the pseudomomentum ${\bf{K}}_L$ reduces to the conserved total canonical momentum ${\bf{P}}$
leading
to an exact separation of the CM motion. For an interacting particle system in the presence of the field
the interpretation of the pseudomomentum possesses an 'interpolating character' between the kinetic and positional representations.
The guiding center approach to the classical dynamics is adequate in the regime where the forces due to the magnetic
field dominate the Coulomb forces (see section \ref{gca}).

A further exact constant of motion of the Hamiltonian is provided by \cite{Johnson49,Mitchell81} 

\begin{equation}
{\mathcal{L}}_{\|} = \sum_i \left(\frac{1}{2 q_i B}\right) \left( {\bf{k}}_i^2 - \pib^2_{i} \right)
\end{equation}

In case of the symmetric gauge for the vector potential, this quantity is identical to the projection of the
total orbital angular momentum onto the magnetic field axis

\begin{equation}
{\mathcal{L}}_{\|} = \frac{1}{B} {\bf{B}} \sum_i \left[ {\bf{r}}_i \times {\bf{p}}_i \right]
\end{equation}

${\mathcal{L}}_{\|}$ is the generator for rotation about the magnetic field
axis, and does not commute with the components of the pseudomomentum perpendicular to the magnetic field
but obeys 

\begin{equation} \left[ {\mathcal{L}}_{\|} , K_{L\|} \right] = \left[ {\mathcal{L}}_{\|} , {\bf{K}}_{L\perp}^2 \right] = 0
\end{equation}

As a consequence,
a complete set of commuting constants of motion for a neutral system, would either be the
components of ${\bf{K}}_L$ or $\{K_{L\|}, {\bf{K}}_{L\perp}^2, {\mathcal{L}}_{\|} \}$. 
For charged systems comprising of heavy and light particles, such as atoms and
molecules, approximate constants of motion exist. These quantities are particularly helpful
for the transformation of the Hamiltonian such that its dependencies on the CM and internal 
degrees of freedom take on a particularly simple and distinct appearance \cite{Baye82,Baye83,Baye86,Schm89a,Schm89b}.

\subsection{Pseudoseparation of the center of mass motion}
\label{pseudsep}

The pseudomomentum is closely related to the center of mass motion of the
underlying particle system. To elucidate this fact, let us perform a transformation from the laboratory
coordinate system to CM and internal coordinates. In the following we focus on atomic systems consisting of a heavy nucleus
with mass $M_0$ and $N$ electrons each with mass $m$. For the symmetric gauge, the pseudomomentum
then takes on the following appearance

\begin{equation} \label{pseudcm}
{\bf{K}} = {\bf{P}} + \frac{Q}{2} {\bf{B}} \times {\bf{R}} +
\frac{e}{2} \alpha {\bf{B}} \times \sum_i {\bf{r}}_i 
\end{equation}

where $({\bf{R}},{\bf{P}})$ are the CM coordinate vector and its canonically conjugate momentum, respectively.
$\{{\bf{r}}_i, {\bf{p}}_i\}$ denote the relative coordinate vectors of the electrons with respect to the nucleus
and their canonically conjugate momenta, respectively, 
$\alpha = (M_0+Zm)/M, Q=(N-Z)e$ and $Z,M,e$ are the nuclear charge number, the total atomic mass and the
electron charge, respectively.  
For $Q=0$, i.e. a neutral atom, the pseudomomentum does not depend on the
CM coordinates. In this case all components of the pseudomomentum commute i.e. 
they can be made sharp simultaneously.
Introducing the pseudomomentum as a canonical momentum eliminates the cyclic CM coordinates from the Hamiltonian
thereby resulting in a so-called pseudoseparation of the CM motion \cite{Lamb59,Avron78,Johnson83,Herold81,Schmelcher94a}.
We emphasize that the pseudoseparation
does not yield a complete separation of the CM and internal motion. Indeed, an intricate coupling
of the CM and internal motion remains, leading to effects such as the classical diffusion of the CM for a 
highly excited Rydberg atom (see section \ref{comc}).

To elaborate the case of a neutral atom in more detail, we observe that the eigenfunctions
of the Hamiltonian can simultaneously be chosen as eigenfunctions of the pseudomomentum. Since the pseudomomentum
is linear
with respect to the coordinates and momenta in Eq. (\ref{pseudcm}), the dependence of the eigenfunctions on the CM has the
following form
\small
\begin{eqnarray} \label{Un}
\Psi \left(\{ {\bf{r}}_i \}, {\bf{R}}; {\mathcal{K}} \right) &=&
\exp \left( i \left( {\mathcal{K}} - \frac{e}{2} {\bf{B}} \times \sum_i {\bf{r}}_i \right) {\bf{R}} \right)
\Psi_0 \left(\{ {\bf{r}}_i \};  {\mathcal{K}} \right) \nonumber \\
& =& \exp \left( i {\mathcal{K}} {\bf{R}} \right) U_n 
\Psi_0 \left(\{ {\bf{r}}_i \}; {\mathcal{K}} \right)
\end{eqnarray}
\normalsize

with the eigenvalue equation $ {\bf{K}} \Psi = {\mathcal{K}} \Psi$. Apart from the plane wave component,
$\exp \left( i {\mathcal{K}} {\bf{R}} \right)$, the unitary transformation $U_n$ represents the only CM coordinate-dependent 
part of the total wave function $\Psi$. The wave function $\Psi_0$ of the internal motion 
depends parametrically on the eigenvalue ${\mathcal{K}}$ of the pseudomomentum. Transforming 
the Hamiltonian with $U_n$ \cite{Johnson83,Schmelcher94a} and replacing the pseudomomentum
with its eigenvalue, provides us with the following pseudoseparated Hamiltonian

\begin{equation}\label{hampseud}
{\mathcal{H}} = \frac{1}{2M} {\mathcal{K}}^2 - \frac{e}{M} \left( {\mathcal{K}} \times {\bf{B}} \right) \sum_i {\bf{r}}_i + {H}_{int}
\end{equation}

where $H_{int}$ represents the part of the Hamiltonian which depends exclusively on the electronic degrees of freedom.
The first term of Eq. (\ref{hampseud}) is a constant energy shift. The second term represents the coupling of the
CM to the electronic motion. It is a Stark term \cite{Lamb59} involving
a motional electric field $\frac{1}{M} {\mathcal{K}} \times {\bf{B}}$: due to the collective motion of the neutral
atom in the magnetic field the atom feels an additional homogeneous electric field perpendicular to the
magnetic field axis which causes a separation of the centers of charge of the atom, i. e. it forms an electric dipole.

There have been a number of investigations addressing the effects of the motional Stark field or finite nuclear mass effects
in general for low-lying electronic states of atoms in strong magnetic fields primarily for the two-body problem
i.e. the hydrogen atom \cite{Herold81,Wunner80a,Wunner81,Ruder94,Vincke88,Cuvelliez92,Potekhin94,Bezchastnov94,Potekhin98,Potekhin97a,Potekhin97b} and more recently also for the helium atom
\cite{Baye90,Vincke90,Al-Hujaj03a,Al-Hujaj03b}. [For early experimental work
on the influence of  motional Stark field on the spectra and transition line shapes, we refer the reader to
Refs. \cite{Crosswhite79,Panock80}]. 
Furthermore, a number of early work relating to various two-body aspects
of particle systems in crossed electric and magnetic fields \cite{Rau79,Bhattacharya82,Burkova76,Gay79,Farelly94}, are
worthy of mention.

The pseudoseparation of the CM motion for neutral systems
yields a natural division of the Hamiltonian into three separate parts involving the CM,
its coupling to the electronic motion and the electronic degrees of freedom, respectively.
The transformed Hamiltonian was obtained for a specific symmetric gauge. 
A manifestly gauge-independent approach has been
developed \cite{Dippel94,Schmelcher01} providing us with a gauge-independent
potential picture for moving atoms in magnetic fields. 

Writing the general gauge as ${\bf{A}} ({\bf{r}}) = \frac{1}{2} {\bf{B}}
\times {\bf{r}} + \nabla \Lambda ({\bf{r}})$, the first step is to perform the
coordinate transformation from laboratory to the CM and internal coordinates. 
By constructing the common eigenfunctions of the
Hamiltonian and the pseudomomentum, and using the specific dependence of the gauge 
function $\Lambda$ on the new variables, it becomes possible to derive a unitary transformation that eliminates
the CM coordinates from the Hamiltonian. Rearranging terms, we arrive at the
Hamiltonian ${\mathcal{H}} = {\mathcal{T}} + {\mathcal{V}}$ with

\small
\begin{eqnarray} \label{gaugeinvT}
{\mathcal{T}} \nonumber   &=&  \frac{1}{2m} \sum\limits_{i}^{} \left( {\bf{p}}_i - \frac{e}{2} {\bf{B}} \times {\bf{r}}_i 
+e \beta {\bf{B}} \times \sum\limits_{i}^{} {\bf{r}}_i  + e {\bf{\nabla}}_i {\it{f}} \right)^2\\ 
&+& \frac{1}{2M_0} \left( \sum\limits_{i} {\bf{p}}_i +e \gamma {\bf{B}} \times \sum\limits_{i} {\bf{r}}_i 
+ e \sum\limits_{i} {\bf{\nabla}}_i {\it{f}} \right)^2 
\\ \label{gaugeinvV}
{\mathcal{V}}  \nonumber &=&  \frac{1}{2M} \left( {\mathcal{K}} -Ne {\bf{B}} \times \left(\frac{1}{N} \sum\limits_{i}
{\bf{r}}_i\right) \right)^2  \\ &-& e {\bf{E}}\sum\limits_{i}{\bf{r}}_{i}
+ V_c\left(\left\{{\bf{r}}_i\right\}\right) 
\end{eqnarray}
\normalsize

where $\beta=\frac{m}{M}, \gamma=\frac{Nm-M_0}{2M}$. In the above Hamiltonian, we have additionally taken into 
account an external homogeneous electric field ${\bf{E}}$.  $f(\{{\bf{r}}_i\})$ is an integration constant
resulting from the differential equation that ensures the total wave function to be an
eigenfunction of ${\bf{K}}$.  ${\mathcal{T}}$ represents the kinetic energy of the
electrons in the presence of the magnetic field and this term is, as expected, explicitly gauge-dependent via the
scalar function ${\it{f}}$. 
${\mathcal{V}}$, on the other hand, is independent of the chosen gauge (no scalar function ${\it{f}}$
occurs in ${\mathcal{V}}$) and can therefore be interpreted as 
a generalized potential. 
Besides the Coulomb interaction terms $V$ and the electric Stark term due to
the external electric field the first quadratic term of the potential ${\mathcal{V}}$ is of particular relevance.
Apart from the trivial constant ${\mathcal{K}}^2/2M$ it contains a motional electric field term
$e/M ({\bf{B}} \times {\mathcal{K}}) \sum {\bf{r}}_i$ and a diamagnetic term $e^2/2M ({\bf{B}} \times \sum {\bf{r}}_i)^2$.
The relevant quantity occuring in the latter two potential terms is the electronic center of mass (ECM),
i.e. ${\bf{R}}= \frac{1}{N}\sum {\bf{r}}_i$ in the internal coordinate frame.
It is therefore the ECM which experiences interactions
beyond the Coulomb potential and which enters the generalized potential for multi-electron
systems. In case of (effective) one-electron systems the ECM reduces (approximately)
to the coordinate vector of the single electron.

We remark, that the first quadratic term in Eq.(\ref{gaugeinvT}) which is 
an important part of the total potential ${\mathcal{V}}$, represents the
kinetic energy of the CM of the atom, which can be verified by
inspecting the Hamiltonian equation of motion for the CM
\begin{equation}
{\dot{\bf{R}}} = \frac{{\partial{\mathcal{H}}}}{\partial{\mathcal{K}}} = \frac{1}{M} 
\left( {\mathcal{K}} -Ne {\bf{B}} \times \left(\frac{1}{N} \sum\limits_{i}
{\bf{r}}_i\right) \right) \label{cmkinenergy}
\end{equation}

Therefore the CM kinetic energy of a neutral atom
provides a potential for the internal motion of the electrons of the atom.
This kinetic energy is due to the vanishing net charge of the system 
independent of any chosen gauge of the vector potential. 

\begin{figure}
\begin{center}
\includegraphics[width=10cm,keepaspectratio,angle=0]{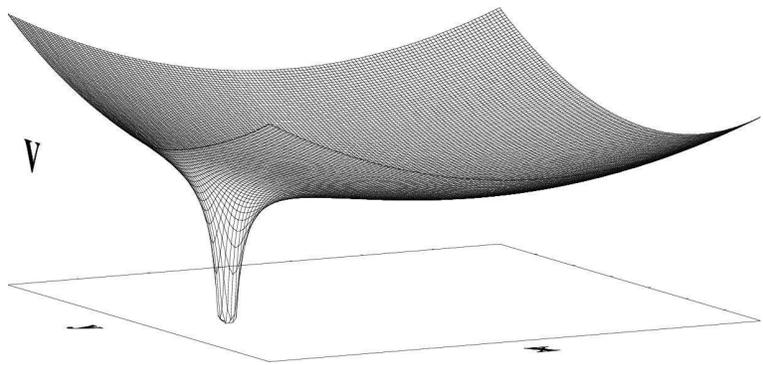}
\end{center}
\caption{A two-dimensional intersection of the potential ${\mathcal{V}}_1$ in the
$(x,y)$ plane perpendicular to the magnetic field \label{intersectV}}
\end{figure}

For
completeness, we give here the Hamiltonian ${\mathcal{H}}_1 = {\mathcal{T}}_1 + {\mathcal{V}}_1$ for
the hydrogen atom. It emerges from Eqs. (\ref{gaugeinvT},\ref{gaugeinvV})
by specializing to $N=1$, subsumming the two quadratic terms of the kinetic energy in Eq. (\ref{gaugeinvT}),
as well as introducing the reduced masses $\mu=\frac{mM_0}{M}$ and $\mu^{\prime}=\frac{mM_0}{M_0-m}$ 
\begin{eqnarray} \nonumber \label{seham}
{\mathcal{T}}_1&=&\frac{1}{2\mu}({\bf{p}}-\frac{e}{2}\frac{\mu}{\mu^{\prime}} {\bf{B}} \times {\bf{r}}
+ e {\bf{\nabla}} {\it{f}} ({\bf{r}}) )^2\\ 
{\mathcal{V}}_1&=&\frac{1}{2M}({\mathcal{K}}-e{\bf{B}}\times{\bf{r}})^2+V_c(r) -e {\bf{E}}{\bf{r}} 
\end{eqnarray}
The generalized potential ${\mathcal{V}}_1$ gives rise to a number of intriguing features
such as a double well potential which accommodates  weakly bound
Rydberg states with huge electric dipole moments, see also 
Sec. (\ref{gidista}) for a discussion of GDS for multi-electron atoms. 
Here we wish to discuss in some more detail the properties of the generalized potential ${\mathcal{V}}_1$
for a single-electron atom \cite{Dippel94,Schmelcher93a,Schmelcher93b}. 
In Figure \ref{intersectV},
we show a two-dimensional intersection of ${\mathcal{V}}_1$ in the
$(x,y)$ plane perpendicular to the magnetic field. 
The saddle, due to a competition between the Coulomb potential and the linear Stark term, is evident. At larger 
values of $x$, the quadratic diamagnetic potential begins to dominate, leading to the formation of a shallow and long-range
potential minimum. Discrete energy states in this potential well are the GDS, see also Sec. 
\ref{gidista}.

The existence of the saddle point and the outer well, depends, 
on the strength of the B field and the pseudomomentum or, respectively, on
the strength of an external electric field oriented perpendicular to the magnetic field.
Indeed, the Stark term due to the external electric field ${\bf{E}}$ may be combined
with the quadratic term that represents the kinetic energy of the CM motion in Eq.(\ref{seham})
by redefining the pseudomomentum ${\mathcal{K}} \rightarrow {\mathcal{K}} + M {\bf{v}}_d$ where
${\bf{v}}_d$ is the drift velocity of charged particles in the crossed external fields $({\bf{E}},{\bf{B}})$.
The equilibrium position of the potential well is obtained from the following conditions:
\begin{equation}
y=z=0~~~~;~~~~~x^3+(K/B)x^2-(M/B^2) = 0~~~~~{\rm{and}}~~~~~x<0 \label{extrema}
\end{equation}
The condition for the existence of both the saddle point and the outer well can be stated as follows:
\begin{equation}
K^3 > \frac{27}{4} B M \label{pseudoextrem}
\end{equation}
The position ${\bf{r}}_0$ of the minimum is then obtained from Eq.(\ref{extrema}),

\begin{equation}
y_0=z_0=0~~~~;~~~~~x_0 \approx (-K/B) + KM/(K^3-2MB) \label{minapprox}
\end{equation}

Due to the confining behaviour of the 
generalized potential ${\mathcal{V}}_1$ perpendicular to the magnetic field, it is evident that
ionization of an isolated atom in a giant dipole state can only take place parallel
to the magnetic field. The quantized states in the outer well (GDS) are discussed in Sec. 
\ref{gidista}.

\subsection{Center of mass coupling to the internal dynamics} \label{comc}

The pseudoseparation of the CM motion and adoption of the symmetric gauge for the hydrogen Hamiltonian in a magnetic 
field results in
\begin{eqnarray}\label{H11}
{\mathcal{H}}_1 &=& {\mathcal{T}}_1 + {\mathcal{V}}_1 \\ \label{H12}\nonumber
&=& 
\frac{1}{2 \mu} \left( {\bf{p}} - \frac{e}{2} \frac{\mu}{\mu^{\prime}} {\bf{B}} \times {\bf{r}} \right)^2
+ \frac{1}{2M} \left( {\mathcal{K}} - e {\bf{B}} \times {\bf{r}} \right)^2 + V_c \\ \label{H13}\nonumber
&=& \frac{1}{2M} {\mathcal{K}}^2 - \frac{e}{M} \left( {\mathcal{K}} \times {\bf{B}} \right) {\bf{r}} 
+ \frac{1}{2 \mu} {\bf{p}}^2 \\ &-& \frac{e}{2 \mu^{\prime}} B L_z + \frac{e^2}{8 \mu} \left( {\bf{B}} \times {\bf{r}} \right)^2
+ V_c 
\end{eqnarray}
The B field is along the z-axis. In the first equation, there is a clear division of kinetic and potential energies, 
the term containing $K$ is an effective potential for the internal motion, the second form of the Hamiltonian does not have
the usual division of the kinetic and potential energies and contains the linear Zeeman and the quadratic diamagnetic 
terms.
The kinetic energy of the CM motion is, as already indicated in section (\ref{pseudsep})
in the context of Eq.(\ref{cmkinenergy}), provided by the quadratic term of the generalized potential for the
internal motion i.e. we have
\begin{equation} \label{cmspot}
 \frac{1}{2M} \dot{\bf{R}}^2 = \frac{1}{2M} ({\mathcal{K}} -e {\bf{B}} \times {\bf{r}})^2
\end{equation}

To understand the effects of the coupling between the CM and electronic motion we focus in the next
two subsections on the classical dynamics of the hydrogen atom. Although being a quantum (or semiclassical) object 
a classical description of the highly excited atom is illuminative and provides relevant insights i.e. a qualitative
description of many of its properties can be obtained via a classical picture (see refs.
\cite{Friedrich89,Friedrich97,Schmelcher98a,Cederbaum97} and refs. therein).

\subsubsection{The classical center of mass motion for a vanishing pseudomomentum}\label{cmmk0}
The Hamiltonian equations of motion for the CM and electronic motion for the
hydrogen atom are 
\begin{eqnarray}\label{cmeom1}
 \dot{\mathcal{K}} &=& 0\\ \label{cmeom2}
 \dot{\bf{R}}  &=& \frac{1}{M} {\mathcal{K}} -\frac{e}{M} \left({\bf{B}} \times {\bf{r}} \right)\\ \label{cmeom3}
 \dot{\bf{r}} &=& \frac{1}{\mu} {\bf{p}} -\frac{e}{2\mu^{\prime}} \left({\bf{B}} \times {\bf{r}} \right)\\ \label{cmeom4}
 \dot{\bf{p}} \nonumber &=& -\frac{e}{M} \left({\bf{B}} \times {\mathcal{K}} \right) - \frac{e}{2\mu^{\prime}} \left({\bf{B}} \times 
                  {\bf{p}} \right) \\  &+& \frac{e^2}{4\mu} {\bf{B}} \times \left({\bf{B}} \times {\bf{r}} \right) 
                  - e^2 \frac{{\bf{r}}}{|{\bf{r}}|^3}
\end{eqnarray}

The CM velocity couples to the relative coordinates, transverse to the B field, even when ${\mathcal{K}} = 0$. The CM
equations of motion are independent of the gauge. 
The Hamiltonian equations of motion
for the electronic degrees of freedom (\ref{cmeom3},\ref{cmeom4}) are for the case ${\mathcal{K}} = 0$ 
independent of the CM motion, i.e. they decouple from the CM motion.

For ${\mathcal{K}} = 0, B \| z$-axis the Hamiltonian possess an azimuthal symmetry i.e. the projection of the orbital 
angular momentum onto the $z-$axis is a conserved quantity. Let us focus here on the case $L_z=0$.
Apart from similarity transformations, the classical dynamics does not depend on the field strength
and energy separately but only on the scaled energy $\epsilon = E B^{-2/3}$. Varying $\epsilon$ from $-3$ to $-0.1$ the
classical phase space belonging to the electronic motion of the hydrogen atom undergoes a transition from (almost)
complete regularity to a fully chaotic situation (see Ref. \cite{Friedrich89} and refs. therein). 

The natural question arises what kinds of center of mass motion are possible if the internal
motion goes through the whole range from regularity to irregularity. Another closely related
question results from the following considerations. The internal motion is restricted to the energy
shell and takes place in a bounded region of phase space. The phase space of the CM motion
however is, at least in principle, unbounded and therefore one may ask whether or not
phase space is filled out by the CM trajectory and how this depends on the regularity or irregularity
of the internal motion \cite{Schmelcher92a,Schmelcher92b,Schmelcher92c}. We remind the
reader that in the absence of the external field the CM motion decouples from the internal motion 
and exhibits a straigth-lined motion.

\begin{figure}
\begin{center}
\includegraphics[width=8cm,keepaspectratio,angle=180]{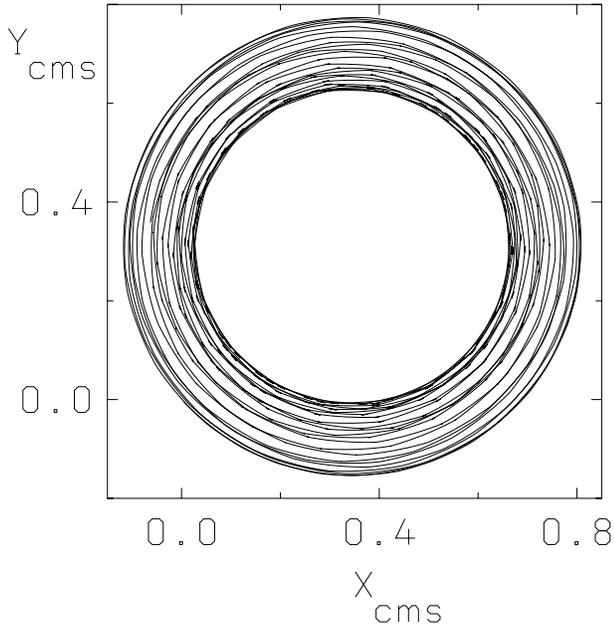}
\end{center}
\caption{A center of mass trajectory in the CM hyperplane perpendicular to
the magnetic field with starting-point ${\bf{R}}(t=0) = (0,0,0)$, for 
$E=-10^{-3},10^{-5}$ (B= 2.35 T), $L_z=0,\epsilon= -2.15$, all in atomic units. From \cite{Schmelcher92c}.
\label{cmtrajreg}}
\end{figure}

Let us begin our discussion with the 'deep' regular region, i.e. the region for which the use
of low order classical perturbation theory with respect to the diamagnetic term is valid. The
classical trajectories for the internal motion of the hydrogen atom in this region have been
classified in ref.\cite{Delos83}: There are the so-called rotators and librators covering the
complete or only part of the interval $[0,2\pi]$ for the corresponding angle variable.
Figure \ref{cmtrajreg} shows a typical CM trajectory if the internal motion is of the rotator type.
The CM performs an approximately smooth circular motion on a ring and is, therefore, limited
to a small range of coordinate space. According to Eq.(\ref{cmeom2}) for ${\mathcal{K}}=0$,
the quasi-periodic behaviour of the electronic degree of freedom ${\bf{r}}$ goes hand
in hand with the quasi-periodic motion of the CM degrees of freedom. 

\begin{figure}
\begin{center}
\includegraphics[width=8cm,keepaspectratio,angle=180]{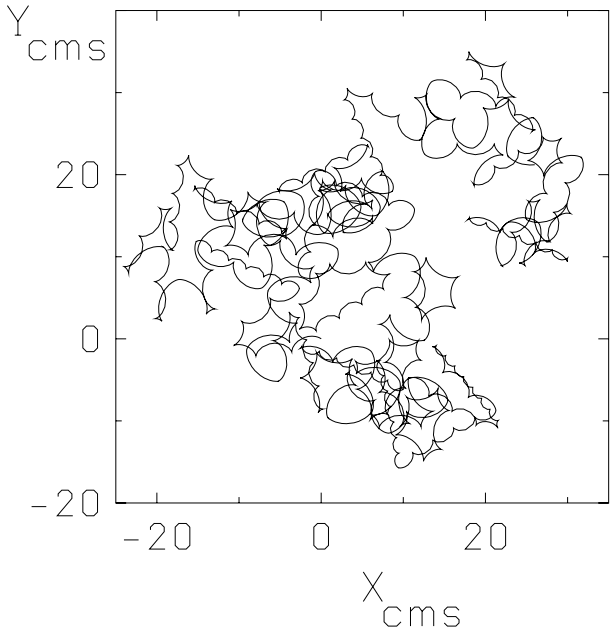}
\end{center}
\caption{A center of mass trajectory in the CM hyperplane perpendicular to
the magnetic field with starting-point ${\bf{R}}(t=0) = (0,0,0)$. The underlying
total energy $E$, magnetic field strength $B$, angular momentum $L_z$ and scaled
energy $\epsilon$ are: $E=-4.7 \cdot 10^{-5},10^{-5}$ (this corresponds to 2.35 Tesla), $L_z=0,\epsilon= -0.101$. 
All values in atomic units. From \cite{Schmelcher92c}. \label{cmtrajirreg}}
\end{figure}

Let us now turn to the situation of a fully chaotic phase space for the internal motion.
Figure \ref{cmtrajirreg} shows a typical CM trajectory for this case. The CM motion is,
independently of the special initial conditions, no more restricted to some bounded volume
of phase space. With increasing time the CM trajectory fills out an increasing volume
of the CM hyperplane. Its propagation closely resembles the random motion of e.g. a Brownian
particle. After all, this is quite natural since the underlying equation of motion (\ref{cmeom2}) 
for the CM is a Langevin-type equation without friction \cite{Chandrasekhar43}.
The corresponding stochastic Langevin force is replaced here by the deterministic
chaotic force $-e \left( {\bf{B}} \times \dot{\bf{r}} \right)$. A characteristic feature of the
random Brownian motion is its diffusion, i.e. the linear dependency of the travelled mean
square distance on time, 
\begin{equation}
\langle \rho_{CM}^2 \rangle = D \cdot t
\end{equation}

where $\rho_{CM}^2 = X_{CM}^2 +  Y_{CM}^2$ and $D$ is the corresponding diffusion constant. The
chaotic deterministic motion of the CM possesses properties of randomness in the sense that it
provides the well-known diffusion law of random walk models (for a discussion of randomness
versus chaoticity see Refs. \cite{Prakash91,Beckmann91,Geisel81}). Let us briefly address the
dependence of the diffusion constant on the energy and magnetic field strength. With increasing
energy (always within a completely chaotic phase space) the mean square distance of the electron
from the magnetic field axis passing through the nucleus increases. As a consequence the mean
CM velocity and therefore also the diffusion constant increases. The diffusion
constant increases with increasing magnetic field strength thereby showing a saturation behaviour.
The latter can be understood by inspecting the explicit as well as implicit field-dependence of the
CM velocity given by Eq.(\ref{cmeom2}) for ${\mathcal{K}} = 0$.
To conclude, the coupling of the CM and internal motion due to the external homogeneous magnetic field causes
a distinct transition in the classical CM motion of the atom from bounded oscillatory to 
unbounded diffusive motion. We emphasize that this transition happens for an isolated
atom and is of inherently different origin than the random motion taking place for Brownian particles
which show diffusive behaviour due to random collision events among different atoms.

\subsubsection{The classical center of mass motion for a nonvanishing pseudomomentum} \label{cmmkn0}

Let us now study the coupled CM and electronic motion for the case of a nonvanishing pseudomomentum.
Inspecting the equations of motion (\ref{cmeom1},\ref{cmeom2},\ref{cmeom3},\ref{cmeom4}), one observes that the
CM velocity now contains an additional constant term proportional to the pseudomomentum. Intuitively,
one would therefore expect a straightlined CM motion being superimposed on the motion resulting from the
corresponding coupling to the internal motion (see previous subsection \ref{cmmk0}). Additionally,
the equations of motion for the internal degrees of freedom depend on the value of the pseudomomentum. 
This is a major difference with the case of a vanishing pseudomomentum, where the internal motion was
decoupled from the CM motion.

\begin{figure}
\begin{center}
\includegraphics[width=9cm,keepaspectratio,angle=0]{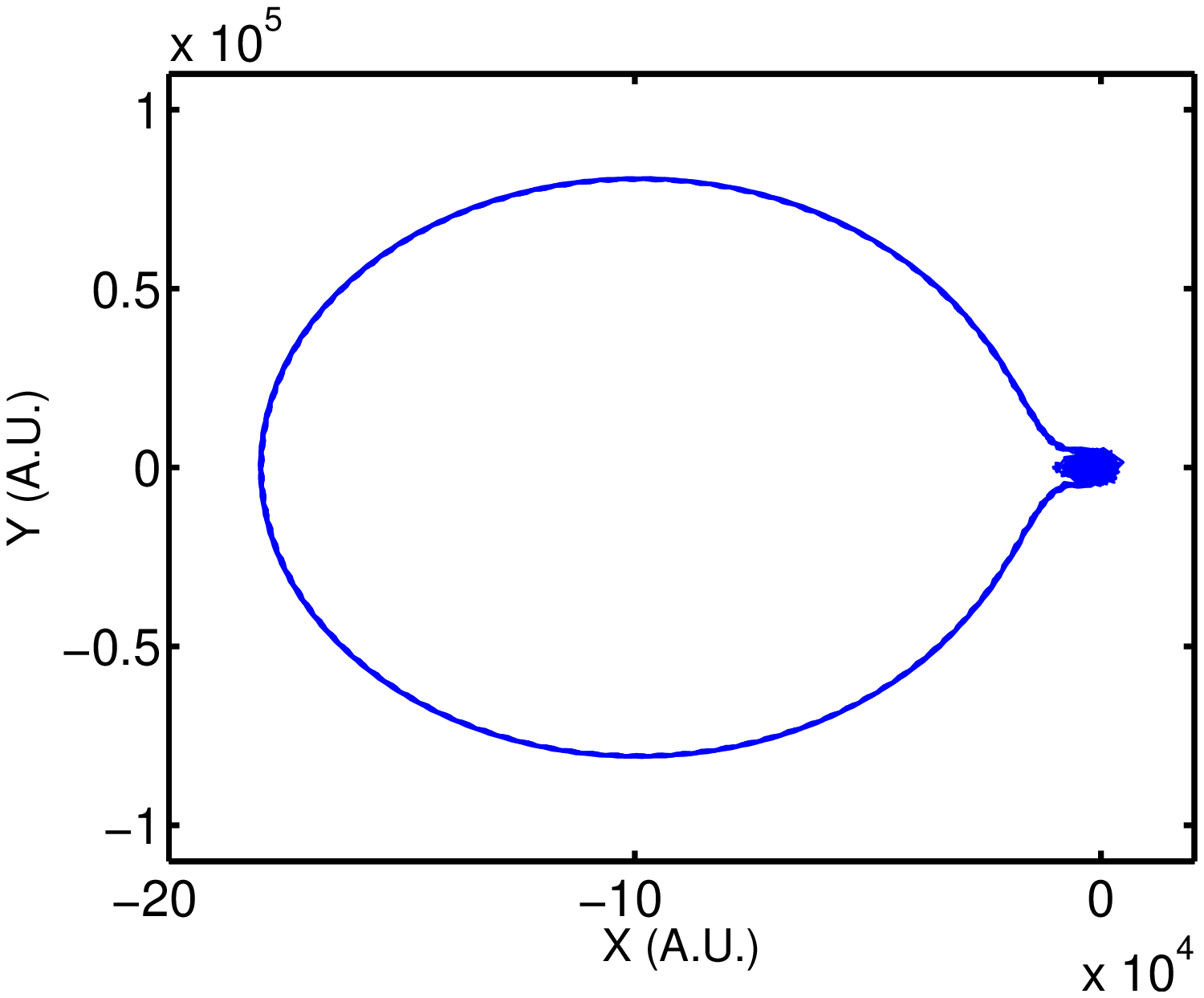}
\includegraphics[width=9cm,keepaspectratio,angle=0]{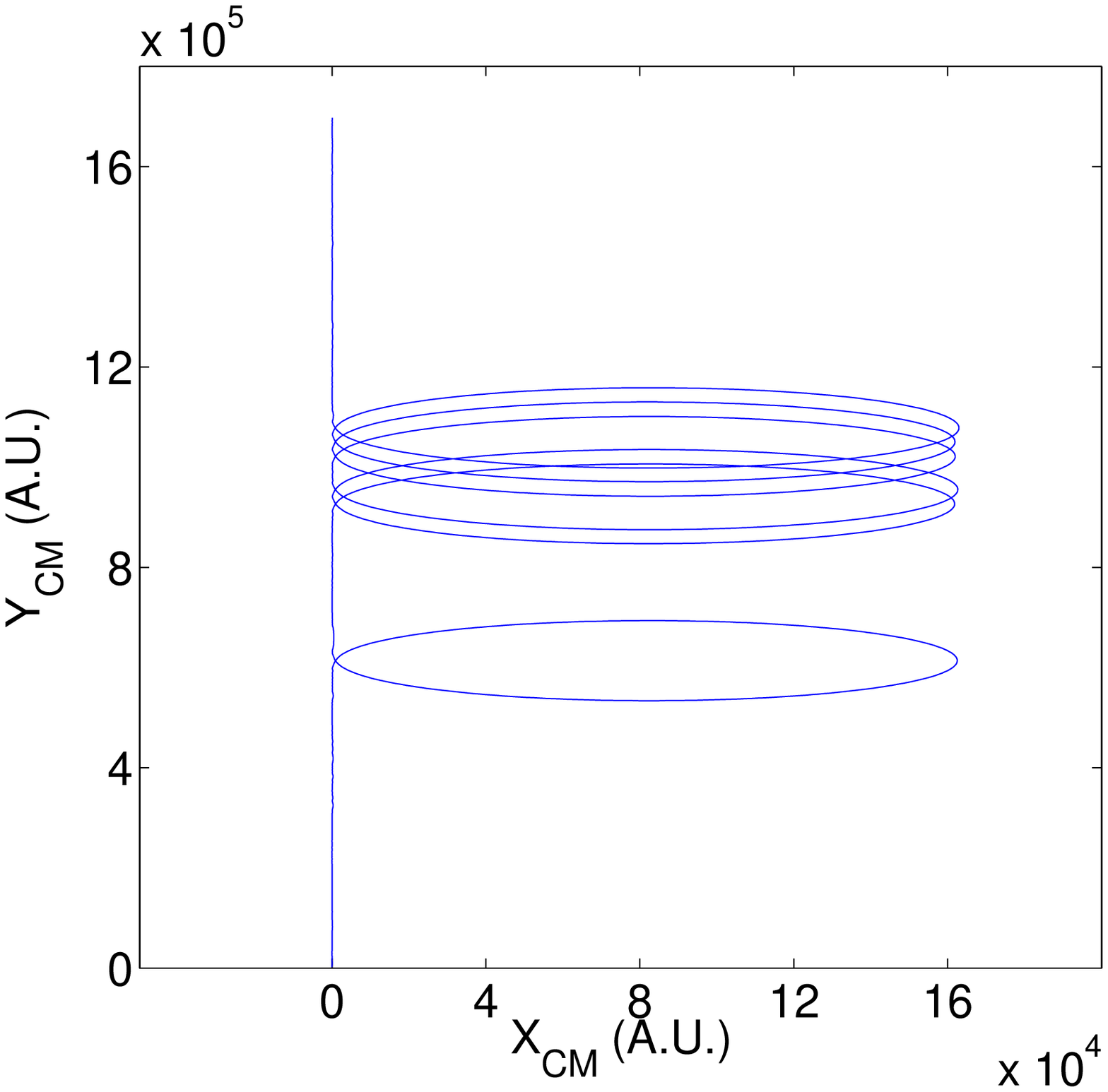}
\includegraphics[width=9cm,keepaspectratio,angle=0]{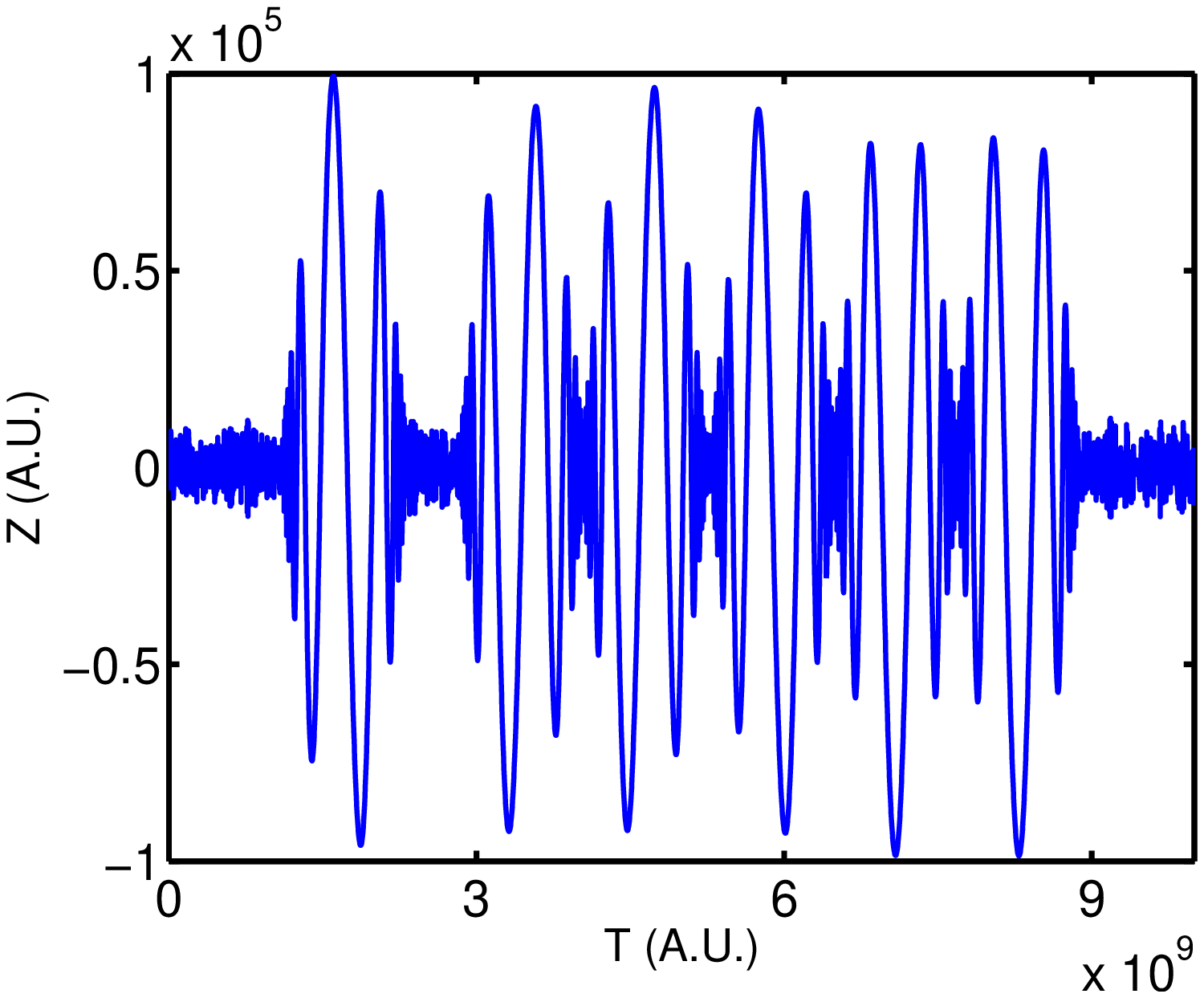}
\end{center}
\caption{Typical intermittent trajectory of the electronic motion for nonvanishing pseudomomentum:
Projection onto the $x,y-$plane - top subfigure, center of mass motion perpendicular to the
magnetic field for ${\bf{R}}(t=0)=(0,0,0)$ - middle subfigure, and internal $z-$coordinate as
a function of time. Field strength, energy and pseudomomentum are
$B=10^{-5}, E = 1.722 \cdot 10^{-4}, {\mathcal{K}} = (0,1,0)$. All values in atomic units. 
\label{intermittxy}}
\end{figure}

A major new consequence for the classical dynamics of the highly excited hydrogen atom with nonvanishing pseudomomentum
is the intermittent behaviour of the corresponding trajectories,
i.e. the appearance of alternating phases of quasiregular and chaotic motion 
\cite{Schmelcher93a}. Figure \ref{intermittxy} shows for a typical trajectory, the projection of the electronic motion on a plane
perpendicular to the magnetic field axis. One immediate observation is that there exists two alternating types
of motion. During one phase of the motion, the electron and the nucleus are in the $x,y-$plane close together
and this shows up through the black dimple close to the origin in Figure \ref{intermittxy}. During this
phase of the motion, the Coulomb and diamagnetic interactions are  comparable and the trajectory
is, therefore, chaotic, having a finite local Lyapunov exponent \cite{Fujisaka83,Grassberger84}
During the other phase,
the electron and the nucleus move far apart from each other. The relative motion in the $x,y-$plane
then approximately takes place on a circle with a large radius. The Coulomb energy provides here only a small
correction to the strong magnetic interaction. This phase of motion has an essentially vanishing
local Lyapunov exponent.

The radius of the circle of the regular phase of the electronic $x,y-$motion of an intermittent
trajectory can be understood in terms of the pseudomomentum ${\mathcal{K}}$. Indeed, according to Eq. (\ref{neutguid}),
the total pseudomomentum is proportional to the cross product of the magnetic field vector and the
distance vector between the two guiding centers for free particles. Since the magnetic field strongly dominates
the Coulomb interaction during the quasiregular circular motion shown in Figure \ref{intermittxy}, we 
encounter the situation of almost free particles and the radius of the circle is approximately
given by
\begin{equation}
r = - \frac{1}{eB^2} |{\bf{B}} \times {\mathcal{K}}| \label{radiusintermittxy}
\end{equation}

i.e. it is completely determined by the magnetic field vector and the pseudomomentum.
On the other hand, we obtain a completely different interpretation of the pseudomomentum if the
electron and the nucleus are close together. In this case, the Coulomb interaction is comparable to or dominates 
the magnetic interaction and hence the pseudomomentum is approximately the linear kinetic momentum of the
translational CM motion. Fig. \ref{intermittxy} shows the CM motion for the trajectory whose
electronic motion is also given in Fig. \ref{intermittxy}. It consists of alternating phases of purely
translational and circular motions. As already mentioned, the electron and the nucleus are strongly
bounded, i.e. close together, during the time interval of chaotic electronic motion. This is precisely
the time period during which the CM performs a purely translation motion. The time period of quasiregular
circular electronic motion corresponds to the period of approximately circular CM motion. Intermittency
therefore shows up in the CM motion by alternating phases of more or less straight-lined and circular motion.

Inspecting the corresponding electronic and nuclear motions in the laboratory coordinate system, it
turns out that the large amplitude motion perpendicular to the magnetic field, 
shown in Fig. \ref{intermittxy} in the case of relative motion,
is performed by the nucleus whereas the electron remains essentially localized close to the origin.
This peculiarity has its origin in the fact that the quasiregular phases of the intermittent motion
are always characterized by large negative $L_z$. Motion with 
$L_z<< 0$, however, can, in the laboratory coordinate system,
only be performed by the nucleus. During the period of large amplitude motion of the nucleus perpendicular to
the magnetic field , the electron performs a large amplitude motion parallel to the magnetic field.
Fig. \ref{intermittxy} shows the electronic relative motion parallel to the magnetic field as a function of time
for the same trajectory whose CM and internal motion perpendicular to the magnetic field are shown in Fig.
\ref{intermittxy}, respectively.

When the nucleus returns to the electron performing large amplitude motion parallel to the
field axis, the two remain close together for the next phase of chaotic motion close to the origin.
During this period of motion the CM performs to a good approximation a purely translational motion (see
Fig.\ref{intermittxy}). We remark that the large amplitude motion perpendicular to the magnetic field
circulates, in case of Eq. (\ref{pseudoextrem}), the outer well whereas for the period of chaotic motion where the
electron and nucleus stay close together the dynamics takes place close to the origin i.e. in the well
created by the Coulomb interaction.

\subsection{Giant dipole states in crossed fields} \label{gidista}

\subsubsection{The hydrogen atom}

The generalized potential (see Eqs. \ref{gaugeinvV},\ref{H11}) 
exhibit, for the case of a sufficiently
large motional or external electric field (see Eq.(\ref{pseudoextrem})), a double well structure. The underlying
classical dynamics has been shown above to be characterized by an intermittent dynamics. Naturally, the question
arises whether there exist quantum states which live exclusively in the outer well and what their properties are
\cite{Dippel94,Schmelcher01,Schmelcher93b,Baye92,Vincke92,Dzyaloshinskii92}.
To explore their properties, let us firstly expand the generalized potential ${\mathcal{V}}_1$ in Eq. (\ref{H11})
around the minimum in the outer well, whose position will be denoted by ${\bf{r}}_0=(x_0,0,0)$ transverse to the 
field which is oriented along the $z-$axis and a motional or external electric field along the $x-$axis.
The expansion of ${\mathcal{V}}_1$ around the minimum ${\bf{r}}_0$ of the outer well has to be accompanied
by a gauge transformation which introduces ${\bf{r}}_0$ as the new gauge center for the kinetic energy ${\mathcal{T}}_1$.
An alternative equivalent path which has been outlined and followed in Refs. \cite{Avron78,Baye92} is to
apply a momentum-dependent unitary transformation that results in a decentered Coulomb potential.
Performing the above expansion up to
the second order and employing the conditions in Eq. (\ref{extrema}), we arrive at the following form ${\mathcal{W}}$ for 
our approximation to the generalized potential ${\mathcal{V}}_1$ close to its outer minimum

\begin{equation}
{\mathcal{W}} = \left(\frac{B^2}{2M} + \frac{1}{x_0^3} \right) x^2 + \left( \frac{B^2}{2M} - \frac{1}{2x_0^3} \right) y^2
-\left(  \frac{1}{2x_0^3} \right) z^2 + C \label{hamapproxV}
\end{equation}

with the constant $C = \left( \frac{2}{x_0} - \frac{B^2}{2M}  x_0^2  \right)$ and $(x,y,z)$ denote now
the elongation coordinates from the minimum ${\bf{r}}_0$. ${\mathcal{W}}$ is the potential
of a three-dimensional anisotropic charged harmonic oscillator.
The underlying
kinetic energy is provided by ${\mathcal{T}}_1$ in Eq.(\ref{H11}) i.e. our approximate Hamiltonian reads
${\mathcal{H}}_a = {\mathcal{T}}_1 + {\mathcal{W}}$. This is the Hamiltonian of a charged anisotropic
oscillator in a homogeneous magnetic field.
Since Eq. (\ref{pseudoextrem}) is fulfilled, all  frequencies belonging to the harmonic potential terms in Eq. (\ref{hamapproxV}) are positive: 

\begin{eqnarray}
\omega_x &=& \left( \frac{2}{\mu} \left( \frac{B^2}{2M} + \frac{1}{x_0^3} \right) \right)^{\frac{1}{2}}\\
\omega_y &=& \left( \frac{1}{\mu} \left( \frac{B^2}{M} - \frac{1}{x_0^3} \right) \right)^{\frac{1}{2}}\\
\omega_z &=& \left( \frac{1}{\mu} \frac{1}{x_0^3} \right)^{\frac{1}{2}}
\end{eqnarray}

The spectrum of this oscillator in the B field is harmonic
\cite{Dippel94}

\begin{eqnarray}
E_{n_+ n_- n_z}  \nonumber &=& C_0 + \left( n_+ + \frac{1}{2} \right) \omega_+ + \left( n_- + \frac{1}{2} \right) \omega_- \\
&+& \left( n_z + \frac{1}{2} \right) \omega_z
\end{eqnarray}

with $C_0 = \left( \frac{2}{x_0} - \frac{B^2}{2M}  x_0^2 + \frac{1}{2M} K^2  \right)$ and the normal
mode frequencies read

\begin{eqnarray}
\omega_{+,-} = \frac{1}{\sqrt{2}} \left( \omega_x^2 + \omega_y^2 + \omega_c^2 \pm 
\sqrt{ \left( \omega_x^2 + \omega_y^2 + \omega_c^2 \right)^2 - 4 \omega_x^2 \omega_y^2} \right)^{\frac{1}{2}}
\label{hogds}
\end{eqnarray}

with the cyclotron frequency $\omega_c = \frac{-eB}{\mu^{\prime}}$. The energetically low-lying GDS
in the outer potential well being well-approximated by the eigenstates of the charged anisotropic harmonic
oscillator in a magnetic field are therefore inherently different from the Rydberg states in the Coulomb
well. In particular, they possess a huge dipole moment along the electric field axis
which is of the order of $-ex_0$. We remark that for large values of $K$,  the ionization 
threshold is much closer to the energy of the minimum of the outer well than to the corresponding energy
of the saddle point.
\begin{figure}
\begin{center}
\includegraphics[width=8cm,angle=0]{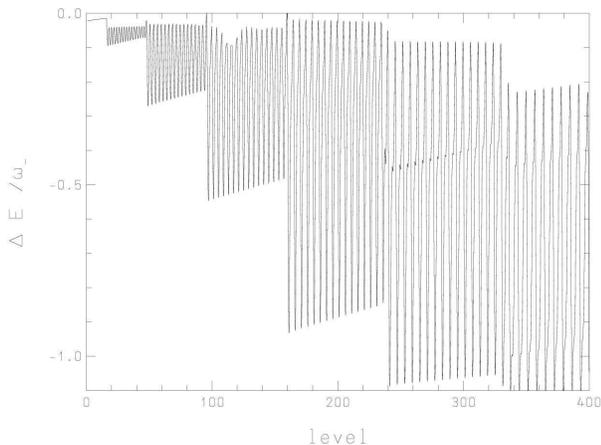}
\end{center}
\caption{Energy difference between the eigenvalues of the anisotropic charged harmonic oscillator in a magnetic
field and the exact eigenenergies of a hydrogen atom in crossed electric and magnetic fields in the outer potential
well in units of $\omega_-$ as a function of the energy level. Parameter values are $B=10^{-5}$a.u.
and $K=0.6$ a.u. From \cite{Dippel94}.} \label{spectrumgds}
\end{figure}

To quantify the deviation from the harmonic spectrum with increasing degree of excitation in the outer
well Fig.\ref{spectrumgds} shows the energy difference of the eigenvalues according to Eq.(\ref{hogds})
and the exact eigenenergies of the GDS of the hydrogen atom in the outer potential well
in units of $\omega_-$.
This difference grows stepwise while neighboring states show very different deviations from the harmonic
approximation. To explain these features let us look e.g. at the energy level 331. We see that the difference
between exact and and approximated energies for this level is much larger than for the levels below 331.
The level 331 has the quantum numbers $n_+=n_-=0$ and $n_z=10$, i.e., the quantum number $n_z=10$ appears for the
first time. Looking at higher levels, there are maxima of energy differences $\Delta E/\omega_-$ every $11$th level
above 331 up to level 397. For these levels (342,353,etc.), the quantum number $n_z=10$ and $n_-=1,2,...$ Between
two levels with $n_z=10$ there are levels with $n_z<10$, and apparently the energy difference for these is smaller.
Therefore, the difference between harmonic and exact energies is mostly determined by the quantum number $n_z$.
Hence, the anharmonicity of the exact potential is most pronounced in the $z$-direction. This can also be seen
in perturbation theory for higher terms of the expansion of the Coulomb potential where the major contributions
to the energy corrections are due to those terms containing high powers of $z$.

In view of the novel properties of the GDS, schemes for their experimental realization are also interesting. 
There have been experiments that indicate the existence of atoms with very large dipole moments
in crossed electric and
magnetic fields \cite{Fauth87,Raithel93} for energies above the saddle-point energy. These were performed in
comparatively weak fields, $B=10^{-6}$a.u., and the resulting estimate for the dipole moment of the Rydberg atoms was
$1.9 \times 10^4$Debye. This is roughly what one should expect for the value of the dipole moment for saddle point states.
The experimental technique employed to determine the atomic dipole moment was to add a slight inhomogeneity
to the electric field and to measure the deflection of the atoms, this deflection being proportional
to the field inhomogeneity and the atomic dipole moment.

Finding deeply-bound GDS is a large experimental challenge. Starting from the ground state,
this requires a transition via Rydberg states to GDS \cite{Averbukh99}.
Alternatively one could think of starting with
free particles in the continuum and forming GDS via three-body recombination in crossed fields,
which is a typical process occuring in ultracold dense plasmas (see section \ref{sec:maryga}). 
The key idea to 'transport' the quantum states of the atom from the centered ground to the decentered GDS
is to combine a laser excitation to a Rydberg state in the Coulomb well, with
a two-step switching procedure of an external electric field which adiabatically transfers the quantum states
to a narrow distribution of strongly bound GDS. This scheme is based on the switching of
an external electric field whereas the corresponding motional electric field should be negligible, as is the
case e.g. in an ultracold environment. Following the laser excitation to a Rydberg state in the Coulomb well
this switching procedure is provided by the time-dependent potential

\begin{equation}\label{switchgds}
V_s = - \frac{e^2}{\sqrt{x^2+y^2+z^2}} + \frac{e^2 B^2}{2M} \left( x^2 + y^2 \right) - e E(t) x
\end{equation}

where 

\[ E(t) = \left\{
\begin{array}{ll}
E_c sin \left( \frac{\pi t}{2 t_1} \right), &~~~~ t \le t_1 \\
E_c, &~~~~t_1 \le t < t_2 \\
E_c + \left( E_f-E_c \right) sin \left( \frac{\pi (t-t_2)}{2 (t_f - t_2)} \right), &~~~~ t_2 \le t \le t_f \\
E_f, &~~~~t_f < t .
\end{array} \right. \]

The value of the intermediate electric field $E_c = 10^{-8} a.u. \approx 5656 V/m$ is reached
after $t_1 = 2.3 \times 10^{8} a.u. \approx 5.6 ns$. The final value of the external electric
field is $E_f = 2.5 \times 10^{-8} a.u. \approx 12855 V/m$. This value is reached at 
$t_f = 9.2 \times 10^{9} a.u. \approx 222 ns$ after a delay of $t_2-t_1 = 5.7 \times 10^{8} a.u. \approx 13.8 ns$.
The period $t_2-t_1$ of a constant
field $E_c$ as well as the subsequent second switching process ending with the final value $E_f$ of the
electric field are chosen to achieve as narrow as possible a final state distribution in the
outer potential well. A single switch to achieve the transfer from the Rydberg to the GDS,
would end up in a broad final distribution of the populated states. 

\begin{figure}
\begin{center}
\includegraphics[width=9cm,keepaspectratio,angle=0]{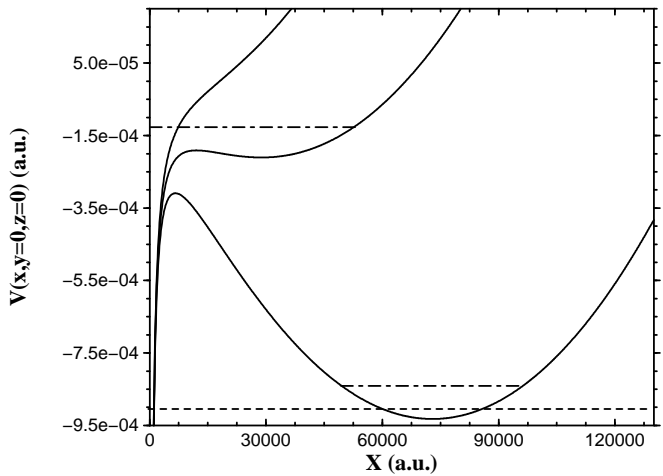}
\end{center}
\caption{Potential energy of the relative motion along the external electric field direction. The upper curve corresponds
to zero electric field $[E(t=0)=0]$. The curve with a shallow well corresponds to $E=E_c (t_1 \le t < t_2)$. The curve
with a deeper well corresponds to $E(t=t_f)=E_f$. The mean energy of the trajectory ensemble at times $t=t_2$ and
$t=t_f$ is shown by dashed-dotted lines joining the turning points of the appropriate potential energy curves. The 
ionization threshold at $E=E_f$ is shown by a dashed line. From \cite{Averbukh99}.} \label{potentialcurvesgds}
\end{figure}

The initial ($t=0$) potential consists exclusively of a diamagnetic and 
Coulomb potential terms (see Eq. (\ref{switchgds}) and Fig. \ref{potentialcurvesgds}).
To simulate the dynamics in the course of the switching process we choose an
ensemble of trajectories according to a single chaotic trajectory 
for a typical energy of a Rydberg state in the (quantum-)chaotic regime thereby ensuring a
random sampling of the energy shell. The first switching-on process from zero to the
electric field strength $E_c$ within the time interval $0< t \le t_1$ leads to the formation of a 
Stark saddle point and a very shallow outer well, as Fig. \ref{potentialcurvesgds}.

\begin{figure}
\begin{center}
\includegraphics[width=9cm,keepaspectratio,angle=0]{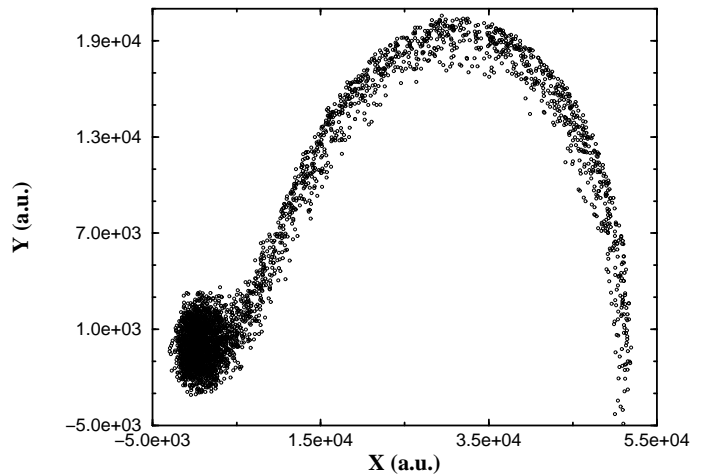}
\end{center}
\caption{Trajectories projected on the $(x,y)$ plane at time $t=t_2$. From \cite{Averbukh99}.} \label{gdstraj1}
\end{figure}

\begin{figure}
\begin{center}
\includegraphics[width=9cm,keepaspectratio,angle=0]{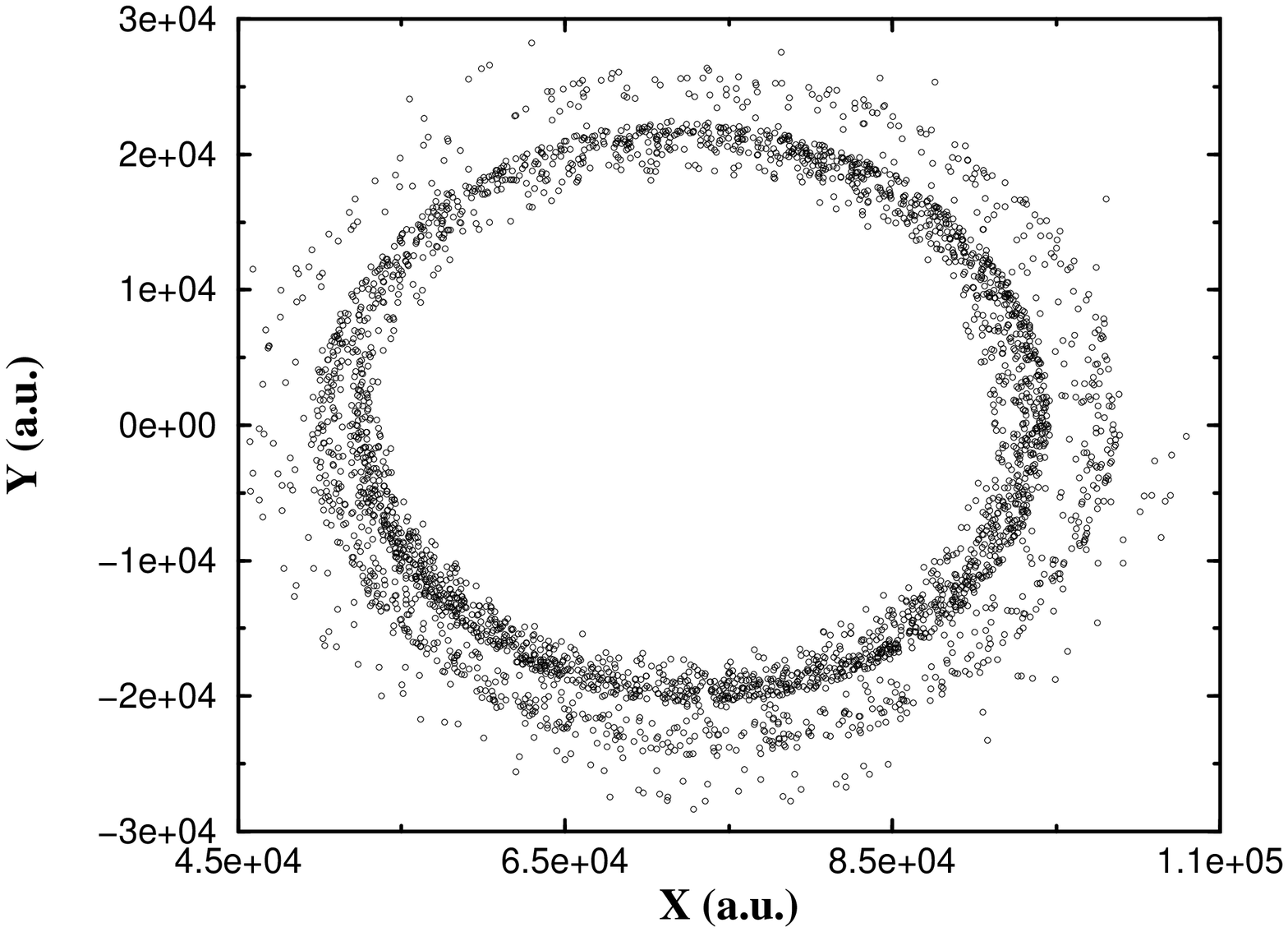}
\end{center}
\caption{Trajectories projected on the $(x,y)$ plane at time $t=t_f$. From \cite{Averbukh99}.} \label{gdstraj2}
\end{figure}

\begin{figure}
\begin{center}
\includegraphics[width=9cm,keepaspectratio,angle=0]{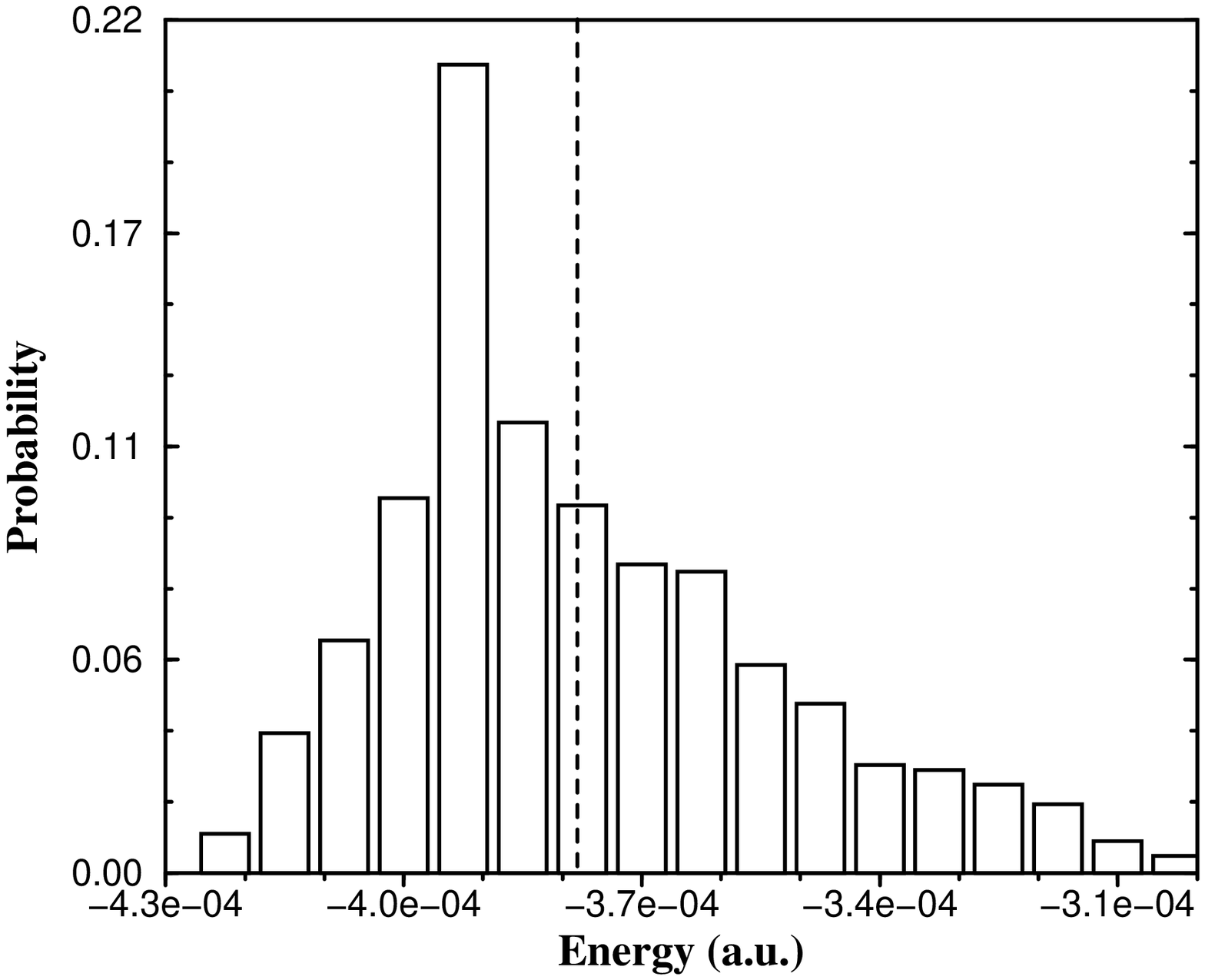}
\end{center}
\caption{Probability distribution as a function of energy in the outer well at $t\ge t_f$
for $B = 6 \times 10^{-5}$. The ionization threshold is shown by a vertical dashed line. From \cite{Averbukh99}.} \label{gdstrajbound}
\end{figure}

One needs to keep the electric field constant for a period of time long enough to enable most of the
atoms to cross the Stark saddle once. 
The positions of trajectories in the $(x,y)$ plane just before the second switching on are shown in Fig. 
\ref{gdstraj1}. A substantial part of the trajectories has passed to the outer well by that time.
They are seen as points scattered around an outer well trajectory. 
Only a tiny fraction of trajectories lead to ionization in the course of the switching on process.
Then one can increase the electric field up to its final value (see Fig.\ref{potentialcurvesgds}).
In such a procedure, the value of $E_c$ determines the most probable final energy in the outer well.
The second switching on has to be sufficiently slow in order to reduce the broadening of the energy
distribution during the trapping of the relative motion in the outer well. Quantum mechanically speaking,
such a broadening occurs due to transitions between the adiabatic outer well states. It cannot be
fully avoided since in the course of trapping the relative motion crosses the separatrix between the
energies above and below the Stark saddle \cite{Clary93}. As shown in Fig. \ref{gdstraj2}, all nonionizing
trajectories have been brought to the outer well at time $t=t_f$. Whether the finally resulting trajectories 
are truely bound ones in the outer well depends sensitively on the parameters chosen. Fig. \ref{gdstrajbound}
shows the probability distribution as a function of energy for the final ensemble for 
$B = 6 \times 10^{-5}$a.u. It represents a distribution with a single dominating peak and a substantial
portion of the energy being below the ionization threshold. The peak is due to the trajectories
which had crossed the Stark saddle before the second switching on phase of the electric field began. 
The trajectories that passed to the outer well during the action of the second field switch show up in the energy
distribution as a long tail for large energies.

\subsubsection{Multi-electron atoms}

We start by discussing the generalized potential in Eq. (\ref{gaugeinvV}) for a multi-electron atom.
Apart from the trivial constant ${\mathcal{K}}^2/2M$, it gives rise to a motional electric field term
$e/M ({\bf{B}} \times {\mathcal{K}}) \sum {\bf{r}}_i$ and a diamagnetic term $e^2/2M ({\bf{B}} \times \sum {\bf{r}}_i)^2$.
The relevant quantity occuring in the latter two potential terms is the electronic center of mass (ECM),
i.e. ${\bf{R}}_e= \frac{1}{N}\sum {\bf{r}}_i$ in the internal coordinate frame.
It is therefore the ECM which experiences interactions
beyond the Coulomb potential and which enters the generalized potential for multi-electron
systems. In case of one-electron systems the above potential reduces to the one derived in
\cite{Dippel94} and in particular the ECM reduces to the coordinate vector of the single electron.

Let us focus on doubly excited two-electron systems. Since major parts of the generalized potential ${\mathcal{V}}$
depend only on the ECM and since both electrons are assumed to be highly excited it is natural to
introduce the ECM as a new coordinate vector. Additionally, we require that the kinetic energy should become
as simple as possible which leads to the relative vector of the two electrons as a good choice for the
second coordinate vector, i.e. ${\bf{R}}_e=({\bf{r}}_1+{\bf{r}}_2)/2$; ${\bf{r}}={\bf{r}}_1-{\bf{r}}_2$.
The transformed Hamiltonian ${\mathcal{H}}_2={\mathcal{T}}_2 + {\mathcal{V}}_2$ therefore decomposes into
\begin{eqnarray}
{\mathcal{T}}_2 \nonumber &=&
\frac{1}{2\mu_2} \left( {\bf{P}}_e -e \frac{\mu_2}{\mu_2^{\prime}} {\bf{B}} \times {\bf{R}}_e \right)^2
+\frac{1}{m} \left( {\bf{p}} -\frac{e}{4} {\bf{B}} \times {\bf{r}} \right)^2
\label{eq2eT}  \\
{\mathcal{V}}_2  \nonumber &=&  \frac{1}{2M} \left( {\bf{K}} -2e {\bf{B}} \times {\bf{R}}_e \right)^2
+ \frac{e^2}{|{\bf{r}}|} \\ &-& Ze^2 \left[ \frac{1}{|{\bf{R}}_e-\frac{1}{2}{\bf{r}}|}
+ \frac{1}{|{\bf{R}}_e+\frac{1}{2}{\bf{r}}|} \right] - 2e {\bf{E}} {\bf{R}}_e \nonumber
\label{eq2eV}
\end{eqnarray}
where $\mu_2=\frac{2mM_0}{M},\mu_2^{\prime}=\frac{2mM_0}{M_0-2m}$. 
As can be seen from Eqs. (\ref{eq2eT}), the coordinate transformation decoupled the kinetic energy terms
belonging to the two electronic  vectors and  simplified the field-dependent 
potential terms in Eq. (\ref{eq2eV}).

\begin{figure}
\begin{center}
\includegraphics[width=9cm,keepaspectratio,angle=0]{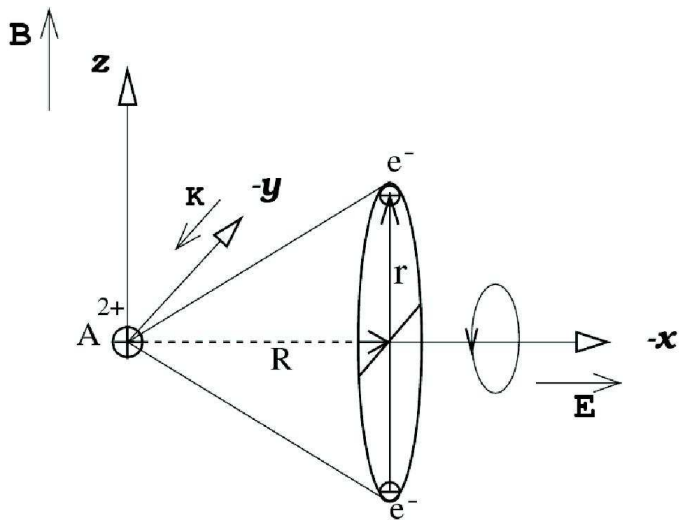}
\end{center}
\caption{Shown is a sketch of the geometrical configuration of the giant dipole
two-electron resonances. The electric and magnetic field vector point along the negative x- and positive z-direction,
respectively. ${\bf{R}}_e$ is the electronic center of mass coordinate and {\bf{r}} the relative
coordinate vector of the electrons. The big circular loop lying in the $yz-$plane
indicates the geometrical position of all extremal configurations. From \cite{Schmelcher01}.} \label{gdsreso1}
\end{figure}

Doubly excited configurations corresponding to the extrema of the six-dimensional potential ${\mathcal{V}}_2({\bf{R}},{\bf{r}})$
in Eq.(\ref{eq2eV}) are good candidates for resonances.
The roots of the six nonlinear coupled equations ${\partial{\mathcal{V}}_2}/{\partial{{\bf{r}}}}=0$ and
${\partial{\mathcal{V}}_2}/{\partial{{\bf{R}}}}=0$ are therefore of immediate interest.
Without loss of generality, we assume in the following again
that the magnetic and electric field vectors point along the positive $z-$axis and negative $x-$axis, respectively.
As previously indicated the Stark term due to the external electric field can be taken into account by redefining the
value of the pseudomomentum. This yields the following geometrical conditions
\begin{eqnarray} \nonumber
\left({\bf{r}}{\bf{R}}_e\right)&= &0 \hspace*{0.5cm} Y_e=Z_e=0 \hspace*{0.3cm};\hspace*{0.3cm}
R_e=\frac{1}{2} \sqrt{3} r \\ 
P(X_e) \nonumber &=& X_e^3+\left(\frac{K}{2B}\right)X_e^2 - \frac{3}{8\sqrt{3}} \frac{M}{B^2} \\ &=& 0 \hspace*{0.3cm} (X_e<0)
\label{extremaconditions}
\end{eqnarray}
where ${\bf{R}}_e=(X_e,Y_e,Z_e)$, ${\bf{r}}=(x,y,z)$ and $r=|{\bf{r}}|, R_e=|{\bf{R}}_e|$.
Accordingly, ${\bf{r}}\cdot{\bf{R}}_e=0$,
the ECM and the interelectronic coordinate vector are orthogonal. Since $Y_e=Z_e=0$, this leads to $x=0$.
Furthermore, the condition $R_e=\frac{1}{2}\sqrt{3}r$ and equally $|{\bf{R}}_e-\frac{1}{2}{\bf{r}}|=
|{\bf{R}}_e+\frac{1}{2}{\bf{r}}|$ leads to the fact that the two electrons and the nucleus form an equilateral
triangle. The remaining nonzero coordinate $X_e$ has to fulfill the corresponding polynomial equation $P(X_e)=0$
in Eq.(\ref{extremaconditions}). This completes the specification of the
extremal configurations which are located on a one-dimensional circular manifold. The electrons form a decentered triangular
configuration and are highly correlated through the fact that they are forced to stay on opposite sides
of a circle. The geometry of the extremal configuration described by the above conditions is illustrated in Fig.
\ref{gdsreso1} in which the circular extremal line as well as the opposite electrons are indicated.
Both electrons are for laboratory field strengths located far from the nucleus, the electron-nucleus
distance scales with $\propto \frac{1}{B}$.
The position of the extrema will in the following be denoted by ${\bf{r}}_0,{\bf{R}}_{e0}$.

If the inequality $K^{3} > \frac{81}{4} \sqrt{3} MB$ is fulfilled, $P(X)=0$ has two real solutions on the negative $x-$axis
(the decentring direction of the atom).
The smaller of these two values (excluding the sign)
corresponds to a maximum (saddle) of the intersection of the potential ${\mathcal{V}}_2$
along the $X-$direction, whereas the larger value yields an outer minimum. Of particular interest is, of course, the
case where the ECM is captured in the outer minimum. To investigate stability criteria for the case when the ECM is located
in the outer well, a normal mode analysis is necessary in the presence of the external field. 
The normal mode analysis
using the second order expansion of ${\mathcal{V}}_2$ around the extremal positions 
yields the following eigenvalue problem
for the harmonic frequencies $\Omega_i$, as the energies of the doubly-excited resonances in
crossed fields
\tiny
\begin{eqnarray*}
\left(- \Omega_i^2 \right) {\bf{V}}_i =
{\it{F}} \times \left( \begin{array}{cccccc}
\frac{2|X_0|}{m} & \frac{-iB\Omega_i}{mF} & 0 & 0 & 0  & -\frac{4|X_0|}{m\sqrt{3}} \\
\frac{iB\Omega_i}{mF} & 0 & 0 & 0 & 0  & 0 \\
 0 & 0 & -\frac{2|X_0|}{m} & -\frac{4|X_0|}{m\sqrt{3}} & 0 & 0 \\
0 & 0 & -\frac{2|X_0|}{\mu_2\sqrt{3}} & -\frac{D_X}{\mu_2} & -\frac{2iB\Omega_i}{\mu_2^{\prime}F} & 0 \\
0 & 0 & 0 & \frac{2iB\Omega_i}{\mu_2^{\prime}F} & -\frac{D_Y}{\mu_2} & 0 \\
-\frac{2|X_0|}{\mu_2\sqrt{3}} & 0 & 0 & 0 & 0 & -\frac{D_Z}{\mu_2}
\end{array}
\right) {\bf{V}}_i
\label{egvp}
\end{eqnarray*}
\normalsize
where we have assumed the specific case $y=0, z \ne 0$ and $F=27 \sqrt{3} / 32 |X_0|^4$ with
\begin{eqnarray}
D_{X} &=& \frac{32 |X_0|^4}{27 \sqrt{3}} \left( \frac{4B^2}{M}-\frac{15 \sqrt{3}}{8|X_0|^3}\right)
\hspace*{0.3cm} \nonumber \\  D_{Y} &=& \frac{32 |X_0|^4}{27 \sqrt{3}} \left( \frac{4B^2}{M}+\frac{3
\sqrt{3}}{2|X_0|^3}\right)\nonumber \\  D_{Z} &=& \frac{4|X_0|}{9}
\label{Ds}
\end{eqnarray}


and ${\bf{V}}_i=(V_1,...,V_6)_i$ are the six-dimensional eigenvectors.
The reader notes that the
matrix on the r.h.s. of Eq. (\ref{egvp}) depends explicitly on $\Omega_i$ which
is due to the appearance of ${\mathcal{T}}_2$ in Eq. (\ref{eq2eT}) in the presence of
the magnetic field.  

The frequency spectrum reads as $E = \sum\limits_{i=1}^{6} \Omega_i (N_i+\frac{1}{2}) + {\mathcal{V}}_2({\bf{r}}_0,{\bf{R}}_0)$
An analysis of  $\Omega_i$ finds that the two largest frequencies are almost degenerate
and are of the order of half the electronic cyclotron frequency $\omega_c$. We call these modes
in the following cyclotron modes. Two of the remaining three frequencies belong to the motion parallel to
the magnetic field which is governed exclusively by the Coulomb interaction, the so-called Coulomb modes.
The remaining frequency is due to the heavy particle dynamics i.e. the CM mode.
The frequencies $\Omega_i$ are different by several orders of magnitude and moreover they are
real for typical laboratory field strengths.
We therefore encounter no decay within our harmonic analysis of the corresponding resonances which indicates that they
should possess a significant life time.

\begin{figure}
\begin{center}
\includegraphics[width=9cm,keepaspectratio,angle=0]{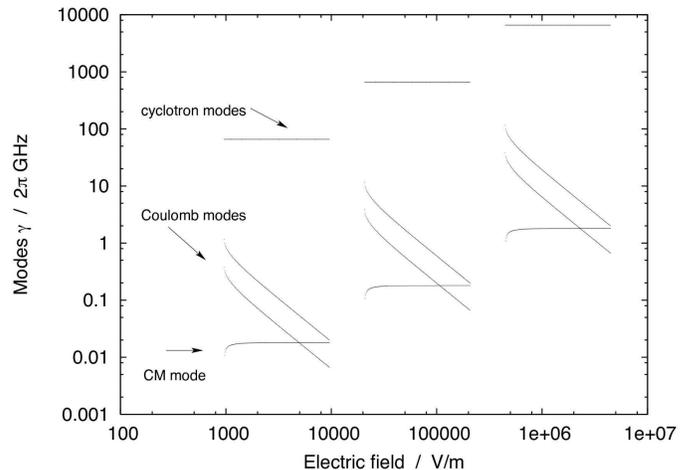}
\end{center}
\caption{Eigenmodes $\{\gamma_{\rho}\}_{\rho}$ for $N=2$ electrons as a
function of the electric field $E\equiv BK/M$ over the range $K/K_{cr}\in[1,10]$
(all modes are imaginary). The three sets of curves refer to 
$B=10^{-5},10^{-4},10^{-3}$ a.u. (from left to right). Each top horizontal
line represents two degenerate cyclotron modes. Below, the
two Coulomb modes fall off quickly and intersect the CM mode (the
nearly horizontal line about four orders below the cyclotron modes).From \cite{Schmelcher01}.}
\label{gdsreso2}
\end{figure}

Fig.\ref{gdsreso2} shows the dependencies of the five nonzero frequencies 
on both the electric as well as the magnetic field strengths. With increasing magnetic field strength
all frequencies increase. For those frequencies which are associated with the cyclotron and CM modes
this behaviour is evident. For the frequencies associated 
with the Coulomb modes parallel to the magnetic field it is a consequence of the fact that the position
$|X_{0}|$ of the outer minimum decreases strongly with increasing field strength. The Coulomb potential
becomes then stronger and the frequency in the corresponding well parallel to the field increases.
As can be seen in Fig.\ref{gdsreso2} the dependence of the frequencies on the electric field strength is twofold:
The frequencies associated with the cyclotron and CM mode show in general
only a very weak dependence on the electric field strength whereas the frequencies associated with the Coulomb modes
generically exhibit a strong dependence on the electric field strength (see Fig. \ref{gdsreso2}).

The above normal mode analysis of the giant dipole resonances provides evidence
that the two-electron case is locally stable, apart from the singular horizontal configuration.
Beyond this analysis, recently a numerical ab initio study of the resonances has been performed 
\cite{Zoellner05a,Zoellner05b}.  Such a six-dimensional resonance study
is both methodologically and computationally  demanding,
especially in view of the fact that our system is governed by dramatically different
time scales: The anticipated order of magnitude is a few picoseconds (ps) for the
electronic cyclotron motion in strong magnetic fields ($B \approx 10^{-4}$a.u.)
and a few thousand up to a few ten thousands ps for the motion parallel
to the magnetic field (the typical time scale for the CM mode is a few thousand
ps). To meet the requirements the multiconfiguration time-dependent
Hartree (MCTDH) method \cite{Meyer90,Beck00,Meyer03,Worth02} has been employed,
known for its outstanding efficiency in high-dimensional
applications.  The MCTDH approach is a multifunctional multidimensional wave packet propagation method which
allows to investigate both time-independent as well as time-dependent problems.
Applying it to a fermionic system like ours finds its sole justification in
the fact that the spatial separation between the electrons is so large
that they are virtually distinguishable.
\begin{figure}
\begin{center}
\includegraphics[width=9cm,keepaspectratio,angle=0]{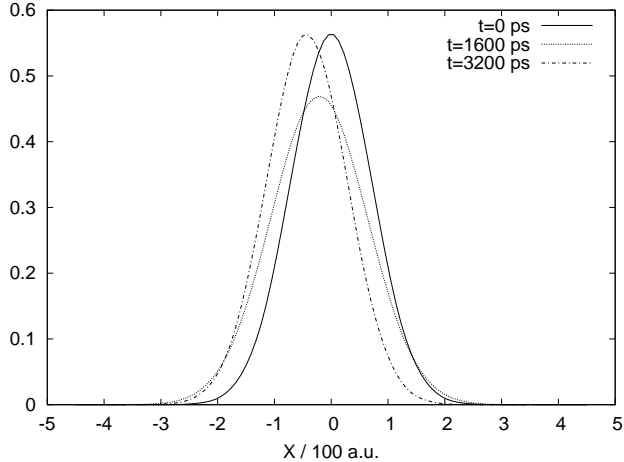}
\end{center}
\caption{The motion in $X$ in the case $K/K_{cr}=1.1$: 
Snapshots of the motion of the one-particle density $\rho_{X}$, reflecting oscillations of
both center and shape of the wave packet. From \cite{Zoellner05b}.} \label{gdsresnum1}
\end{figure}

Both the stability as well as the spectral properties of the giant dipole two-electron
resonances have been studied this way. Moreover, following the evolution of wave packets
with an initial displacement from the extremum, the robustness of the resonances has been
tested \cite{Zoellner05a,Zoellner05b}. For $K=1.1 K_{cr}$, i.e. very close to the critical
point of the existence of the outer well ($K_{cr}=\frac{81}{4} \sqrt{3} MB$), the observed
motion display an instability with respect to some of the degrees of freedom.
The original wave packet (see Ref.\cite{Zoellner05a,Zoellner05b} for its preparation)
shows minor oscillations with respect to the degrees of freedom $X,Y$ which are essentially
due to an oscillation of its center and shape (see Fig.\ref{gdsresnum1}). The relative stability in these two directions
is readily explained in terms of the generalized potential. Apart from a stabilization
by the $\mathbf{B}$ field, they experience an additional confinement
via the quadratic term of the potential ${\mathcal{V}}_2$ in Eq.(\ref{eq2eV}). However, it is
evident that the packet broadens over time, thus slowly delocalizing. Even though the
relative motion $(x,y)$ perpendicular the magnetic field is also gyrationally stabilized,
there is no confining term for it in the generalized potential ${\mathcal{V}}_2$ as for the ECM.
The $y$ direction is, for the vertical configuration considered here, the zero mode, while
the cut through the $x$ direction refers to a maximum. In this light,
we cannot expect the system to be arbitrarily stable in these two
degrees of freedom. This is a general fact, but for $K=1.1K_{cr}$
it is very pronounced. The reason being, near the extremum, the Coulomb
interaction (the source of these instabilities) becomes negligible
for higher $K$. For the case $K=1.1K_{cr}$ the zero mode spreading is substantial on a time
scale of a few thousand ps. The reduced density for $Z$ broadens
and continually leaks outward. Although in the $Z$ direction,
the particle lives in a fairly harmonic well it is not only lifted in energy to way above the bottom of the well,
but also affected by the instability in $x$ via coupling. Turning
to the last degree of freedom, $z$, an overall stable behavior is obtained.

\newcommand{\exv}[1]{\langle#1\rangle}

Considering the cases of larger $K$-values specifically $K= 2 K_{cr}, 10 K_{cr}$ the vertical
configuration is found to be stable on a time scale of $10 ns$ and $100ns$, respectively,
whereby the latter value was the maximum propagation time for $K= 10 K_{cr}$.
A closer look at the response of the system for $K = 2 K_{cr}$ upon displacing
$Z$ and $z$ by $2.000\,\mathrm{a.u.}$ unveils that for the excited
degrees themselves, $Z$ and $z$, the wave packet is simply reflected
between two positions $\pm\exv{z}_{0}$ with minor ($Z$) or more
pronounced ($z$) deformations and smearing-out due to competing modes.
The most interesting question may be the effect on the $(xy)$ modes. In fact, $x$ is rendered
slightly unstable by the excitation of the parallel motion. On the
time scale of several ten thousands of ps the wave packet slowly
but inevitably starts leaking to outer regions. For the case $K = 10 K_{cr}$
it is only the $y-$motion associated with the zero mode which shows
a minute broadening being the first sign of a possible decay which is
extrapolated to take place for time scale of many microseconds.

Thus far, we have focused on the vertical two-electron configuration,
which is defined by the alignment of the interelectronic vector with
the $z-$axis. Let us now briefly comment on the impact of rotations
of the two-electron configuration around the $x-$axis. The horizontal configuration,
which corresponds to an alignment along the $y-$axis indeed adds
instability, which is discernible even for very high $K$, if less
distinct. The in-between diagonal configuration is partly unstable on a timescale
comparable to that of the vertical case.
Finally, for more than two-electron atoms, the reader is encouraged to follow the discussion in Ref.
\cite{Zoellner05a,Zoellner05b}.

\subsection{Matter-antimatter systems in crossed fields} \label{matantimat}

The lowest order decay rate for positronium $\Gamma = \sigma v \rho$ in field-free space, where
$\sigma$ is the plane-wave cross section for free pair annihilation, $v$ is the relative velocity  of
the electron and positron, and $\rho$ is the square of the wavefunction evaluated at contact,
amounts to a tenth of a nanosecond (parapositronium) or roughly 140 nanoseconds (orthopositronium) \cite{Stroscio75,Adkins83}.
In view of the transition from a single to a double well structure of the potential ${\mathcal{V}}$ and
${\mathcal{V}}_1$ (see Eqs.(\ref{gaugeinvV},\ref{seham}),respectively)
with increasing (motional and/or external) electric field strength and the associated emergence of
giant dipole states, it is an intriguing question to investigate scenarios 
for a combined decentered matter-antimatter system \cite{Ackermann97,Shertzer98,Schmelcher98b,Shertzer02}.
Does the intuitive picture of putting
one particle into the outer well and the other one to the Coulomb well lead to an enhanced stability
of these microscopic matter-antimatter systems? This seems to be a natural conjecture particularly in view of the
separation of the two wells by a wide and high potential barrier for sufficiently strong fields.

Let us begin by inspecting the underlying Hamiltonian for a positronium atom $e^+/e^-$ in crossed
electric and magnetic fields

\begin{equation}
{\mathcal{H}}_p=\frac{1}{m}{\bf p}^2 + \frac{1}{4m} ({\bf K} + e {\bf B} \times {\bf r})^2
- \frac{e^2}{r} - e {\bf E} \cdot {\bf r} \label{posham}
\end{equation}

\begin{figure}
\begin{center}
\includegraphics[width=8.6cm,keepaspectratio,angle=0]{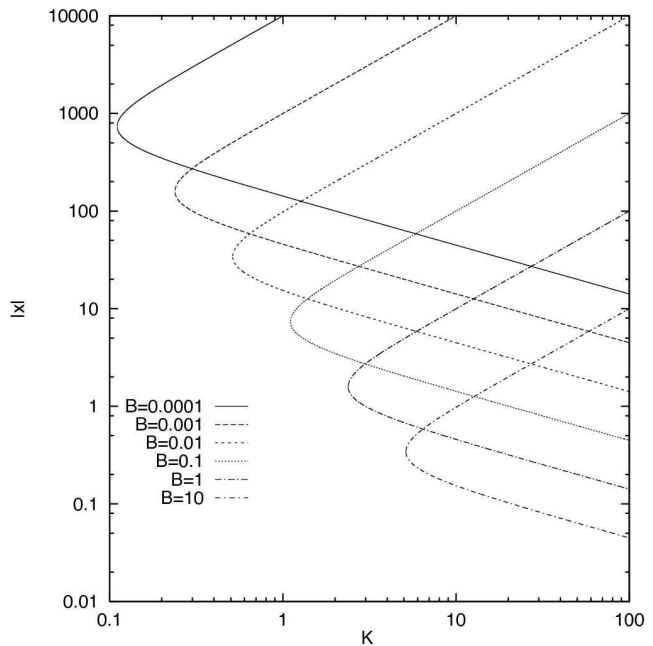}
\end{center}
\caption{The location of the saddle point $x_s$ and the outer
well minimum $x_o$  as a function of the pseudomomentum $K$  for various values
of the magnetic field strength $B$. The lower side of each curve is the
value of $x_s$;  $x_s(K\rightarrow\infty)\rightarrow 0$.
The upper side of each curve is the value of $x_o$;  $x_o(K\rightarrow\infty)\rightarrow -\infty$.
The two segments of the curve coalesce at the critical point $x_c=x_s(K_c)=x_o(K_c)$.
All quantities are in atomic units. From \cite{Shertzer98}}\label{minsa-k}
\end{figure}

We observe that because for Ps, $\mu=\frac{m}{2}$, and $\mu' = \infty$, the kinetic energy becomes the
free-particle kinetic energy, see Eq. (\ref{gaugeinvT},\ref{seham}).
Second, we encounter a generalized potential term 
$\frac{1}{4m} ({\bf K} + e {\bf B} \times {\bf r})^2$ which scales according to $\propto \frac{1}{m}$.
This is in contrast to the hydrogen case (see Eq.(\ref{seham})), where the scaling is $\propto \frac{1}{M}$. 
However, because the 
minimum of the CM kinetic energy i.e. of the generalized potential excluding the Coulombic potential is
independent of the total mass and the condition for the existence of the outer well $K^3 > \frac{27}{2}Bm$
is indicative that the formation of the well takes place for much smaller values of the pseudomomentum
and/or electric field strength for Ps atom, much smaller sizes of the GDS 
are possible.
Typically GDS of positronium can be of the size of a few
thousand \AA~(see Fig.\ref{minsa-k})~whereas GDS of hydrogen amount to many ten thousands \AA.

In Fig.~\ref{minsa-k}, 
we show  $|x_o|$ and $|x_s|$ for different values of the B field.
The position of the minimum, $x_0$, increases and the position $x_s$ of the saddle point
decreases with increasing value of the pseudomomentum (or the electric field strength), 
common to both the hydrogen and positronium atoms. At $K=0.4, B= 5 \times 10^{-5}$a.u. the ionization
energy is much closer to the bottom of the well than to the saddle point. The barrier width
at the ionization energy is over $7000$ a.u. This suggests that the bound state spectrum of the exact Hamiltonian
should be nearly identical to the spectra from the two isolated potential wells: as set of Coulomb states
localized at the origin and a set of outer well states centered around the minimum.

The validity of this isolated potential approach depends on the amount of tunneling through the barrier, or in other
words, the amount of mixing between the Coulomb states and the outer well states. If the barrier
is sufficiently high and wide, the eigenstates of the the two isolated wells will not mix.
The probability density will reach nearly zero near the middle of the barrier. If positronium
is in an outer well state, the electron (or positron) cannot tunnel through the barrier and
annihilate. An estimate for the tunneling probability from the outer well ground state into the Coulomb well, can be 
obtained from
\begin{equation}
P= \exp \biggl [-2 \int_{x_1}^{x_2} \; \vert p(x) \vert \; dx \biggr ]
\end{equation}
where  $x_1$ and $x_2$ are the classical turning points and $|p(x)| = \sqrt{V(x,0,0) - (\omega_x/2+C) }$.
Lifetimes for tunneling are expected to be of the order of some characteristic time $\tau \approx 2\pi/\omega_x$ divided
by the probability $P$. For $B=5\times 10^{-5}$ and $.0878 \ge K \ge .0908$,
the harmonic ground state lies above $V_s$; there are no outer well states, only Coulomb states and
saddle states which have  probability in both wells.  (Saddle states can occur any time the saddle point
lies  below the ionization energy; for certain choices of $B$ and $K$, all three types
of states - Coulomb, outer well, and saddle - can be found in the spectrum.)
The tunneling probability drops dramatically with increasing $K$ and the outer well state(s) of positronium
can be  considered stable for $K>0.1$.  The corresponding tunneling lifetime is greater than one year.
Finite element calculations fully confirm this simple estimate of the tunneling probabilities.
Giant dipole states of e.g. positronium offer therefore a unique pathway of preventing microscopic matter antimatter
systems from annihilation i.e. for their preparation in long-lived quasi stable states.

\subsection{The guiding center approximation} \label{gca}
In the limit of large magnetic fields the previous analysis can be greatly simplified within the so-called guiding center 
approximation \cite{GCA1,GCA2}.
The premise for this approximation rests on the decoupling of the different degrees of freedom in a strong B field due to their
widely separated frequencies. If the field is sufficiently strong, the cyclotron frequency $\omega_{\rm c}=eB/m$ is the largest dynamical frequency and the cyclotron radius $r_{\rm c}$ is the smallest length scale in the system. In this limit, the rapid cyclotron motion can be separated from the remaining degrees of freedom, which then describe the motion of the guiding center (see Eq.(\ref{guidingcenter})) of the electron.

\begin{figure}[tb]
\begin{center}
\includegraphics[width=7.5cm]{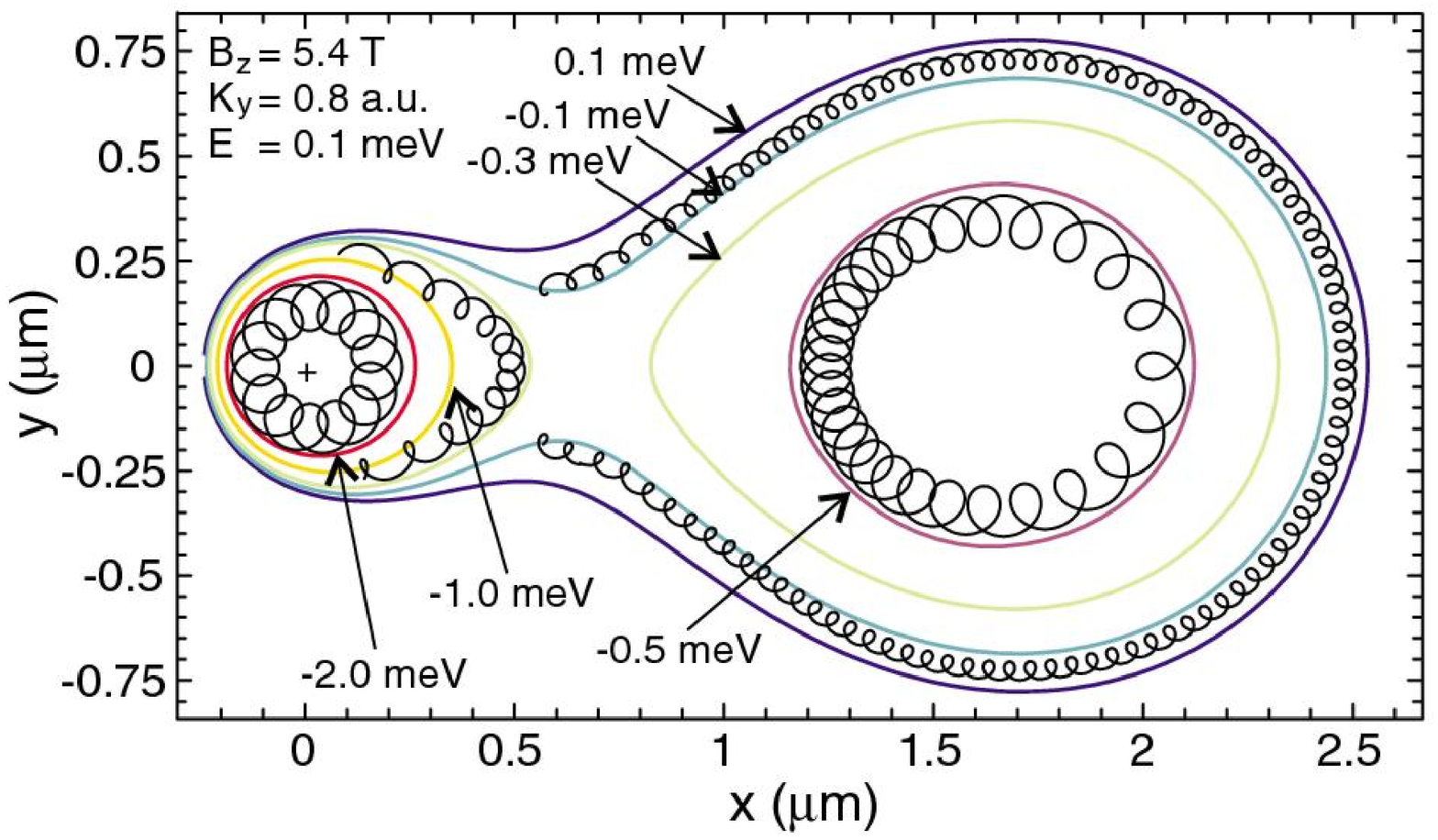}\hfill
\includegraphics[width=4.5cm]{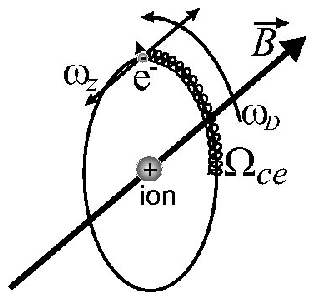}
\end{center}
\caption{Left: Classical trajectories of different bound electron trajectories for a finite pseudomomentum of $K_{\perp}=1$a.u.. Cyclotron radii have been increased and the cyclotron frequency decreased to make them visible. From \cite{vrinceanu}. Right: Schematics of a circular guiding center atom. In descending order of the frequencies, the electron performs cyclotron oscillations, oscillates axially along the field and executes ${\bf E}\times{\bf B}$ drift motion around the ion. From \cite{kuz04}.}\label{fig2_6_1}
\end{figure}

Fig. \ref{fig2_6_1} illustrates typical classical trajectories of guiding center atoms. The guiding center approximation holds when the electron's cyclotron radius is much smaller than the distance to the ion. This condition can be fulfilled at sufficiently weak binding and also holds for the giant dipole states (see section \ref{gidista}), characterized by strongly de-centered electron trajectories in the outer potential well. 
Such trajectories, however, are loosely bound to the parent ion, and are, thus, expected to have limited lifetimes in collisional environments 
such as cold plasmas. More deeply bound trajectories are confined to the central Coulomb well and follow regular orbits around the ion. A sketch of the particularly illustrative and simple case of a circular, nearly planar guiding center atom \cite{kuz04} is shown on the right 
in Fig. \ref{fig2_6_1}.
Confined by the ionic Coulomb field, the electron's guiding center oscillates along the direction of the magnetic field and performs an ${\bf E}\times{\bf B}$ drift motion around the ion. If the amplitude $z_{\rm m}$ of the longitudinal oscillations is smaller than the radius $\rho$ of the drift motion around the ion, the frequency of the field-aligned oscillations is approximately $\omega_z=\sqrt{e^2/(m\rho^3)}$, and the frequency of the ${\bf E}\times{\bf B}$ drift motion is given by $\omega_D=e/(B\rho^3)$. Requiring the cyclotron frequency to be 
the largest in the system also implies that $\omega_z\gg\omega_D$ and leads to the following hierarchy
\begin{equation}\label{frequ_hierarchy}
\omega_c\gg\omega_z\gg\omega_D\;.
\end{equation}
These inequalities are fulfilled for large radii of the drift motion, $\rho\gg(m/B^2)^{1/3}$. At the same time, the hierarchy (\ref{frequ_hierarchy}) implies that the cyclotron radius is much smaller than the drift radius $\rho$, which is fulfilled if the cyclotron energy $E_{\rm cyc}$ is less than or on the order of the potential energy $e^2/\rho$ \cite{kuz04}.

\subsubsection{Equations of motion for guiding center atoms}
To simplify the discussion, let us consider the limiting case of a stationary atom $M/m\rightarrow\infty$. Within the 
guiding center approximation, the electronic cyclotron energy $E_{\rm cyc}$ appearing in the total Hamiltonian
\begin{equation}
H=E_{cyc}+\frac{p_z^2}{2m}-\frac{e^2}{r_{ei}}
\end{equation}
represents an approximate constant of motion. Consequently, the fast cyclotron motion of the electron can be averaged out, which yields the following dynamical equations for the electron's guiding center momentum and position
\begin{eqnarray}
\dot{x}&=&\frac{e}{B}\frac{y_{ei}}{r_{ei}^3}\nonumber \\
\dot{y}&=&-\frac{e}{B}\frac{x_{ei}}{r_{ei}^3}\nonumber\\
\dot{z}&=&v_z\nonumber\\
m\dot{v_z}&=&-e^2\frac{z_{ei}}{r_{ei}^3}\;.
\end{eqnarray}
The circular guiding-center atom solution is readily obtained from these equations, and allows to read off the different motional frequencies introduced above. In this case $r\approx\rho\approx{\rm const. }$ such that the ($x,y$) motion follows a circular orbit with 
radius $\rho$ around the ion and with the drift frequency $\omega_{\rm D}=e/(B\rho^3)$. The frequency $\omega_{z}=\sqrt{e^2/(m\rho^3)}$ can be directly read off the last equation.
To leading order in $z/\rho\ll1$, the $z$-motion is decoupled from the remaining degrees of freedom, such that one can define
 the binding energy 
\begin{equation} \label{Ebind}
E_{b}=H_z=\frac{p_z^2}{2m}-\frac{e^2}{\left(\rho^2+z^2\right)^{3/2}}
\end{equation}
of a GCA atom. If $E_{\rm b}<0$, the electron is bound to the ion, irrespective of its 
cyclotron energy $E_{\rm cyc}$, i.e. even if the total energy is positive. The guiding center approximation has proven valuable in gaining 
useful insights into the properties of strongly magnetized Rydberg atoms, permitting for example the 
explicit derivation of their magnetic moments \cite{Choi06}, radiative decay rates \cite{Guest03c} 
and response to electric fields \cite{kuzmin,vrinceanu}. The latter will be discussed in the next section.

\subsubsection{Electric field effects on guiding center atoms}\label{sect_efield}
In sections \ref{gidista} and \ref{matantimat} some effects of additional electric fields on the behavior of strongly magnetized Rydberg atoms and their utility in preparing and stabilizing particular Rydberg states have been discussed. 
From a different perspective, ionization of Rydberg atoms by electric fields has proven to be a 
powerful tool to probe their internal states. The magnetic field free case has been studied in great detail theoretically and experimentally (see \cite{GallagherBook}). Additional magnetic fields can significantly complicate the ionization process, and produce 
chaotic dynamics \cite{imf98,tr07}. 
On the other hand, in very strong magnetic fields the electron motion regains  comparably simple character, such that the guiding center approximation provides an intuitive picture for field ionization of strongly magnetized Rydberg atoms.

\begin{figure}[tb]
\begin{center}
\includegraphics[width=8.6cm]{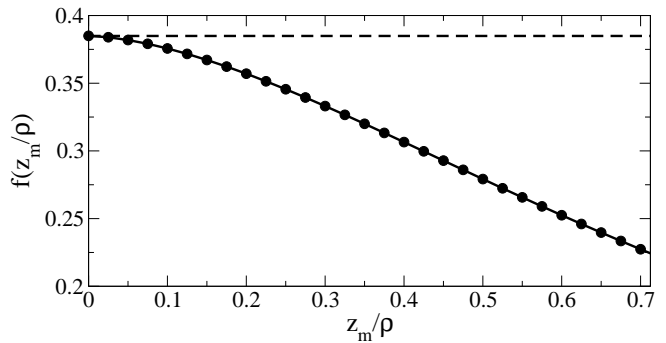}
\end{center}
\caption{Scaled critical ionization field $f=F/(e/\rho^2)$ of a cricular guiding center atom as a function of the scaled axial bounce amplitude $z_{\rm m}/\rho$. From \cite{kuz04}}\label{fig2_6_2}
\end{figure}

Let us consider a circular Rydberg atom with binding energy $E_{\rm b}$ (see Eq. (\ref{Ebind})) and $M\rightarrow\infty$. 
The electronic guiding center follows a circular drift orbit with radius $\rho$, and longitudinally oscillates with an 
amplitude $z_{\rm m}\ll\rho$, determined by
\begin{equation}
E_b=\frac{e^2}{\sqrt{\rho^2+z_{\rm m}^2}}\;.
\end{equation}
The application of a parallel electric field lowers the potential well that confines the longitudinal $z$-motion. 
If the field exceeds a critical value $F$, the electron is able to escape and the atom ionizes. For $z_{\rm m}=0$,
ionization occurs as the well vanishes, and the ionization field $F$ is straightforwardly obtained as \cite{vrinceanu}
\begin{equation}
F=\alpha\frac{e}{\rho^2}\;,\quad\alpha=\sqrt{\frac{4}{27}}\;.
\end{equation}
For finite axial energies, i.e. for $z_m\neq0$, the calculation is somewhat more complicated \cite{kuz04} and has to be performed numerically. However, one can use some scaling-arguments for the axial Hamiltonian $H_z$, and write the critical ionization field as
\begin{equation} \label{ionfield}
F=\frac{e}{\rho^2}f(z_{\rm m}/\rho)\;.
\end{equation}
Apparently, the critical field depends only on the drift radius $\rho$ and the relative axial amplitude $z_{\rm m}/\rho$, with $f(0)=\sqrt{(4/27)}$. Kuzmin et al. \cite{kuz04} determined the functional dependence of $f$ on the relative amplitude from a saddle-point analysis. The result is shown in Fig.\ref{fig2_6_2}. Eq.(\ref{ionfield}) provides a useful expression to analyze state-selective field ionization measurements and will be applied 
in section \ref{sect_twostep} to calculate field spectra of Rydberg atoms, which were observed in antihydrogen experiments \cite{atrap2,GabRev05}.

\begin{figure}[tb]
\begin{center}
\includegraphics[width=6.8cm]{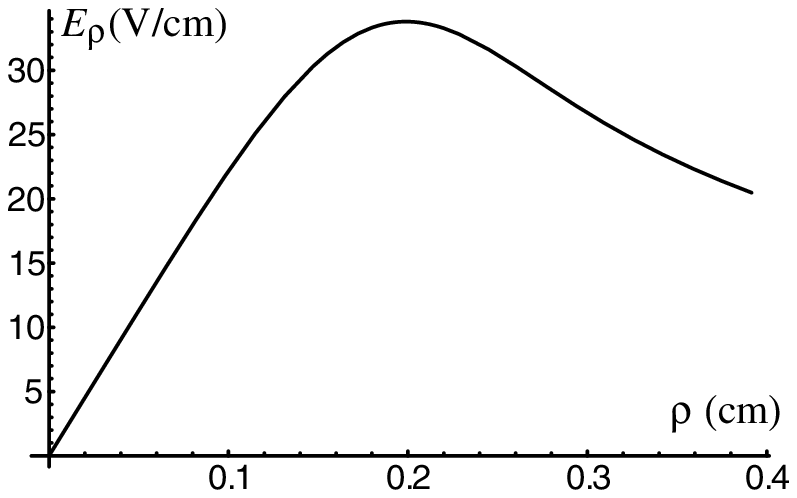}\hfill
\includegraphics[width=6.8cm]{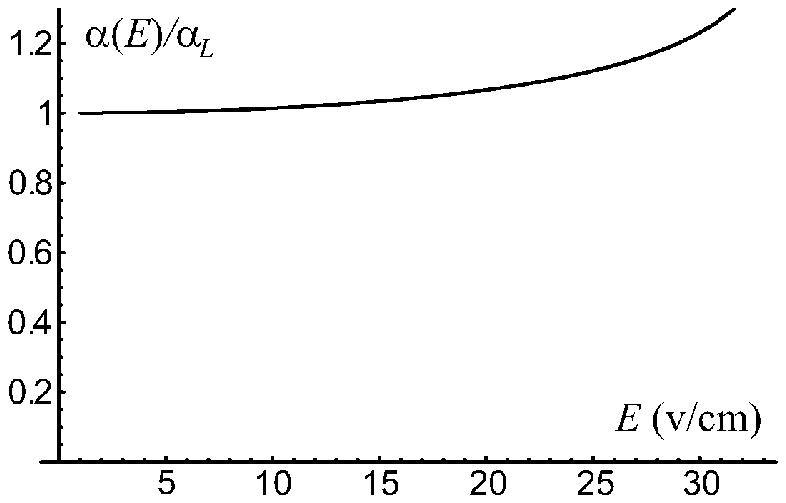}
\end{center}
\caption{Left: Space charge electric field produced by a long plasma column of radius $0.2$cm and density 
$2.5\times 10^8$cm$^{-3}$. Right: Electric field dependence of the polarizability, scaled by its zero-field value $\alpha_{\rm L}$ for 
$\rho=0.4 \mu$m. From \cite{kuz04b}.}\label{fig2_6_3}
\end{figure}

\begin{figure}[tb]
\begin{center}
\includegraphics[width=6.5cm]{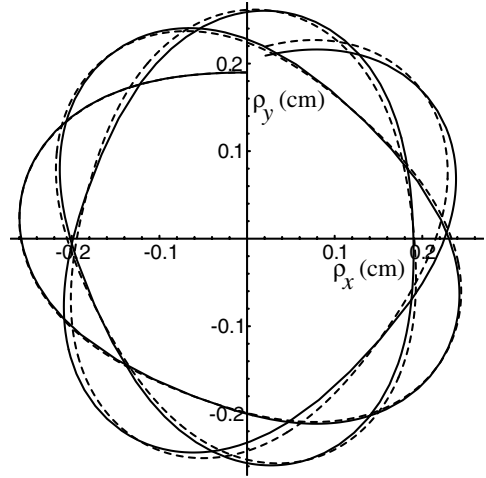}
\end{center}
\caption{Trapping of a guiding center atom with a initial kinetic energy of $4$K and binding energy of $40$K in the radial electric field shown in Fig. \ref{fig2_6_3}a. The dashed line shows the result of an exact propagation compared to the prediction of Eq.(\ref{com_motion}) for $\alpha=\alpha(25 {\rm V/cm})=1.18\alpha_{\rm L}$. From \cite{kuz04b}.}
\label{fig2_6_4}
\end{figure}

Small electric fields, that do not cause ionization, nevertheless polarize the atom and thereby affect its center-of-mass motion. In situations, where magnetized Rydberg atoms are formed through recombination in cold plasmas (see section \ref{sec:ryatffp}), 
the plasma's space-charge electric field (Fig. \ref{fig2_6_3}a) can significantly alter their motion and, as suggested in \cite{kuzmin,kuz04b}, even lead to trapping of weakly bound, magnetized atoms. Averaging  out the rapid cyclotron motion of the electron, one can
derive a simple and instructive equation of motion for the atomic center-of-mass velocity 
${\bf V}$ \cite{kuzmin,kuz04b}
\begin{equation}\label{com_motion}
M\dot{\bf V}=\nabla\frac{\alpha F^2}{2}+\alpha\left(\nabla_{\perp}\cdot {\bf F}\right)\left({\bf V}\times{\bf B}\right)\;,
\end{equation}
which is valid in the guiding center limit. The electric polarizability $\alpha$ is a function of the magnetic field strength $B$, the drift radius $\rho$, the relative axial bounce amplitude $z_{\rm m}/\rho$ and the strength $F$ of the electric field. For 
$F\rightarrow0$ and $z_{\rm m}\ll\rho$, the zero-field polarizability $\alpha_{\rm L}$ is obtained as \cite{kuzmin,kuz04b}
\begin{equation} \label{polarizability}
\alpha_{\rm L}=\frac{5}{2}\rho^3\frac{1+\frac{2B^2}{5M}\rho^3-\frac{33}{40}\frac{z_{\rm m}^2}{\rho^2}}{\left(1+\frac{B^2}{M}\rho^3-\frac{3}{4}\frac{z_{\rm m}^2}{\rho^2}\right)^2}\;.
\end{equation}
Fig. \ref{fig2_6_4} compares this simplified approach to the result of exact classical calculations that 
follow the coupled ion and electron motion and resolve the fast cyclotron oscillations. 
The figure shows a trajectory of a circular atom in an axial magnetic field of $B=3$T and the radial electric field of 
Fig. \ref{fig2_6_3}a, as produced by a long plasma column. The result of the simplified equations of motion (\ref{com_motion}) nicely agrees with the exactly calculated center-of-mass dynamics over several orbital periods. Moreover, Fig. \ref{fig2_6_4} shows that already moderate electric fields can promote radially bound atom motion for rather large binding energies of $\sim 40$K and center-of-mass energy of $4$K $-$ typical values realized in antihydrogen experiments \cite{athena}.

However, such trapping is efficient only for a limited range of Rydberg states and kinetic energies \cite{kuz04b} and lacks axial confinement. Several approaches to achieve reliable three-dimensional confinement have been proposed \cite{Dutta00,mhn05,hog08}. As will be discussed in the next section, strongly magnetized Rydberg atoms may, apparently, be most effectively confined in magnetic traps, i.e. through the application of inhomogenous magnetic fields.

\section{Ultracold Rydberg atoms in inhomogeneous fields} \label{sec:atproinhomag}
Processing ultracold atoms with static magnetic fields requires inhomogeneous field configurations,
i.e. spatially varying magnetic fields. Since the hyperfine interaction is typically much stronger than
the interaction of the atoms with the external magnetic field, the total angular momentum and its resulting
magnetic moment provide the key quantity which couples to the local magnetic field. Atoms in their electronic
ground state and some hyperfine substate are within this picture treated as point-like magnetic moment
carrying particles. For most investigations and applications on ultracold atomic ensembles in magnetic
traps, an adiabatic approach is employed i.e. it is assumed that the atomic magnetic moment is always 
aligned with the local magnetic field and therefore we can replace the vector coupling 
{\boldmath{$\mu_J$}}$\cdot {\bf{B}}$ by the scalar one {\boldmath{$|\mu_J|$}}$|{\bf{B}}|$.
This is a simplification which becomes increasingly valid the more the ultracold motion
of the atoms becomes adiabatic i.e. the change of the magnetic field strength
seen by the slowly moving atoms on the time scale of the precision of their magnetic momentum
around the local magnetic field should be small compared to the local field strength.

In spite of the frequent use of the adiabatic magnetic coupling,
theoretical studies of the quantum properties of neutral ground state atoms
in magnetic traps using vector coupling, are of relevance not only for obtaining the lifetimes
but also to reveal potential new classes of (trapped or untrapped) eigenstates. 
Such studies, which we will review in the following, have been performed since the late eighties. 
Several magnetic field configurations have been in the focus of numerous publications,
e.g. the quadrupole field \cite{Bergeman89},
the wire trap \cite{Hau95,Berg-Soerensen96,Burke96} or the magnetic guide and
the Ioffe-Pritchard trap \cite{Hinds00,Hinds00E,Potvliege01}. However, only
a few field configurations allow for stationary solutions. Bound
states of spin-$\frac{1}{2}$-particles trapped by a wire have been
found analytically \cite{Bluemell91,Hau95}. The latter authors
employed a super-symmetric approach in order to derive the Rydberg
series of bound states around an infinitely thin wire. Finite wire
sizes were accounted for by introducing a quantum defect in the
Rydberg formula. A numerical derivation of a 
similar series of energies for particles with spin up to 2 was given \cite{Burke96} by combining
the finite element method with multichannel quantum defect theory. However, in
most cases atoms are trapped in meta-stable states which have 
finite lifetimes. In an early work \cite{Bergeman89}, dozens of spin-$\frac{1}{2}$ resonances 
inside a three-dimensional magnetic quadrupole field have been studied by
determining the phase-shift of scattered waves. A similar approach \cite{Hinds00,Hinds00E} was chosen 
while studying the dynamics of neutral particles in a magnetic guide and corresponding
results on resonant states were reproduced by a complex scaling calculation \cite{Potvliege01}. 
A comprehensive study of the spectral properties and lifetimes of neutral spin-$\frac{1}{2}$-fermions
in a magnetic guide has been performed recently \cite{Lesanovsky04d}. A wealth of unitary
and antiunitary symmetries has been revealed giving in particular rise to a twofold degeneracy of the
energy levels. The dependence of the lifetimes of the resonances on the angular momentum 
and in particular the existence of so-called quasibound states has been analyzed in detail.
Quantum states of neutral fermions and bosons in a magnetic quadrupole configuration have been explored
\cite{Lesanovsky05b}: The distribution of the energies and decay widths of the resonances
were studied identifying the conditions under which states with long lifetimes can be achieved.
A peculiar class of short-lived negative energy resonances in a magnetic quadrupole field
has been found recently \cite{Saeidian06} and a mapping of the two branches of positive and negative
energy resonances was derived. Global properties of the resonant quantum dynamics of 
neutral spin-1 atoms in a magnetic guide have been studied \cite{Bill06} including the effect of a Ioffe field.
For certain parameter regimes the ground state resonance was shown to exhibit a longer lifetime
than the energetically neighbored excited states. The case of a comparable hyperfine 
and field interaction was explored very recently \cite{Saeidian07}.

The past few years have seen substantial progress toward 
the development of stable trapping and guiding mechanisms
and trapping on microscale levels, as in optical lattices and on atom chips.
There are two domains
of interest: The case of a highly excited CM motion
of the Rydberg atoms which can be described in terms of classical dynamics or semiclassical approaches 
and the case of a quantized ground and excited CM motion. We will review in the remaining part
of this section the status in case of a semi-classical CM motion (see in particular the review
in Ref. \cite{ckll}). The subsequent section will be devoted
to the quantum Rydberg regime i.e. the case of a coupled quantum CM and electronic Rydberg motion.

Trapping of long-lived strongly magnetized Rydberg atoms being collected in a superconducting magnetic
trap with a strong bias field (2.9 T) has been explored experimentally \cite{ChoiRaithel2005}. Here so-called drift or guiding center
Rydberg atoms (see subsection \ref{gca}) are produced by Rydberg collisions. Observation of oscillatory motion
of Rydberg atoms in the resulting magnetic potential has provided strong evidence that long-lived
Rydberg atoms have been trapped. Trapping periods up to 200 ms have been found. In a follow-up work \cite{Choi06}
the picture of the guiding center drift atoms in a strong magnetic field, based on a hierarchy
of interactions and time scales, has been used to derive simple models of the effective magnetic moment
and to perform classical trajectory calculations (see also refs.\cite{Guest03b,Guest03c} for
the background on circular Rydberg atoms in strong magnetic fields).
An anisotropic response of hydrogen atoms
to inhomogeneities longitudinal and transverse to the magnetic field has been observed.
A proposition to use the ponderomotive energy of Rydberg electrons in standing-wave light fields
to form an optical lattice for Rydberg atoms has been worked out in ref.\cite{Dutta00}.
Atom-chip technique proposals have recently been worked out for the trapping of a single Rydberg atom
in a circular state \cite{mhn05}. The small size of microfabricated structures allows for trap
geometries with microwave cut-off frequencies high enough to inhibit the spontaneous emission of the
Rydberg atom, paving the way to complete control of both external and internal degrees of freedom
over very long times.

\subsection{Atomic structure and quantum dynamics in an inhomogeneous magnetic field}

Our objective is to develop an ab initio approach of moving Rydberg atoms in inhomogeneous
magnetic field configurations. We thereby provide a consistent description of the
quantum behaviour for the center of mass and electronic motions: both types of motion will be
treated on equal footing. A major goal is to work out and analyze configurations which allow
for the controlled processing of Rydberg atoms in the regime where both the CM and electronic
motion are governed by quantum effects. This includes traps and waveguides for ultracold Rydberg
atoms.

As we have seen in chapter \ref{sec:atprohomag}, the case of a moving atom
in a homogeneous magnetic field exhibits a number of intriguing phenomena. In particular, we
encounter an intricate coupling of the CM and electronic motion which gives rise to both new structural
properties and dynamics.
It has to be expected that moving atoms in an inhomogeneous field will exhibit
their own unique behavior. In view of the high dimensionality and large density of levels,
it is advisable
to firstly study the electronic structure for fixed nucleus 
\cite{Lesanovsky04a,Lesanovsky04b,Lesanovsky04c,Lesanovsky05a,Lesanovsky05c}
and subsequently investigate the complete problem of a moving Rydberg atom 
\cite{LesanovskySchmelcherPRL2005,Bock05,Lesanovsky05e,Hezel06,Schmidt07,Mayle07,Hezel07}.
Concerning the magnetic field configurations, we will focus here on the 3D quadrupole field
\cite{Lesanovsky04a,Lesanovsky04b,Lesanovsky05a,Lesanovsky05c}. For the
2D quadrupole field (side guide) and the Ioffe-Pritchard configurations, we refer
the reader to the literature \cite{Lesanovsky04c}.

\subsubsection{Electronic structure of Rydberg atoms for fixed nucleus} \label{esryfn}

Keeping the position of the nucleus fixed and studying exclusively the electronic structure
of the atom in an inhomogeneous external magnetic field is certainly a special but instructive case of 
a Rydberg atom in a magnetic field. It corresponds to the situation where the confinement length of the 
CM motion is much shorter than the extension of the electronic Rydberg wave function.
One of the motivations to study the impact of the inhomogeneity of the field on the
electronic structure is the experimental availability of extremely strong gradients; current carrying
wires on atom chips can reach gradients up to $10^4$ {\mbox{T/cm}} \cite{Folman02}. It is thus expected that
such large field gradients have sizeable impact on the Rydberg wave functions.

The 3D magnetic quadrupole field 
\cite{Lesanovsky04a,Lesanovsky04b,Lesanovsky05a,Lesanovsky05c} is characterized by 
${\bf{B}}({\bf{r}}) = b (x,y,-2z)$ with the gradient $b$. This vector field is rotationally
symmetric around the $z$-axis and invariant under the $z$-parity
operation. A corresponding vector potential in the Coulomb gauge
reads $ {\bf{A}}({\bf{r}})=\frac{1}{3}\left[{\bf{B}}({\bf{r}})\times{\bf{r}}\right]$.
Adopting atomic units ($b=1a.u.=4.44181\times 10^{15} \frac{T}{m}$) the
nonrelativistic spinor Hamiltonian in case of a single active Rydberg electron reads
\begin{eqnarray} \label{hamfn3dquad}
{H_{fn}} \nonumber &=&-\frac{1}{2}\triangle-\frac{1}{\sqrt{x^2+y^2+z^2}}
-b\,z{L}_z \\ &+&\frac{b^2}{2}z^2\left(x^2+y^2\right) 
+\frac{b}{2}\left(\sigma_xx+\sigma_yy-2\sigma_zz\right)\label{eq:hamiltonian_rc=0_cart}
\end{eqnarray}
with $L_z=-i\left(y\partial_x-x\partial_y\right)$ and $\sigma_x$,
$\sigma_y$, $\sigma_z$ are the Pauli spin matrices 
(${\bf{S}}=\frac{1}{2}${\boldmath{$\sigma$}}).
The paramagnetic ($\propto b$) or orbital Zeeman-term depends, in
contrast to the situation of the atom in a homogeneous field, 
not only on $L_z$ but also linearly on the
$z$-coordinate. The diamagnetic term ($\propto b^2$)
represents a quartic oscillator coupling term between the
cylindrical coordinates $\rho=\sqrt{x^2+y^2}$ and $z$. In a
homogeneous field the diamagnetic interaction is a pure harmonic
oscillator term proportional to $\rho^2$ and yields a confinement
perpendicular to the magnetic field. 
The coupling of the electronic spin to the quadrupole
field depends linearly on the Cartesian coordinates. The latter
prevents the factorization of the motion in the spin and spatial
degrees of freedom and renders the corresponding Schr{\"o}dinger
equation into a spinor equation. This is again in contrast to the
case for a homogeneous magnetic field where the spin component
along the field is a conserved quantity. 

As a consequence of the rotational invariance of the quadrupole
field the $z$-component of the total angular momentum
$J_z=L_z+S_z$ is conserved, i.e. $\left[H_{fn},J_z\right]=0$.
Additionally we have the discrete symmetry represented by the
unitary operator $P_\phi OP_z$, i.e. $\left[H_{fn},P_\phi
OP_z\right]=0$. Here $O\equiv\sigma_x$ exchanges the components of
a $\frac{1}{2}$-spinor and $P_\phi$ represents the 'reflection'
$P_\phi:\phi\rightarrow 2\pi-\phi$. Apart from these symmetries
the Hamiltonian possesses two generalized anti-unitary time
reversal symmetries namely $TOP_z$ and $TP_\phi$
involving the conventional time reversal operator $T$ ($T^2=1$).
The operators $TOP_z$, $TP_\phi$
and $P_\phi O P_z$ form an invariant Abelian sub-group. Together
with $J_z$ they obey the following (anti-)commutation rules:
\small
\begin{eqnarray}
\left[J_z,TP_\phi\right]=\left\{J_z,TOP_z\right\}=\left\{J_z,P_\phi
OP_z\right\}=0\\
\left[TP_\phi,TOP_z\right]=\left[TP_\phi,P_\phi
OP_z\right]=\left[P_\phi
OP_z,TOP_z\right]=0\label{eq:top_pop_commutator}
\end{eqnarray}
\normalsize

Apparently the spin-spatial symmetry operations form a non-Abelian
symmetry group. This group generated by $P_\phi OP_z$ and $J_z$ is
isomorphic to $C_\infty \bigotimes C_s$.

The interplay of the above symmetries leads to a degeneracy of the eigenvalues of the
Hamiltonian (\ref{hamfn3dquad}). Indeed, if we consider a state
$\left|E,m\right>$ which is an eigenstate of the Hamiltonian
(\ref{hamfn3dquad}) with the energy $E$ and to $J_z$
with the half-integer quantum number $m$, then since
the state $TOP_z\,\left|E,m\right>$ is also an energy eigenstate
with the energy $E$, we obtain
\begin{eqnarray}
J_z\,TOP_z\,\left|E,m\right>=-TOP_z\,J_z\,\left|E,m\right>=-m\,TOP_z\,\left|E,m\right>.\nonumber
\end{eqnarray}
Thus the state $TOP_z\,\left|E,m\right>$ can be identified with
$\left|E,-m\right>$. Hence the states with the eigenvalues $m$ and
$-m$ are degenerate. This two-fold degeneracy of each energy level
in the presence of the inhomogeneous magnetic field is a
remarkable feature reminiscent of the Kramers degeneracy of
spin $\frac{1}{2}$ systems in the absence of external fields
\cite{Haake01}.

Let us now discuss on the basis of some examples new features that occur for
the electronic structure of Rydberg atoms in inhomogeneous magnetic fields using
the specific example of the 3d quadrupole field.
To demonstrate the deformation effects on the electronic cloud
we consider for reasons of illustration an extremely high gradient.
\begin{figure}[htbp]\center
\includegraphics[angle=0,width=7cm]{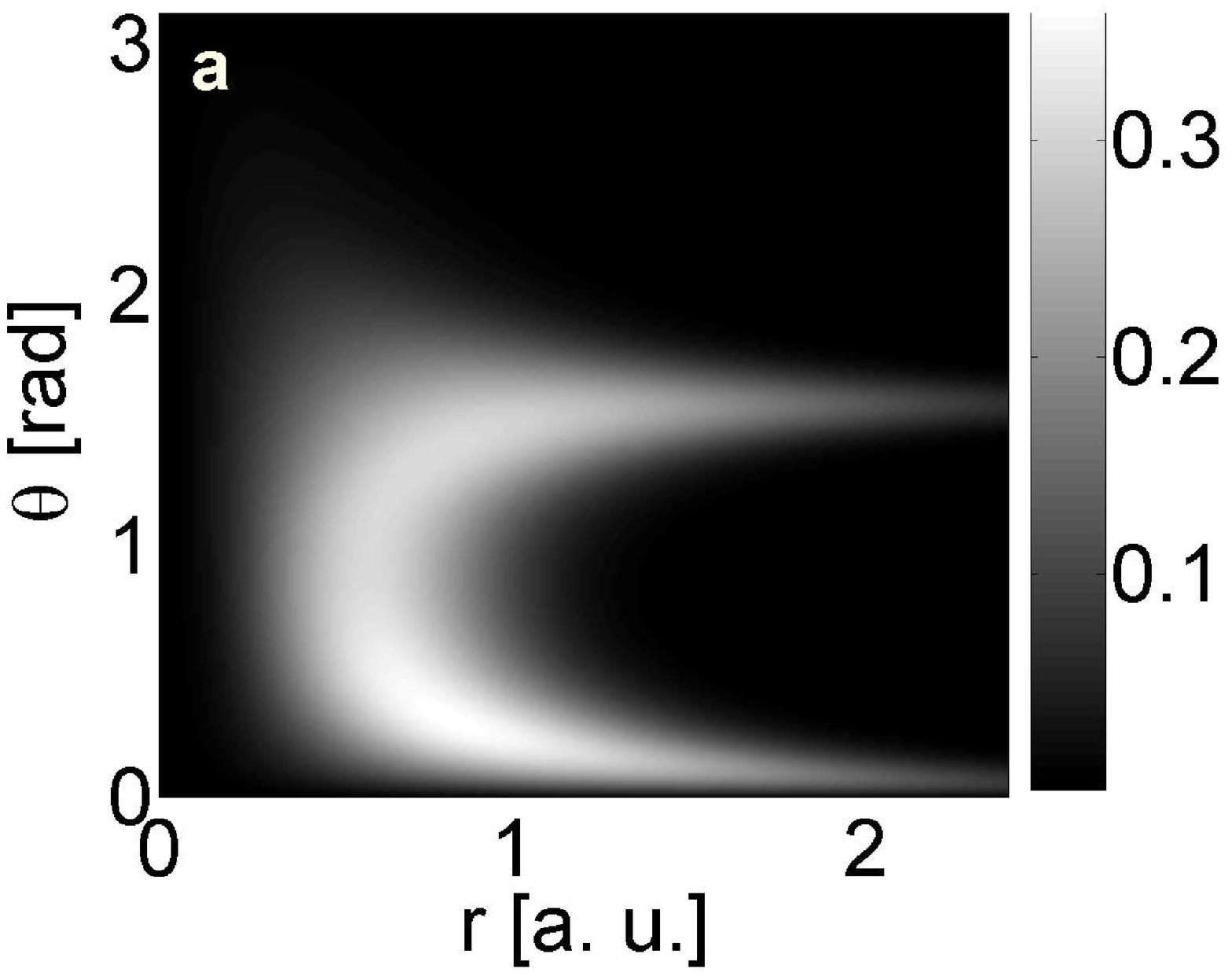}
\includegraphics[angle=0,width=7cm]{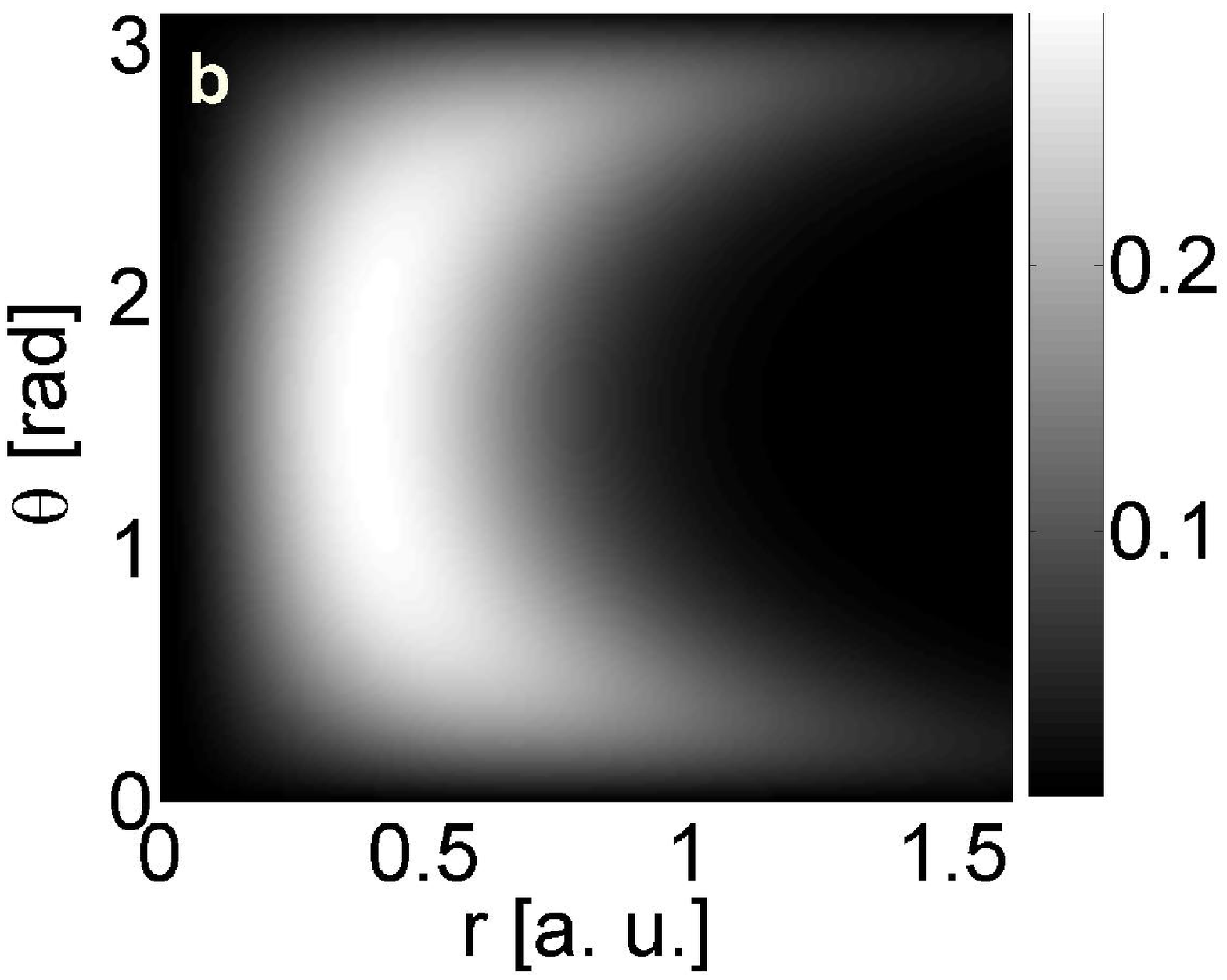}
\caption{Spatial probability density of the ground state in the
quadrupole field (\textbf{a}) and a homogeneous field pointing in
the $z$-direction (\textbf{b}) for $b=B=10$a.u. and
$m=\frac{1}{2}$. From \cite{Lesanovsky04b}.} \label{fig:groundstate}
\end{figure}
Figure \ref{fig:groundstate} shows the spatial probability density
of the ground state in the quadrupole field
(\ref{fig:groundstate}a) at $b=10$ and the homogeneous field
(\ref{fig:groundstate}b) at $B=10$ ($m=\frac{1}{2}$ in both
cases).
For the quadrupole field we observe an asymmetric deformation with
respect to the $\theta=\frac{\pi}{2}$-plane: the electronic
wavefunction is almost completely confined to the upper
half-volume ($\theta<\frac{\pi}{2}$) which is a consequence of the
symmetries: $z$-parity
is not conserved and eigenstates appear in pairs one being the
mirror image of the other with respect to the reflection at the
$x-y$-plane. Furthermore we observe that the electronic motion is
localized particularly along the two 'channels' for $\theta=0$
corresponding to the lower $z$-axis and $\theta=\frac{\pi}{2}$
being the $x-y$-plane. This property as well as the detailed shape
of the electronic probability density in the individual
half-volume is determined by the diamagnetic term which is
dominant in the high gradient regime. For the quadrupole field it
is proportional to $\sin^2\theta\cos^2\theta$ reaching its maximum
value at $\theta=\frac{\pi}{4},\frac{3\pi}{4}$. The probability
density in a homogeneous field (see figure \ref{fig:groundstate}b) exhibits the
above-mentioned corresponding reflection symmetry due to the
invariance of the corresponding Hamiltonian with respect to
$z$-parity. Here we observe the maximum of the probability density
at $\theta=\frac{\pi}{2}$ and a deformation towards $\theta=\pi$
and $\theta=0$ leading to a cigar-like shape. The diamagnetic term
is proportional to $\sin^2\theta$ having its maximum value at
$\theta=\frac{\pi}{2}$ thus coinciding with the regions possessing
the strongest deformation of the probability density. For both
field configurations the probability density vanishes at $r=0$.

\begin{figure}[htbp]\center
\includegraphics[angle=-90,width=7cm]{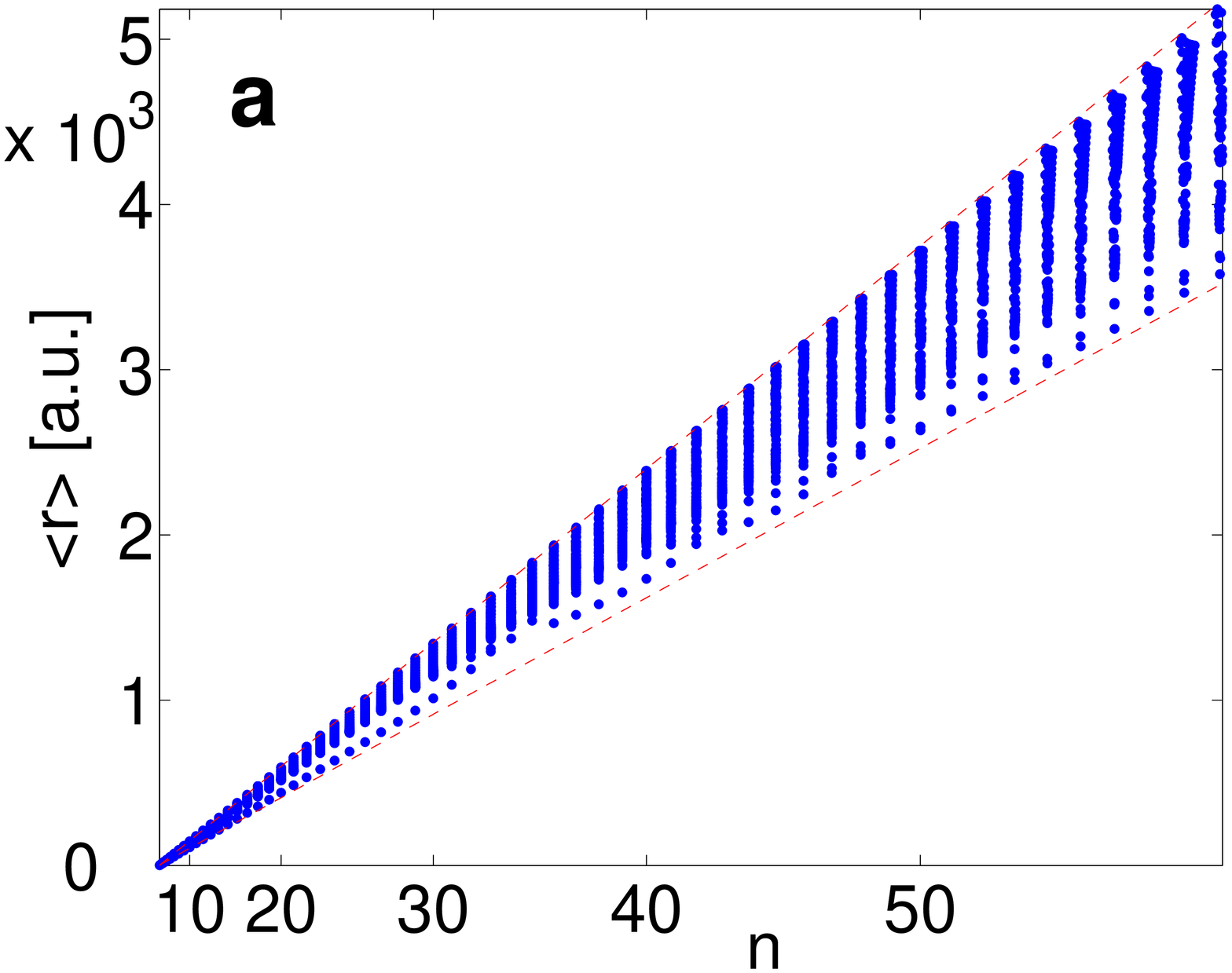}
\includegraphics[angle=-90,width=7cm]{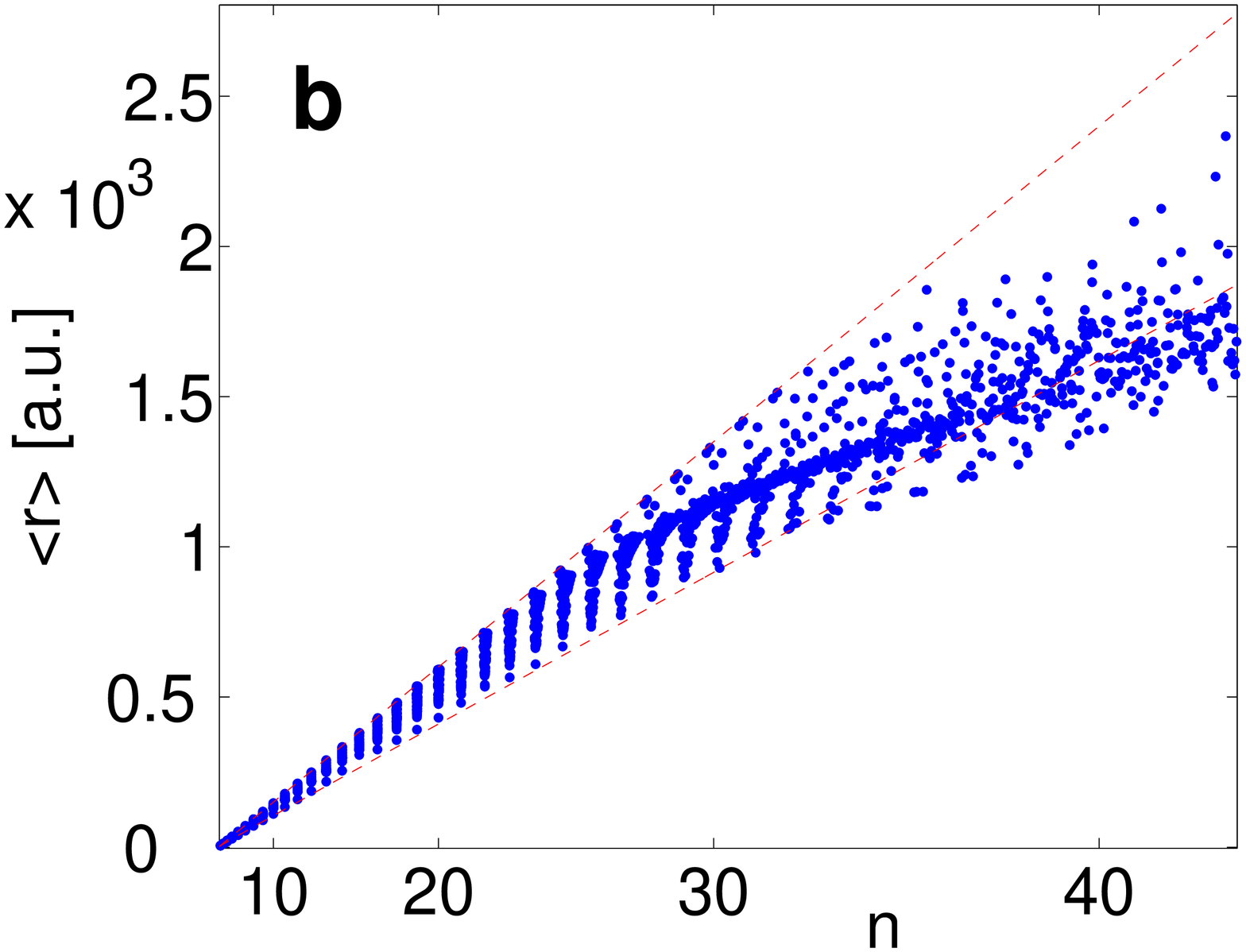}
\caption{Expectation value of the radial coordinate $r$ plotted
against $n$ at different gradients
(\textbf{a:} $b=10^{-10}$, \textbf{b:}
$b=10^{-8}$). From \cite{Lesanovsky04b}.}\label{fig:radial_expectation_value}
\end{figure}

To explore the compression of Rydberg states in strong (atom-chip) laboratory gradients
we first remark that 
\begin{eqnarray}
\left<r\right>=\left<r\right>^\pm_{TOP_z}=\left<r\right>_{J_z}.
\end{eqnarray}
where the lower index at the expectation values indicates the type of eigenstates
employed. 
For the one-electron problem without external field we have
$\left<r\right>^H=\frac{1}{2}\left(3n^2-l(l+1)\right)$. Since $l$
satisfies $0\leq l \leq n-1$ one obtains corresponding upper and
lower bounds for the expectation value of $r$:
$n\left(n+\frac{1}{2}\right)\leq \left<r\right>^H \leq
\frac{3}{2}n^2$. In figure \ref{fig:radial_expectation_value} this
expectation value is shown for the atom in the quadrupole field as
a function of $n$ for two different 
gradients ($n$ serves here as a label for the energy levels and is
not a good quantum number!).
The radial expectation values lie in between the boundaries given
by the field-free inequality indicated by the dashed lines (see figure
\ref{fig:radial_expectation_value}a).
Expectation values of states belonging to the same $n$-multiplet
are arranged in vertical lines expressing their energetical
degeneracy (see figure \ref{fig:radial_expectation_value}a for
$b=10^{-10}$). Here states with large expectation values of $r$
possess small expectation values of the angular momentum and vice
versa. The situation is different for $b=10^{-8}$. Here a
systematic decrease of the $r$-expectation values takes place
for states with an energy corresponding to $n> 30$. This is also
the energy regime where the inter $n$-manifold mixing due to the
diamagnetic term sets in.

As a next step let us consider properties of the Rydberg atoms that
involve the spin degrees of freedom.
In a homogeneous magnetic field, the projection of the spin
operator onto the direction of the magnetic field
is a conserved quantity. In this case one can choose the energy
eigenstates to be also eigenfunctions with respect to $S_z$. 
In the quadrupole field, $S_z$ is not conserved, and we therefore consider
the expectation value 
\begin{eqnarray}
\left<S_z\right>_{J_z}=\frac{1}{2}\left[\left<u\mid
u\right>-\left<d\mid d\right>\right].
\end{eqnarray}
\begin{figure}[h!]\center
\includegraphics[angle=0,width=7cm]{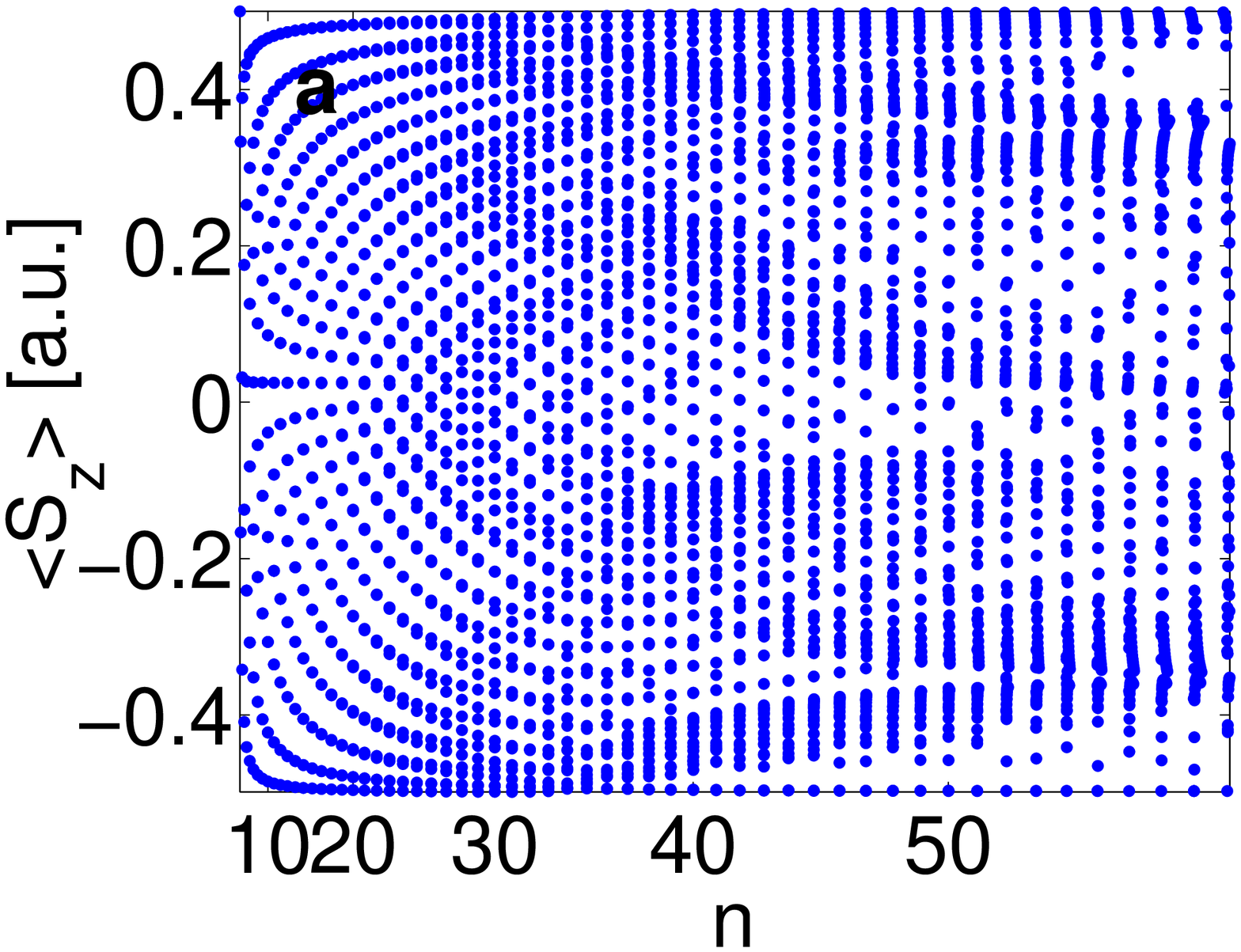}
\includegraphics[angle=0,width=7cm]{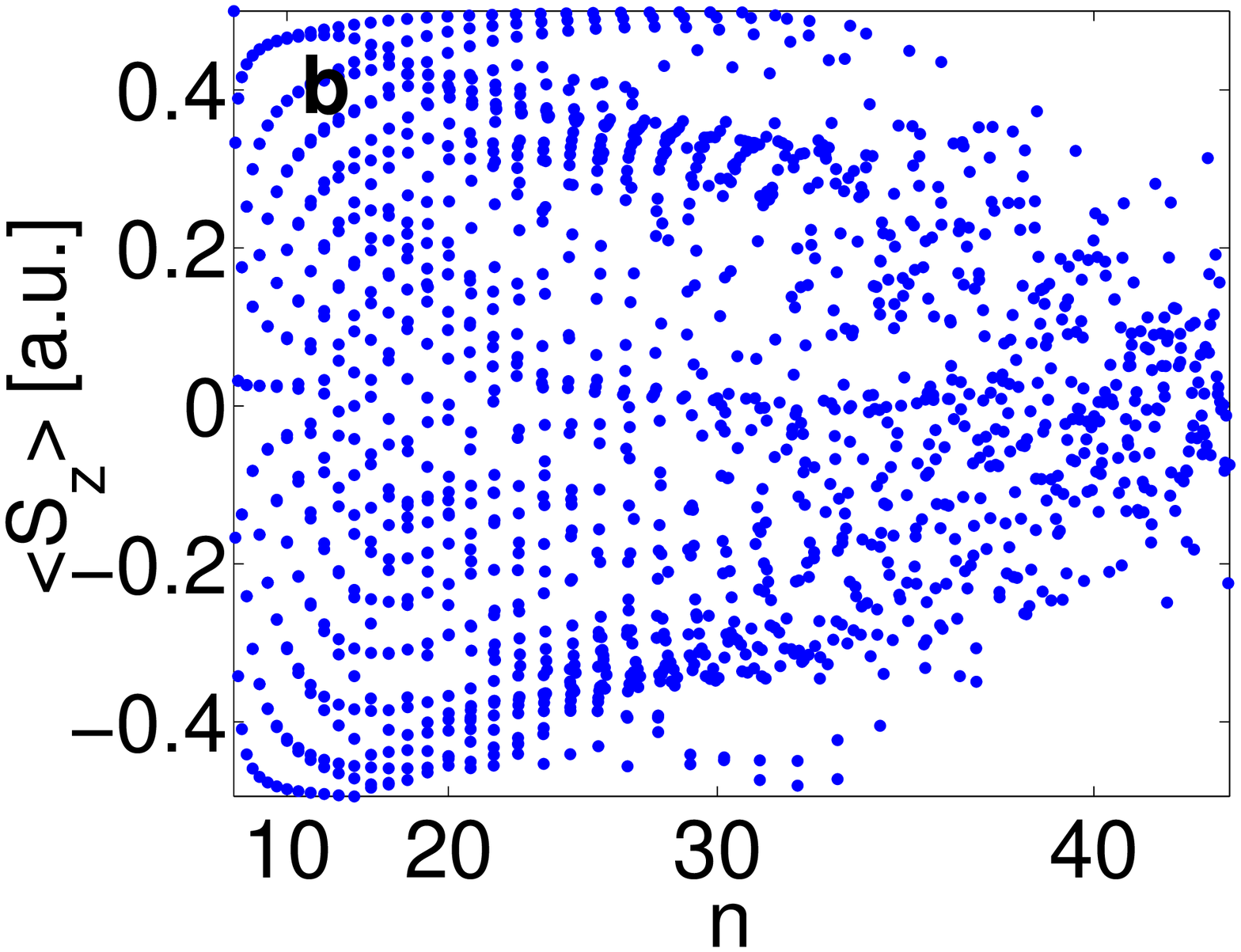}
\caption{Expectation values of the $z$-component of the spin
operator as a function of the quantum number $n$ for different
gradients (\textbf{a}: $b=10^{-10}$, \textbf{b}: $b=10^{-8}$). From \cite{Lesanovsky04b}.
}\label{fig:sz_expectation_value}
\end{figure}
Figure \ref{fig:sz_expectation_value} shows the distribution of
$\left<S_z\right>_{J_z}$ for electronic states of the $m=\frac{1}{2}$
subspace as a function of $n$. Since
$S_z$ is not conserved, the values of $\left<S_z\right>$ are
allowed to cover the complete interval
$\left[-\frac{1}{2},\frac{1}{2}\right]$. For
$b=10^{-10}$ (figure \ref{fig:sz_expectation_value}a) and a low
degree of excitation the expectation values are evenly distributed
over the interval. When reaching highly excited states this
pattern becomes increasingly distorted. The expectation values
agglomerate at $-0.35$ and $0.35$ for $n\geq 50$. Due to the
approximate degeneracy of the energy levels at low energies, i.e.
small $n$, the values of $\left<S_z\right>$ form vertical lines.
These lines widen for higher $n$. Since at $b=10^{-10}$ no
significant $n$-mixing up to our maximum converged energy levels
takes place, neighboring lines are well separated.
For a higher gradient $b=10^{-8}$ (figure
\ref{fig:sz_expectation_value}b) the above properties are equally
present for low-lying states. However, with increasing
excitation energy we now observe a complete $n$-mixing  regime
where the regular patterns disappear and we obtain an irregular
distribution of $\left<S_z\right>$. Overall the distribution
narrows, e.g. for $n=40$ the occupied interval is approximately
$\left[-0.3,0.3\right]$.
We remark that due to the fact that the $S_z$-operator
anti-commutes with $TOP_z$, we have
$\left<S_z\right>_{TOP_z}^\pm=0$. Apparently there is no preferred
direction for the electronic spin in a state obeying the $TOP_z$
symmetry.

Due to the coupling of the spatial and spin degrees of freedom
the spin orientation becomes spatially dependent and it is instructive
to consider the spin $S_z$-polarization $W_S$. For a $J_z$-eigenstate
$\left|E,m\right>$ it reads
\begin{eqnarray}
W_S(\bf{r})&=& \nonumber \frac{\left<E,m\mid \bf{r}
\right>{S}_z\left<\bf{r}\mid E,m\,\right>}{\left<E,m\mid
\bf{r}\right>\left<\bf{r}\mid E,m\,\right>} \\
&=& \frac{1}{2}\frac{\left|\left<u\mid
\bf{r}\right>\right|^2-\left|\left<d\mid
\bf{r}\right>\right|^2}{\left|\left<u\mid
\bf{r}\right>\right|^2+\left|\left<d\mid
\bf{r}\right>\right|^2}.\label{eq:spin_density}
\end{eqnarray}

\begin{figure}[h!]\center
\includegraphics[angle=0,width=7cm]{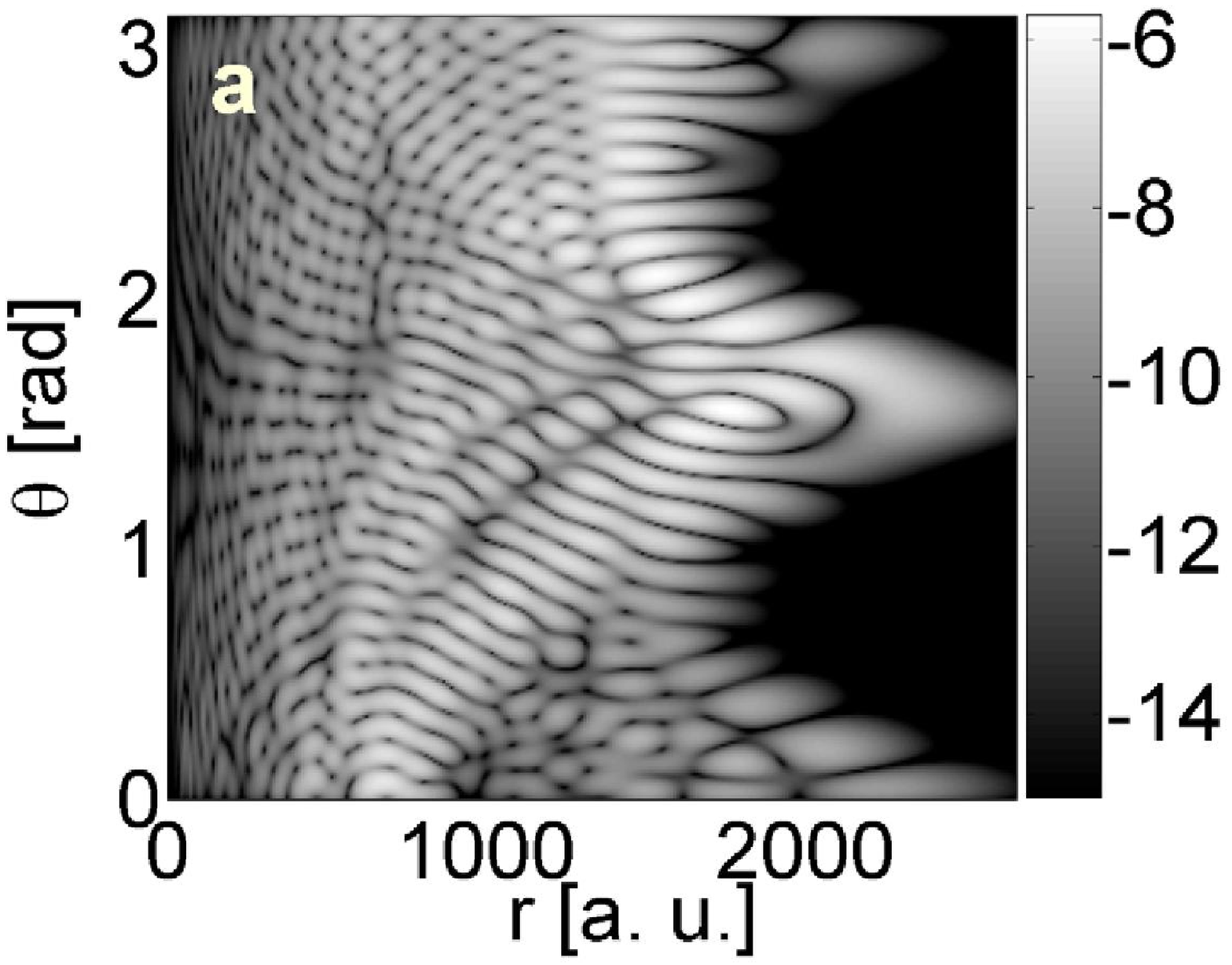}
\includegraphics[angle=0,width=7cm]{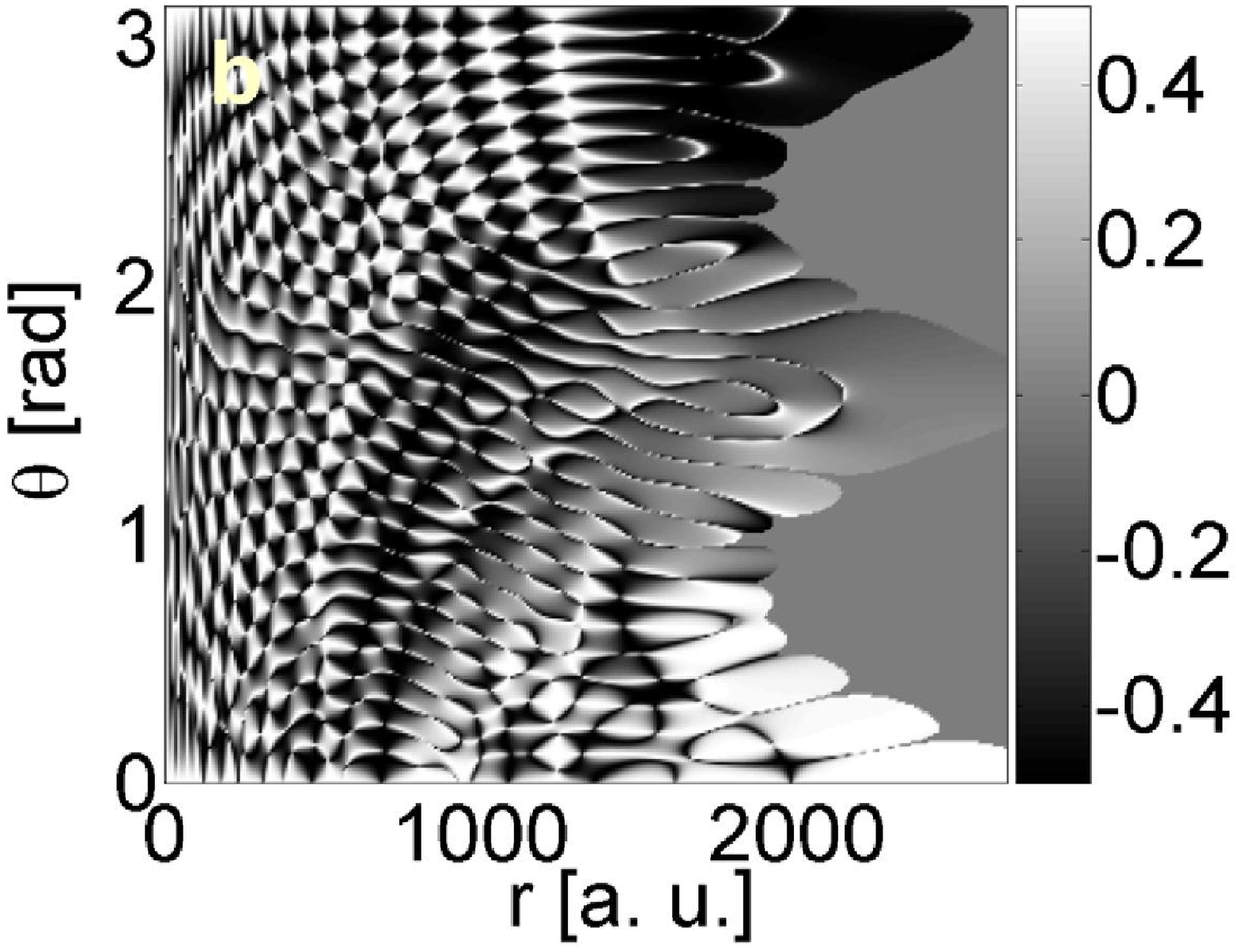}
\caption{Spatial probability density (\textbf{a}) and
$S_z$-polarization (\textbf{b}) for the $1117th$ excited state for
$m=\frac{1}{2}$ and $b=10^{-8}$. At large $r$ the $S_z$-polarization
becomes similar to $W_S^+$ indicating an antiparallel alignment of
the electronic spin to the magnetic
field. From \cite{Lesanovsky04b}.}\label{fig:sz_density_2}
\end{figure}
Figure \ref{fig:sz_density_2} shows the spatial probability
distribution (\ref{fig:sz_density_2}a) and the $S_z$-polarization
(\ref{fig:sz_density_2}b) for the $1117th$ excited state for
$m=\frac{1}{2}$ and $b=10^{-8}$ In contrast to the constant
$S_z$-polarization we would encounter in the absence of a field or
a homogeneous field we observe a complex pattern of domains
exhibiting different spin orientation (white: spin up, black: spin
down). At low $r$ values these domains form a pattern similar to
that of a chess board. The junctions where four spin domains meet
each other coincide with the nodes of the spatial probability
density. The Coulomb interaction as well as the spin-Zeeman term
are responsible for the interwoven network of island of different
spin orientation. The additional presence of the orbital Zeeman
and the diamagnetic term leads to a deformation of this network.
For low radii the spin orientation changes locally from
island to island generating an appealing pattern whereas we observe an overall tendency
of the electronic spin polarization in the region characterized by stripes ($r \approx 2000$).
Here, independently of the nodal structure, the spin orientation
changes smoothly from downwards at $\theta=\pi$ to upwards at $\theta=0$.
This feature can be understood by inspecting the spin Zeeman term only.

In a homogeneous magnetic field parity is a symmetry and therefore
atomic electronic eigenstates (for fixed nucleus) do not possess a
permanent electric dipole moment. Let us investigate the
electric dipole moment of the electronic states in the quadrupole
field. First of all we remark that
for $\sigma^\pm$ dipole transitions the corresponding matrix element 
$\left<E^\prime,m^\prime\right|r\sin\theta\,e^{\pm i\phi}\left|E,m\right>$ is only non-zero if
$m^\prime-m=\pm1$.  The expectation value of $D_{\sigma^\pm}$ in the $J_z$-eigenstates vanishes
\begin{eqnarray}
\left<D_{\sigma^\pm}\right>=\left<E,m\right|r\sin\theta\,e^{\pm
i\phi}\left|E,m\right>=0
\end{eqnarray}
\begin{figure}[h!]\center
\includegraphics[angle=0,width=7.3cm]{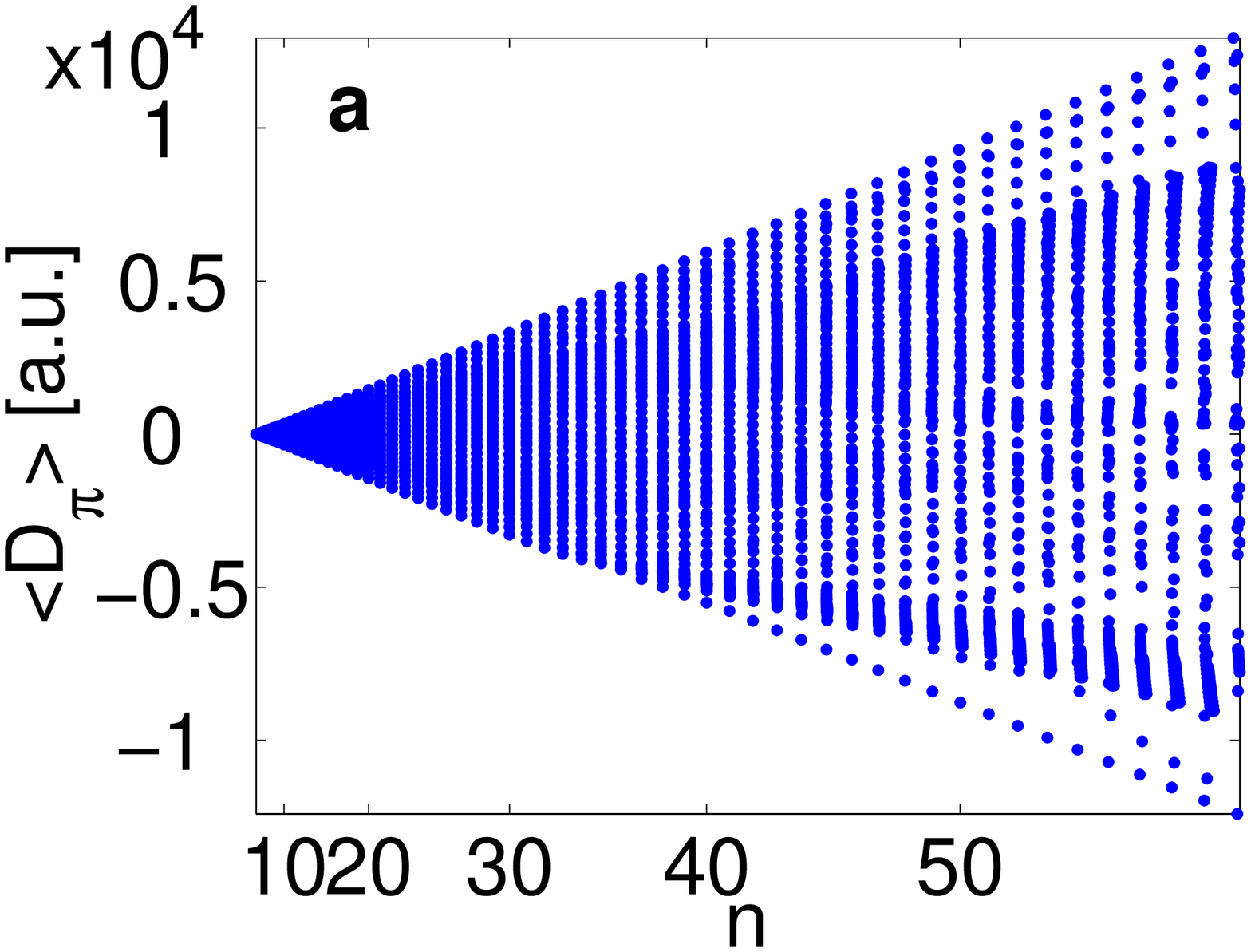}
\includegraphics[angle=0,width=7.3cm]{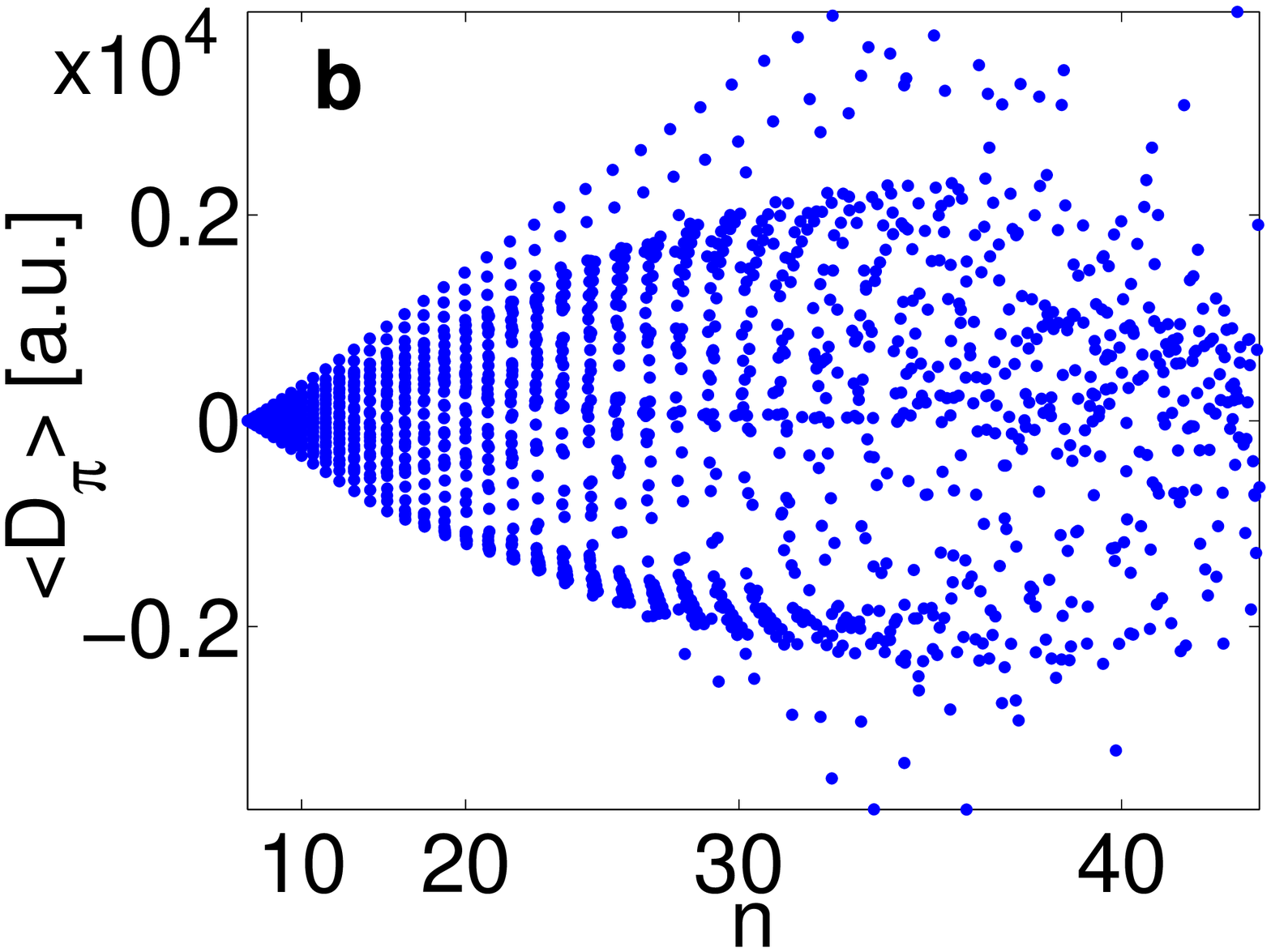}
\caption{Expectation value of the dipole operator $D_\pi$ plotted
against $n$ for different gradients
(\textbf{a}: $b=10^{-10}$, \textbf{b}: $b=10^{-8}$). From \cite{Lesanovsky04b}.}\label{fig:dipolemoment}
\end{figure}
However, the expectation value of $D_\pi$ is in general non-zero.
$\left<D_\pi\right>$ is shown in figure
\ref{fig:dipolemoment}(a,b) for the two gradients $b=10^{-10}$ and
$b=10^{-8}$, respectively. For $b=10^{-10}$ the electric dipole
moments belonging to the same $n$-multiplet are arranged along
vertical lines, which is a result of the approximate degeneracy of
the energy levels. The substates for fixed $m$ of a given
$n$-multiplet exhibit different dipole momenta spreading between
two (upper and lower) bounds that depend linearly on $n$. With increasing gradient
and degree of excitation the $n$-mixing starts and disturbs the
observed regular pattern. For $b=10^{-8}$ and $n>35$, the
distribution of the dipole moments becomes completely irregular.
Therefore we encounter the remarkable effect that the external
magnetic quadrupole field induces a state dependent permanent
electric dipole moment.  This is the result of the asymmetric form
of the wavefunction in the quadrupole field (see discussion of the
probability density in a strong gradient field above).

\subsubsection{Rydberg atoms motion}

\paragraph{\it 3D quadrupole trap:}
We will start with some conceptual ideas and remarks on the 3D quadrupole case, refering further
details to the literature \cite{LesanovskySchmelcherPRL2005,Bock05,Lesanovsky05e}, and will then work out the Ioffe-Pritchard
configuration in more detail. Finally the addition of an electric field, i.e.
a magneto-electric trap, is outlined. 

Similar to the case for a fixed nucleus (see subsection \ref{esryfn}), a moving Rydberg atom in a 3D quadrupole
field possesses a rotational symmetry around the $z$-axis. The corresponding conserved quantity is the projection
$J_z$ of the total (spin and orbital) angular momenta onto the symmetry axis of the quadrupole field.
To exploit this constant of motion there are essentially two pathways
that could be pursued \cite{Bock05} in order to derive effective equations of motion. In both cases the first step is
to transform the two-body problem from laboratory to CM and internal electronic coordinates and
to simplify the resulting Hamiltonian and its coupling terms with a suitable Unitarian.
The first approach eliminates two of the six (CM plus electronic) spatial degrees of freedom and leads to
an (infinite) set of coupled channel equations for the spin and spatial degrees of freedom.
The second approach introduces $J_z$ as a canonical momentum, thereby eliminating the corresponding cyclic angle.
The latter approach has subsequently been used \cite{LesanovskySchmelcherPRL2005,Lesanovsky05e} to explore the case
of a 3D quadrupole field. Since many of the concepts and ideas developed will be described below
for the case of the Ioffe-Pritchard configuration we provide here only the key result for the 3D quadrupole field.
It turns out that trapping is possible for a certain class of coupled CM and electronic quantum states:
Both the CM and electronic motions have equally large angular momentum states. The requirement that specifically
the CM motion bears a large angular momentum is due to the fact that the electronic interaction terms
in the Hamiltonian lead to a clustering of adiabatic potential energy surfaces and therefore strong nonadiabatic 
effects. The latter, however, leads to an uncontrollable unstable CM motion.
Consequently a spatial regime where the energy is dominated by the electronic energy has to be avoided
altogether and this is achieved by increasing the angular momentum which leads to an increase of the
angular momentum barrier terms which in turn shrinks the spatial regime where the electronic interaction
exerts influence on the CM motion. 

To summarize, trapping of Rydberg atoms in the 3D quadrupole field
is possible, but only for
a class of high-angular momentum CM and electronic states, while
control and manipulation of these states are difficult. 
Of course, the experimental preparation of high angular momentum CM states 
requires additional effort. As indicated the electronic part of the trapped states possesses also a large
angular momentum i.e. they are close to circular states \cite{GallagherBook}
for which well-defined experimental recipes of preparation exist \cite{Lutwak97}.

\paragraph{\it Ioffe-Pritchard trap:}
The starting-point is the two-body non-relativistic Hamiltonian in the laboratory frame, 
which incorporates the Rydberg electron and the 
nucleus or effective (closed shell) ionic core of an alkali atom \cite{Hezel06,Hezel07}. 
\begin{eqnarray} \label{eq:Hinit}
    H_{init}&=&\frac{1}{2M_1}\left(\bm{p}_1-q_1\bm{A}(\bm{r}_1)\right)^2
    +\frac{1}{2M_2}(\bm{p}_2-q_2\bm{A}(\bm{r}_2))^2 \nonumber\\
    &&+V(\left|\bm{r}_1-\bm{r}_2\right|)
    -\bm{\mu}_1\cdot\bm{B}(\bm{r}_1)-\bm{\mu}_2\cdot\bm{B}(\bm{r}_2) \; .
\end{eqnarray}
where {\boldmath{$\mu_1$}}, {\boldmath{$\mu_2$}} are the magnetic moments of the particles 1 and 2, respectively.
First one introduces CM and electronic relative coordinates and performs a unitary
transformation reminescent of the transformation applied in the case of a 
homogeneous magnetic field (see $U_n$ in Eq. (\ref{Un})). Retaining only leading order terms
with respect to the inverse masses, neglecting the diamagnetic interaction, and 
inserting the Ioffe-Pritchard field configuration (up to the linear order) for the vector potential, respectively,
give
\begin{gather}
  \label{eq:overallpotfield}
  {\bm{A}}=\frac{B}{2} \left( \begin{array}{ccc} -y \\ x \\ 0 \end{array} \right)
  +G\left( \begin{array}{ccc} 0 \\ 0 \\ x y \end{array} \right) \hspace*{1cm} \\
  {\bm{B}}= B \left( \begin{array}{ccc} 0 \\ 0 \\ 1 \end{array} \right)
  +G \left( \begin{array}{ccc} x \\ -y \\ 0 \end{array} \right) 
\end{gather}
while the Hamiltonian in a.u. is   
\begin{align}\label{hamiop}
H_{IP} =\nonumber H_A +\frac{{\bm{P}^2}}{2M} +\frac{1}{2}{BL_z} +G(x+X)(y+Y)p_z \\- \bm{\mu}_1 \bm B(\bm R + \bm r)-\bm{\mu}_2\bm{B}(\bm R) \; 
\end{align}
where $H_A$ is the operator for the field-free atom. This allows for the separation of
the free CM motion in $Z$-direction.
The Zeeman term comes from the uniform Ioffe field generated by the Helmholtz coils.
The next term, involving the field gradient $G$, arises from the linear field generated by the Ioffe bars and
couples the relative and CM dynamics. The last term couples the spin of particle 2 to the magnetic field.
Since the electronic spins of closed shells combine to zero, the spin of particle two is the nuclear spin only.
Even though $\bm \mu_2\bm B$ scales with ${1}/{M_2}$, this term is important for a proper symmetry analysis,
but will be omitted later on.

The above Hamiltonian $H_{IP}$ is invariant under a number of symmetry transformations $U_S$ 
that are composed of the elementary operations listed in Tab.~\ref{t:sym}. The unitary symmetries are
\begin{subequations}\label{eq:sym}
  \begin{gather}
    P_x P_y S_z \Sigma_z                                    \label{eq:symI} \\
    P_y P_z I_{xy} S_{xy} \Sigma_{xy}                                 \label{eq:symII} \\
    P_x P_z I_{xy} S^{*}_{xy} \Sigma^{*}_{xy} \; .                    \label{eq:symIII}
  \end{gather}
\end{subequations}
The Hamiltonian is also left invariant under the antiunitary symmetry transformation
$T P_y$.   
By consecutively applying the latter operator and the unitary operators Eqs. (\ref{eq:symI}),
(\ref{eq:symII}) and (\ref{eq:symIII}), it is possible to create further antiunitary symmetries:
\begin{subequations}\label{eq:symAall}
  \begin{gather}
    T P_x S_z \Sigma_z                                        \label{eq:symAI} \\
    T P_z I_{xy} S_{xy} \Sigma_{xy}                                     \label{eq:symAII} \\
    T P_x P_y P_z I_{xy} S^{*}_{xy} \Sigma^{*}_{xy} .                   \label{eq:symAIII}
  \end{gather}
\end{subequations}
Considering that $S_{xy}^2=- S_z$ and $\Sigma_{xy}^2=- \Sigma_z$
and that $T$ neither commutes with $S_y$ nor with $S_{xy}$ and $\Sigma_{xy}$,
one finds that the operators (\ref{eq:symI}-\ref{eq:symAIII}) form a symmetry group.

\begin{table}[tb] 
  \caption[Symmetry operation nomenclature]{\label{t:sym} Symmetry operation nomenclature. 
    $P_j$, $S_j$, and $\Sigma_j$ are exemplified by $j=x$, but hold of course also for $j=y,z$.}
  \begin{center}
    \begin{tabular}{lll}
      \hline \\[-12pt]
      operator &            & operation\\
      \hline \hline \\[-9pt]
      $P_{x}$ & $x$ parity & $x\rightarrow -x$, $X\rightarrow -X$ \\
      $S_x$ & electronic spin $x$ op. & $S_y\rightarrow -S_y$, $S_z\rightarrow -S_z$ \\
      $\Sigma_x$ & nuclear spin $x$ op. & $\Sigma_y\rightarrow -\Sigma_y$, $\Sigma_z\rightarrow -\Sigma_z$ \\
      $I_{xy}$ & coordinate exchange & $x\leftrightarrow y$, $X\leftrightarrow Y$ \\
      $S_{xy}$ & el.\ spin component exc.& $S_x\rightarrow -S_y$, $S_y\rightarrow S_x$ \\
      $\Sigma_{xy}$ & nuclear spin comp. exc.&  $\Sigma_x\rightarrow -\Sigma_y$, $\Sigma_y\rightarrow \Sigma_x$ \\
      $T$ & conventional time reversal & $A\rightarrow A^{*}$ \\
      \hline \hline \\[-25pt]
    \end{tabular}
  \end{center}
\end{table}

If no Ioffe field is present ($B=0$), eight additional symmetries can be found leaving the Hamiltonian invariant.
For an effective one particle approach (and the corresponding one particle symmetries) 
this was discussed in Ref.~\cite{Lesanovsky04c}.

For the parameter regime we are focusing on (typical experimental Ioffe field strength and gradients)
there is no inter-manifold mixing and each $n$-manifold can therefore be separately considered.
We introduce the scaled CM coordinates, 
$\mathbf{R}\rightarrow \gamma^{-\frac{1}{3}} \mathbf{R}$, with $\gamma=G M$, 
and the scaled energy  $\epsilon=\gamma^\frac{2}{3}/M$ to remove explicitly the dependence on $B$, $G$, and $M$. 
Introducing the effective magnetic field
\begin{equation}
  \bm{G}(X,Y) = \left( \begin{array}{ccc} X \\ -Y \\ \zeta \end{array} \right) \; ,
  \quad \zeta =BM\gamma^{-\frac{2}{3}}\; ,
\end{equation}
and omitting the zero-field energy offset $E_A^n$, 
the Hamiltonian matrix within an  $n$-manifold has the simple form 
\begin{equation}                                                                                        \label{eq:Hwork}
  \mathcal H= \frac{P_x^2+P_y^2}{2}
  + \bm{\mu}\cdot\bm{G}(X,Y) 
  + \gamma^{\frac{1}{3}} (xyp_z+x S_x-y S_y).
\end{equation}
The first term is the CM kinetic energy.
$\bm \mu$ is the $2n^2$-dimensional matrix representation of the total magnetic moment of the electron,
$\frac{1}{2}(\bm L_{\bm r} +2\bm S)$,
and the second term in eq.(\ref{eq:Hwork}) describes its coupling to the effective magnetic field $\bm{G}$.

The large differences in  masses and velocities in the two body system allows for an adiabatic separation of 
the electronic and the CM motions. This is true regardless of the slow
motion of the Rydberg electron compared to the ground state electron. 
It remains to be investigated how the spatial field inhomogeneity will affect density of levels and 
level crossings. The goal is therefore to find conditions under which isolated adiabatic potential 
energy surfaces might exist. The adiabaticity will be provided by the CM motion.

The adiabatic approximation is introduced by subtracting the transversal CM 
kinetic energy, $\mathcal T={(P_x^2+P_y^2)}/{2}$, 
from the total Hamiltonian (\ref{eq:Hwork}).
The remaining electronic Hamiltonian for fixed position of the CM reads 
\begin{equation}                                                                                       \label{eq:BOHe}
  \mathcal H_e=  \bm{\mu}\cdot\bm{G}(X,Y)   + \gamma^{\frac{1}{3}} (xyp_z+x S_x-y S_y).
\end{equation}
The electronic wave function $\varphi_\kappa$ depends parametrically on $\bm R$ 
and the total atomic wavefunction can be written as 
\begin{equation}
  |\Psi(\bm r,\bm R)\rangle =
  |\varphi_\kappa(\bm r;\bm R)\rangle
  \otimes |\psi_\nu(\bm R)\rangle \; ,
\end{equation}
where $|\psi_\nu(\bm R)\rangle$ is the CM wave function, and the electronic wave functions are solutions of
\begin{equation}                                                                                         \label{eq:internal}
  \mathcal H_e \; |\varphi_\kappa(\bm r;\bm R)\rangle = E_\kappa(X,Y)\; |\varphi_\kappa(\bm r;\bm R)\rangle
\end{equation}
akin to the electronic wave functions in the Born-Oppenheimer approximation. We emphasize that this adiabatic separation
of the CM and electronic coordinates for the Rdyberg atom in intrinsically tied to the field inhomogeneity. The electronic
potential energy surfaces on which the CM moves are $E_\kappa(X,Y)$.
The equation of motion for the CM wave function now reads
\begin{equation}                                                                                         \label{eq:BOcm}
  \left( \mathcal T + E_\kappa(X,Y) \right) \; |\psi_\nu(\bm R)\rangle = \epsilon_{\nu} \; |\psi_\nu(\bm R)\rangle \; .
\end{equation}
A discussion of the nonadiabatic coupling terms can be found in Ref. \cite{Hezel07}.

It can be demonstrated that the energy surfaces $E_\kappa$, exhibit three mirror symmetries.
If we apply the corresponding restricted symmetry operation $U_{P} = P_x P_y \hat S_z \hat \Sigma_z$ in Eq. (\ref{eq:symI}),
that was already shown to leave the Ioffe-Pritchard Hamiltonian in Eq. (\ref{hamiop}) invariant,
to the (unscaled) electronic Hamiltonian $H_e$, we find 
\begin{equation}
U_{P}^\dagger  H_e(\bm r; X,Y)  U_{P} =  H_e(\bm r; -X,-Y) \; .
\end{equation}
i.e. the energy surfaces are inversion symmetric. The symmetry operator $U_{Y}=T P_y$, 
and the operator that is composed of $U_{Y}$ and $U_{P}$, namely
$U_{X}=T P_x \hat S_z \hat \Sigma_z$, mirror the energy surfaces at the axes, 
\begin{align}
  U_{Y}^\dagger H_e(\bm r; X,Y) U_{Y} &=  H_e(\bm r; X,-Y) \; ,\\
  U_{X}^\dagger H_e(\bm r; X,Y) U_{X} &=  H_e(\bm r; -X,Y) \; .
\end{align}

The electronic Hamiltonian in Eq. (\ref{eq:internal}), with the core fixed at an arbitrary position, is three-dimensional. 
No symmetry arguments can be exploited to reduce its dimensionality.

Let us now analyze the properties of the electronic adiabatic potential energy surfaces resulting
from the diagonalization of the electronic Hamiltonian for different regimes of the
Ioffe field strengths and field gradients. These two parameters can be employed
to change the appearance of the potential in which the center of mass dynamics takes place.
A careful inspection of the electronic Hamiltonian shows that its
first term, $\bm{\mu}\cdot\bm{G}(X,Y)$, 
includes the terms $X(\frac{1}{2}L_x+S_x) - Y(\frac{1}{2}L_y+S_y)$, 
that are of the order of $\langle L_i \rangle \approx n$ for high angular momentum states,
and the Zeeman term $\zeta(\frac{1}{2}L_z+S_z)$, which can be as large as $\zeta n$. 
The second term, $\gamma^{{1}/{3}}(xyp_z+x S_x-y S_y)$ can be estimated as $\gamma^{{1}/{3}} n^3$.
In a nutshell, we have the following relative orders of magnitude, 
\begin{equation} \label{eq:factors}
  1  \; , \quad  \zeta \quad \textrm{and} \quad  \gamma^{\frac{1}{3}} n^2 \; .
\end{equation}
Due to the special form of the electronic Hamiltonian, 
changing the magnetic field parameters~$B$ and~$G$ while keeping their ratio~$\zeta/ \gamma^{{1}/{3}}=B/G$
(and $n$) constant results in a mere scaling of the CM~coordinates (we remark that the 
characteristic length scale of the CM dynamics is of the order of one in scaled atomic units). 

To understand the impact of the Ioffe field strength $B$ on the adiabatic energy surfaces, 
we isolate its effect by suppressing other influences. This can be done by choosing a 
relatively low field gradient $G$ and/or a comparatively low $n$.
such that  $\gamma^{\frac{1}{3}} n^2$ is small, and the last term in 
Eq.~(\ref{eq:BOHe}) will become negligible. Within this regime
approximate analytical expressions for the electronic adiabatic energy surfaces can be derived. 
We therefore diagonalize the resulting electronic Hamiltonian 
\begin{equation}            \label{eq:Hetilde}
  \tilde H_e= \frac{1}{2} \bm{G} \, (\bm L + 2\bm S) \; .
\end{equation}
by applying the spatially dependent unitary transformation
\begin{equation}            \label{eq:diagtrafo}
  U_D(X,Y)=  e^{i\phi(L_z+S_z)}  e^{i\beta(L_y+S_y)} \; , 
\end{equation}
with $\phi=\arctan\frac{Y}{X}$, $\cos\beta=\gamma^{-\frac{2}{3}}M_2 B|\bm{G}(X,Y)|^{-1}$ 
and $\sin\beta=-\sqrt{X^2+Y^2}|\bm{G}(X,Y)|^{-1}$. This yields 
\begin{equation}
  U^\dagger_D \tilde H_e U_D = \frac{1}{2} ( L_z + 2 S_z) |\bm G(X,Y)| \; 
\end{equation}
for the transformed approximate electronic Hamiltonian. 
The spatially dependent transformation~$U_D$ locally rotates the magnetic moment of the
electron, which includes its spin and its angular momentum, such that it is parallel to
the local direction of the magnetic field. The operators~$L_z$ and~$S_z$ are not identical
to the ones before the transformation in Eq. (\ref{eq:diagtrafo}); they are rather
related to the local quantization axis defined by the local magnetic field 
direction~\cite{LesanovskySchmelcherPRL2005}.

The approximate adiabatic potential surfaces are now 
\begin{align} \label{eq:Ekappasmallgradient}
  E_\kappa(X,Y) &= \frac{1}{2} (m_l + 2m_s) |\bm G(X,Y)| \nonumber\\
  &= \frac{1}{2} (m_l + 2m_s) \sqrt{X^2+Y^2+\zeta^2} \; .
\end{align}
The possible combinations of $m_l$ and $m_s$ yield $2n+1$ energy surfaces. The surfaces highest and
lowest in energy correspond to circular states, ($|m_l|=l_{max}=n-1$, $m_l+2m_s=\pm n$), 
and they are the only non-degenerate ones. For the other surfaces ($|m_l+2m_s| < n$),
by $2n - |m_l+2m_s+1| - |m_l+2m_s-1|$. Starting from the highest energy surface, 
the levels of degeneracy are thus 0, 2, 4, 6, \dots. 

The approximate surfaces $E_\kappa$ are rotationally symmetric
around the $z$-axis. An expansion around this axis ($\rho = \sqrt{X^2+Y^2} \ll \zeta$) 
yields a harmonic potential, 
\begin{equation} \label{eq:Eksmallrho}
  E_\kappa(\rho) \approx (\zeta + \frac{1}{2\zeta} \rho^2)\cdot\frac{1}{2} (m_l + 2m_s) \; ,
\end{equation}
while we find a linear behavior, 
\begin{equation} \label{eq:Ekbigrho}
  E_\kappa(\rho) \approx \frac{\rho}{2}\cdot(m_l + 2m_s) \; ,
\end{equation} 
when the center of mass is far from the $z$-axis ($\rho \gg \zeta$). 

\begin{figure}[tbh!]
  \centering
  \includegraphics[width=8cm]{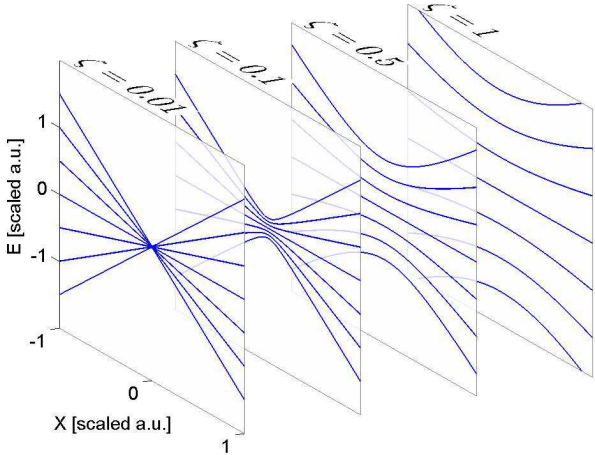}
  \caption[Sections for increasing Ioffe field, $n=3$.]{Sections along the $X$-axis through the electronic adiabatic energy surfaces 
    of an entire $n=3$ manifold. The field gradient is fixed at $G=1$ Tesla/m in order to suppress the influence 
    of the last term in $H_e$ (Eq. \ref{eq:BOHe}). 
    From left to right, $\zeta=B M \gamma^{-{2}/{3}}$ increases due to an increasing Ioffe field. From \cite{Hezel07}.} 
    \label{f:differentzetas}
\end{figure} 

Fig.~\ref{f:differentzetas} shows sections $(\gamma^{{1}/{3}} n^2=0.003, \zeta = 0.01 - 1)$  through all the surfaces 
$n=3$. This principal quantum number has been chosen in order to keep the 
sections simple while displaying the entire $n$-manifold. 

 \begin{figure}[tb!]
   \centering
     \includegraphics[width=7.5cm]{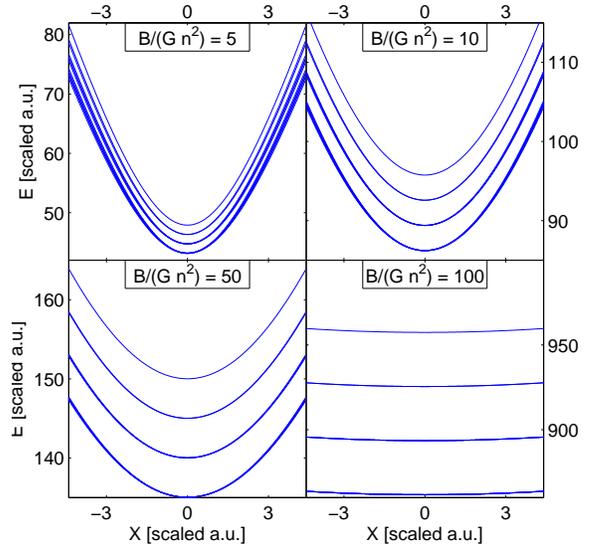} 
     \caption[Sections for increasing Ioffe field, $n=30$.]{Sections along the $X$-axis through the uppermost $21$ surfaces 
       of the $n=30$ manifold of $^{87}$Rb for increasing ratios $B/(G n^2)$. 
       The field gradient is fixed at $G=10$ T/m while the Ioffe field is increased from top left to bottom right. 
       ($B=24$ mG, $B=48$ mG, $B=0.24$ G, $B=0.48$ G). 
       For small ratios $B/(G n^2)$,  the influence of the second term in Eq. (\ref{eq:BOHe}) is not completely suppressed 
       as can be seen from the lifted degeneracies in the upper subfigures. From \cite{Hezel07}.}                      \label{f:n30differentzetas}
 \end{figure} 

Fig.~\ref{f:n30differentzetas} shows for  $n=30, G=0.1$~T/m ($\rightarrow \gamma^{{1}/{3}} n^2=0.14$)
the uppermost $21$ energy surfaces for different values of $B$.
The harmonic behavior around the origin is clearly visible and the
minimal distance between the surfaces becomes larger with increasing $\zeta$.  
Since $\zeta$ and $\gamma^{{1}/{3}} n^2$ are of the same order of magnitude in subfigure (a), 
the contribution of the last term in Eq. (\ref{eq:BOHe}), that lifts the degeneracy of the curves, is visible. 

The minimum energy gap between two adjacent surfaces is at origin 
\begin{equation}      \label{eq:mindist}
  | E_\kappa(O) - E_{\kappa\pm 1}(O) | 
= \frac{B}{2} M\gamma^{-\frac{2}{3}} 
= \frac{\zeta}{2} \; . 
\end{equation}

  \begin{figure}[bt!]
    \centering
    \includegraphics[width=8.5cm]{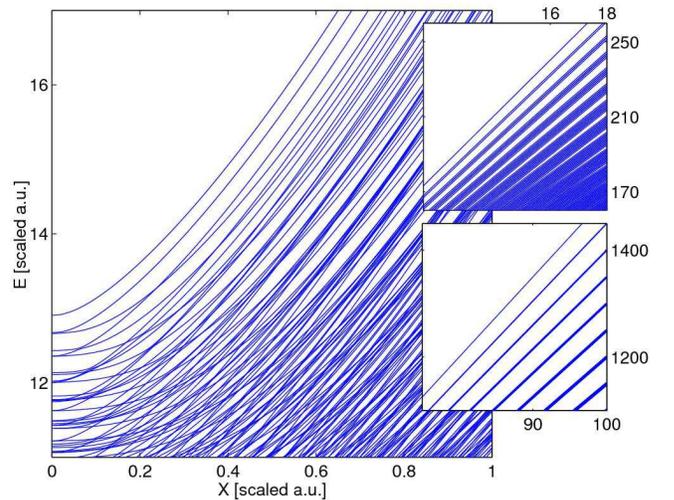}
    \caption[Section $0.01$ G $20$ T/m, $n=30$.]{
      Section through the $n=30$ manifold for a field strength of $0.01$ Gauss and a field gradient of $20$ T/m ($^{87}$Rb).
      A large number of avoided crossings can be observed.
      The uppermost curve, however, stays isolated from the other curves.
      The insets show the linear behavior of the surfaces far away from the $z$-axis. From \cite{Hezel07}.
    }
    \label{f:001G20T}
  \end{figure} 

The parameter $\zeta$ (and hence the field strength $B$) is the tool to control the energetic
distance between the adiabatic surfaces. The uppermost energy surface proves to be very robust:
It is energetically well-isolated from the other adiabatic surfaces and is therefore best-suited
for the trapping of the ultracold Rydberg atoms. This holds even in the regime of small Ioffe field
strength $B$ and comparatively large gradients $G$, illustrated by the corresponding
sections of the adiabatic electronic energy surfaces shown in Fig.~\ref{f:001G20T} for
a Ioffe field strength of $0.01$~G and a field gradient of $20$~T/m.
For these parameters, the contributions of all terms in the electronic Hamiltonian are of the 
same order of magnitude around $X=1$.  One immediate observation is that a large number of avoided crossings occur
between the surfaces, while
the uppermost curve however remains isolated from the rest of the curves.
Far away from the trap center, i.e.~for large $\rho=\sqrt{X^2+Y^2}$, the coupling term in Eq. (\ref{eq:BOHe}),
$X(\frac{1}{2}L_x+S_x) - Y(\frac{1}{2}L_y+S_y)$, becomes dominant.

The energetically uppermost adiabatic electronic energy surface 
is the most suitable for achieving confinement. It does not suffer significant deformation
when the field gradient is increased and it stays energetically isolated from lower surfaces
for a wide range of parameters. The motion along the Z direction is unrestricted. 

By choosing large gradients and appropriate bias fields,
tight confinement for highly excited atoms can be obtained.
For a Ioffe field strength of $B=0.1$ G and a field gradient of $G=100$ T/m, for instance,
the ratio of $\langle \rho \rangle$ and~$\langle r \rangle$ for the ground state ($\nu=1$) is as small as
${\langle \rho \rangle}/{\langle r \rangle} = 0.4$.
The extension of the CM wave function is thus smaller than the extension of the electronic cloud, and the 
Rydberg atoms can no longer be considered as point-like particles.
Moreover, it is possible to enter a regime where the CM and the electronic wave functions do not even overlap
\cite{Hezel07}.

We conclude that the Ioffe-Pritchard field configuration provides a strong confinement 
for ultracold Rydberg atoms in two dimensions.
A relatively weak longitudinal confinement along the $z$-axis
could additionally be provided by employing a non-Helmholtz configuration. The preparation of a 
quasi-stable 1D Rydberg gas in the wave guide provided by the Ioffe-Pritchard field configuration
has been worked out very recently \cite{Mayle07} by adding a homogeneous electric field component
perpendicular to the Ioffe field. The undesirable collisional autoionization process of ultracold Rydberg
atoms is prevented here by the repulsive dipole-dipole interaction among the atoms.

\subsection{Lifetime of trapped Rydberg atoms}\label{lifetime}
Thus far, we were mainly concerned with static properties of Rydberg atoms in strong inhomogeneous magnetic fields. However, when confined in magnetic traps for sufficiently long times, spontaneous decay of excited levels starts to play a crucial role for their dynamics.
Owing to the large excursions of highly excited electrons, the Rydberg state decay rate can be small, scaling 
in the magnetic field-free case as \cite{GallagherBook}
\begin{equation}\label{decay_noB}
\Gamma_{nlm}\sim n^{-3}
\end{equation}
with $n$ for a fixed $l$. 
The decay rate of circular states with $l\sim n-1$ has an even stronger $n$-dependence, $\sim n^{-5}$, 
resulting in extremely long lifetimes of several milliseconds for typical excitations of $n>40$.

\begin{figure}[tb]
\begin{center}
\includegraphics[width=9.0cm]{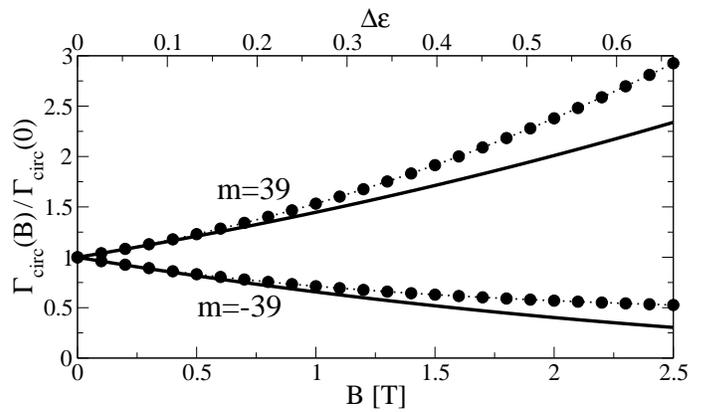}
\end{center}
\caption{Spontaneous decay rate of circular Rydberg states with $m=\pm39$ as function of the magnetic field. Exact numerical results (dots) are compared to the prediction of Eqs. (\ref{decay_B1}) and (\ref{decay_B2}) (solid lines). The upper x-axes shows the relative Zeeman shift $\Delta\varepsilon=|m|\mu_{\rm B}B/E_{nlm}$.}\label{fig3_3_1}
\end{figure}

External magnetic fields strongly affect spontaneous decay \cite{pohl2,Guest03c,robicheauxDecay}, and, depending on $m$, 
can significantly enhance or reduce the lifetime of Rydberg atoms. To illustrate this effect, we consider first 
the simple limit of weak $B$, in which the binding energies experience a linear Zeeman-shift 
$E_{nlm}(B)=E_{nlm}(0)+m\mu_{\rm B}B$, while the transition dipole matrix elements remain unaltered. 
Here $\mu_{\rm B}=e\hbar/2m$ denotes the Bohr magneton. Since the spontaneous decay rate is proportional to the 
third power of the energy difference between the initial and final states, we obtain for the decay from  $(n,l,m)$ to 
$(n^{\prime},l^{\prime},m\pm1)$, the magnetic field dependence 
\begin{equation} \label{decay_B1}
\Gamma^{nlm}_{n^{\prime}l^{\prime}m\pm1}(B)=\left(1\mp\frac{n^2n^{\prime 2}}{n^2-n^{\prime 2}}\frac{\mu_{\rm B}B}{{\mathcal R}}\right)^3\Gamma^{nlm}_{n^{\prime}l^{\prime}m\pm1}(0)\;,
\end{equation}
where ${\mathcal R}=13.6$ eV.
For circular Rydberg states ($|m|=l=n-1$), the transition $(n,m,l)\rightarrow(n-1,m\pm1,l)$ provides the only dipole-allowed decay channel. The decay rate $\Gamma_{\rm circ}^{\pm}$ of circular Rydberg states with $m=\pm|m|$, as obtained from eq.(\ref{decay_B1}) \cite{pohl2}
\begin{equation}\label{decay_B2}
\Gamma_{\rm circ}^{\pm}(B)=\left(1\pm\frac{n^2(n-1)^2}{2n-1}\frac{\mu_{\rm B}}{\mathcal{R}}B\right)^3\Gamma_{\rm circ}(0)
\end{equation}
shows that the decay of positive-$m$ states is enhanced and the decay of negative-$m$ states is reduced by an external magnetic field as compared to the field-free decay rate $\Gamma_{\rm circ}(0)$. In Fig. \ref{fig3_3_1}, 
we compare these simple expressions to exactly calculated  decay rates for circular Rydberg states with $n=40$ and $m=\pm 39$. Eqs.
(\ref{decay_B1}) and (\ref{decay_B2}) yield a rather good description for magnetic fields smaller than $\sim1$T, for which the unperturbed energies are Zeeman shifted by up to $25\%$ and the positive-$m$ decay rates are enhanced by more than $50\%$. At larger fields, $B\gtrsim1T$, Eqs. (\ref{decay_B1}) and (\ref{decay_B2}) systematically underestimate the exact decay rates. This partially originates from the quadratic Zeeman shift, which increases the energy difference between initial and final states for both negative and positive $m$ states. Consequently, at fields that mark the strongly magnetized regime the above perturbative treatment breaks down. 

\begin{figure}[tb]
\begin{center}
\includegraphics[width=9.0cm]{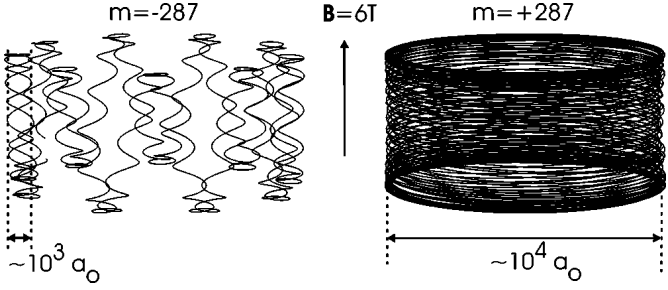}
\end{center}
\caption{Classical electron trajectories of strongly magnetized Rydberg atoms with negative (left) and positive (right) projection of the angular momentum. In both cases, the transverse size is $10^4$ a.u. 
The negative-$m$ state shown in the left panel corresponds to a circular guiding center atom as discussed in section \ref{gca}. 
From \cite{Guest03c}.}\label{fig3_3_2}
\end{figure}

The qualitative decay dynamics of such strongly magnetized atoms can be understood within a classical picture. Rydberg states with large negative 
$L_z$ fall into the class of circular guiding center atoms (see section \ref{gca}). Fig. \ref{fig3_3_2} 
shows a typical trajectory of such an atom \cite{Guest03c}. As will be discussed below, the atom's lifetime is limited by the decay of the ${\bf E}\times{\bf B}$ drift motion, which (see section \ref{gca}) has a frequency $\omega_{\rm D}=e/B\rho^3$. The  electron orbit produces an electric dipole that oscillates with frequency $\omega_{\rm D}$ and amplitude $\rho$. 
The drift motion, thus, radiates with the rate $\sim\omega_{\rm D}^3\rho^2=e^3/B^3\rho^7$. 
For comparison, a field-free, circular atom of the same size has a 
classical electron orbit with  frequency $\omega_0=\sqrt{e^2/m\rho^3}$. 
Note that this expression coincides with the axial bounce frequency $\omega_z$, which we found to be much smaller than $\omega_{\rm D}$ (see Eq. (\ref{frequ_hierarchy}), section \ref{gca}). This simple estimate, hence, yields decay a rate $\Gamma_{\rm circ}^{-}$, suppressed by a factor of 
\begin{equation}\label{decay_negm}
\frac{\Gamma_{\rm circ}^{-}(B)}{\Gamma_{\rm circ}(0)}\sim\frac{{m\rho}^{3/2}}{B^3}\;.
\end{equation}
The corresponding classical orbit with the same radius $\rho$ but positive angular momentum $L_z$ is  also shown in Fig.\ref{fig3_3_2} \cite{Guest03c}. Flipping the angular momentum completely changes the character of the classical electron motion. 
The slow magnetron motion is replaced by a "giant" cylotron oscillation around the ion. 
Consequently, the decay rate $\Gamma_{\rm circ}^{+}(B)\sim\omega_{\rm c}^3\rho^2$ of circular, positive-$m$ states is drastically enhanced by a factor of
\begin{equation}\label{decay_posm}
\frac{\Gamma_{\rm circ}^{+}(B)}{\Gamma_{\rm circ}(0)}\sim B^3\rho^9\;,
\end{equation}
with respect to the field-free decay rate $\Gamma_{\rm circ}(0)$.

\begin{figure}[tb]
\begin{center}
\includegraphics[width=6.7cm]{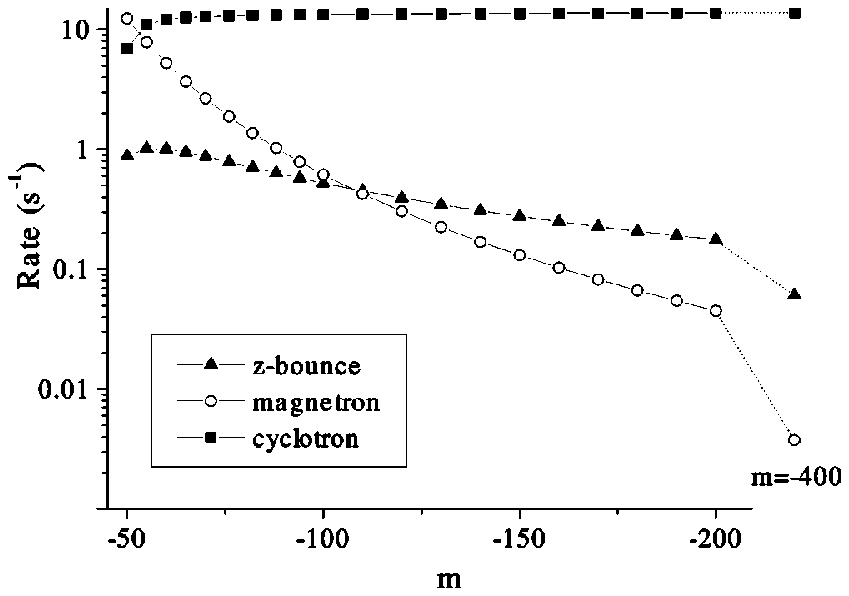}
\includegraphics[width=6.7cm]{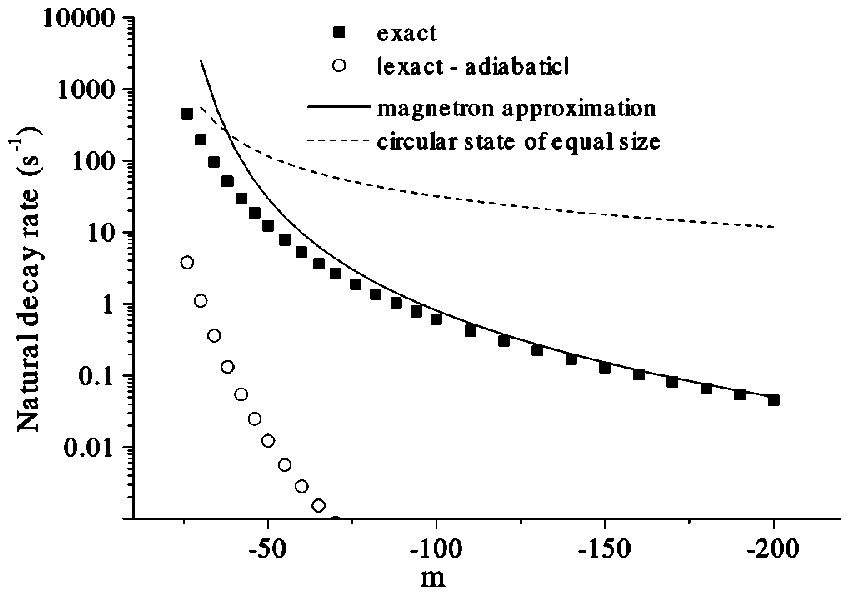}
\end{center}
\caption{Left: $m$-dependence of the cyclotron, $z$-bounce and magnetron decay rates of Rydberg atoms in a $6$T magnetic field. The rates are caluclated for the lowest states of the $m$ manifolds for which the respective transition can occur. Right: Decay rate for the lowest states within the $m$-manifold. The adiabatic approximation (squares) is compared to exact results, showing agreement for $|m|>4B^{-1/3}\approx140$ (see Eq.(\ref{BO_cond})). The field-free decay rates of circular Rydberg atoms with a similar size demonstrates the strong suppression due to the magnetic field. From \cite{Guest03c}.}\label{fig3_3_3}
\end{figure}

An insightful quantum treatment of strongly magnetized Rydberg atoms in the spirit of the guiding center approximation has been put forward in \cite{Guest03b}. Using a Born-Oppenheimer approximation, the $z$-bounce and 
transverse $\rho$ dynamics are decoupled adiabatically \cite{Guest03b,Guest03c}. If this Born-Oppenheimer approximation holds, which is the case for \cite{Guest03c}
\begin{equation} \label{BO_cond}
|m|>4B^{-1/3}\;,
\end{equation}
the energies associated with the cyclotron, the magnetron drift, and the axial bounce motion are adiabatic constants of motion, such that respective approximate quantum numbers can be defined. This permits to categorize the radiative decay in 
cyclotron transitions, bounce, and magnetron decays. Fig. \ref{fig3_3_3} shows the corresponding decay rates as a 
function of $m<0$. If the condition 
(\ref{BO_cond}) is fulfilled, the three decay rates form a hierarchy, just like the corresponding oscillation frequencies (see Eq.(\ref{frequ_hierarchy}), section \ref{gca}). 
Consequently, a guiding center atom quickly dissipates its cyclotron energy, followed by a decay of the axial degree of freedom to its ground state, such that the overall decay rate of the Rydberg state is limited by the magnetron decay. As shown in Fig. \ref{fig3_3_3}, 
the latter can be suppressed by several orders of magnitude for highly excited states.

Such guiding center atoms (with $m<0$), when sufficiently deeply bound, typically have high-field seeking character, and, thus, are 
removed from the magnetic traps. 
A trapped Rydberg gas, hence, consists of atoms in positive-$m$ states, such that its spontaneous decay is enhanced compared to the field-free case. However,  the magnetic moment of strongly magnetized, negative-$m$ states reverses its 
sign with increasing atom size ($\rho>2E_{\rm cyc}/e^2$), such that very highly excited states can be trapped 
despite having $m<0$. It is this sign reversal at high excitations that has enabled trapping of extremely long-lived, strongly-magnetized Rydberg gases \cite{ChoiRaithel2005,ckll}. While this property is certainly appealing for studies of magnetized Rydberg gases, the opposite is the case for antihydrogen experiments, which aim at the production of ground state atoms through recombination in cold plasmas. It appears that the production of positive $m$ is essential, since these states remain low-field seekers during their radiative cascade to lower-lying states. Here, the field-enhanced decay rates of these low-field seeking states  is beneficial, as it accelerates the production of cold ground state antihydrogen atoms from radiatively decaying Rydberg gases.
A more detailed account of the decay dynamics of such trapped Rydberg gases will be given in the next section.

\subsection{Long-time dynamics and cooling in strong field traps}\label{selfcooling}
The preceding discussion of the magnetic moments  
and spontaneous decay rates (section \ref{lifetime}) elucidates 
how to separately treat either the translational center-of-mass dynamics in weak-gradient traps or the radiative cascade of Rydberg states. However, since the magnetic trapping potential depends on the atom's internal state, both of these processes are intimately connected. 
In fact, it was shown in Refs. \cite{pohl2,tjr06}  that this entangled dynamics can lead to significant cooling of the 
trapped Rydberg gas during the radiative cascade of the atoms to their internal ground states. This effect may be of potential importance for antihydrogen experiments, as it strongly enhances the trapping efficiency of ground state atoms produced through radiative decay of initially recombined (see section \ref{sec:ryatffp}) highly excited Rydberg atoms.

Several different magnetic and electric field configurations are currently used to simultaneously confine charged and neutral particles, in order to synthesize neutral antihydrogen atoms within a two-component plasma and trap the produced atoms. Both the ATRAP and the ALPHA 
collaborations use a  nested Penning trap (see section \ref{HbarExp}) to create antihydrogen atoms. 
This plasma trap is superimposed by either a quadrupole (ATRAP, \cite{atrap07,atrap08}) or an octupole (ALPHA, \cite{alpha07}) magnetic field to provide confinement for eventually formed neutral atoms. This additional field breaks the cylindrical symmetry of the total magnetic field. A lack of cylindrical symmetry implies a lack of a confinement theorem \cite{dun99} for the charged particles and, thus, ultimately causes loss of plasma charges \cite{alpha}. However, both trapping configurations were experimentally demonstrated to provide sufficiently long confinement times \cite{atrap07,atrap08,alpha07}. Alternatively, the ASACUSA collaboration developed a so-called MCEO trap, which consists of an electric octupole field and a  magnetic quadrupole (cusp) field to simultaneously confine positrons, antiprotons and antihydrogen  atoms in the same spatial region \cite{mohri}. Producing a cylindrically symmetric magnetic field, this configuration promises long plasma confinement, but might cause atom loss at the magnetic field minimum where $B=0$ \cite{mohri}. In the following, we provide a general description of the dynamics of Rydberg gases in weak-gradient magnetic multipole traps. We will further assume that low-field seeking atoms, i.e. positive-$m$ states, 
are present in the trap. The preceding formation of such Rydberg atoms in cold plasmas along with the characteristic internal states resulting from their formation will be discussed in  section \ref{sec:ryatffp}.

Since we consider slowly moving atoms in weak-gradient magnetic traps, it suffices to determine the atoms'  
field- and state-dependent magnetic moments $\mu(B,{\bm \eta})$ and radiative decay rates 
$\Gamma_{{\bm \eta}^{\prime}}^{{\bm \eta}}(B)$. One can then employ a semiclassical treatment of the gas dynamics described by the kinetic equation  \cite{pohl2}
\footnotesize
\begin{eqnarray}\label{kineq}
\left(\frac{\partial}{\partial t}+{\bf v}\frac{\partial}{\partial {\bf r}}+\frac{\partial U_{\bm \eta}({\bf r})}{\partial {\bf r}}\frac{\partial}{\partial {\bf v}}\right)f({\bf r},{\bf v},{\bm \eta},t)&=&  \sum_{{\bm \eta}^{\prime}}\Gamma_{{\bm \eta}}^{{\bm \eta}^{\prime}}f({\bf r},{\bf v},{\bm \eta}^{\prime},t)\nonumber\\&&  - \Gamma_{{\bm \eta}^{\prime}}^{{\bm \eta}}f({\bf r},{\bf v},{\bm \eta},t)\;
\end{eqnarray}
\normalsize

determining the time-evolution of the phase-space density $f({\bf r},{\bf v},{\bm \eta},t)$, i.e. the probability to find an atom at a position ${\bf r}$, with a velocity ${\bf v}$ and in an atomic state ${\bm \eta}$, denoting a set of quantum numbers that defines the internal state. The trapping potential is given by $U_{\bm \eta}({\bf r})=\mu(B,{\bm \eta})B({\bf r})$. It is the ${\bm \eta}$-dependence of the confinement potential that couples the translational center-of-mass dynamics to the internal state dynamics, as  described by the right-hand side of Eq. (\ref{kineq}). It can be efficiently solved using a test-particle treatment. Within this approach, 
one randomly generates an ensemble of atomic states (${\bf r}$, ${\bf v}$ and ${\bm \eta}$) and propagates the classical coordinates according to the equations of motion. After each propagation time step, 
the internal state is changed according to the radiative decay rates by using a Monte Carlo procedure.

\begin{figure}[tb]
\begin{center}
\includegraphics[width=9.0cm]{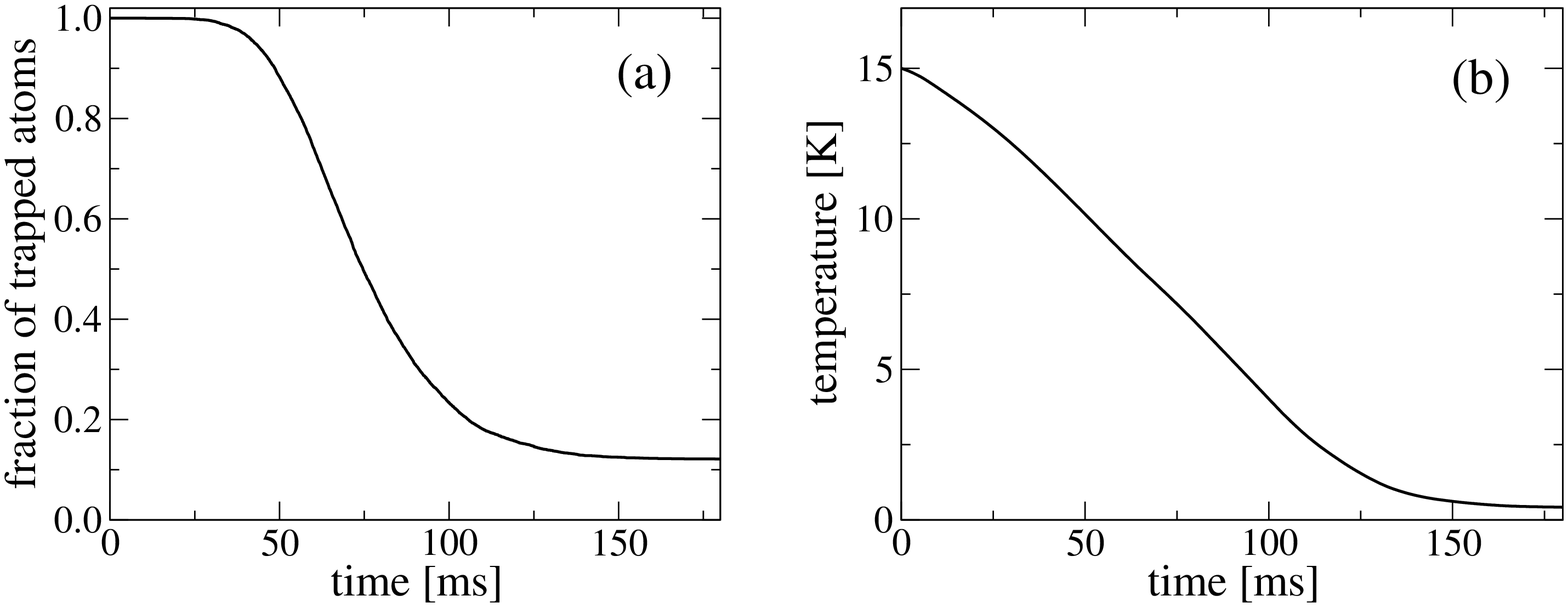}
\end{center}
\caption{Time evolution of the number of trapped atoms (a) and the atomic
  temperature (b) for $n_0=44$, $m_0=43$. Calculations have been performed for a magnetic cusp trap with a $B$-field difference of $4$T.}\label{fig3_4_1}
\end{figure}

Fig. \ref{fig3_4_1} shows an example of the calculated time evolution of the atomic temperature and the fraction of 
trapped atoms in a magnetic cusp trap with a depth of $4T$. The temperature drastically decreases from its initial value of $15$K to below $\sim1$K, and the fraction of finally trapped ground state atoms is as large as $10$\%. 
Considering that the effective trap depth decreases to about $1/10$ of the initial temperature, this is a remarkably large trapping efficiency, achieved only through the additional cooling.

The basic mechanism can be understood in rather simple terms. Whenever an atom at position ${\bf r}_{0}$ decays to a state with a lower magnetic moment $\mu^{\prime}<\mu$, the local trapping potential decreases by a factor $\mu^{\prime}/\mu$. Consequently, the atom looses potential energy and, therefore, decreases its total energy by $(\mu^{\prime}-\mu)B({\bf r}_0)$. Subsequently, the total energy loss is distributed between potential and kinetic energy as the atom moves through the trap. This causes a net cooling effect and thereby increases the trapping efficiency of atoms following the radiative cascade.

\begin{figure}[tb]
\begin{center}
\includegraphics[width=9.0cm]{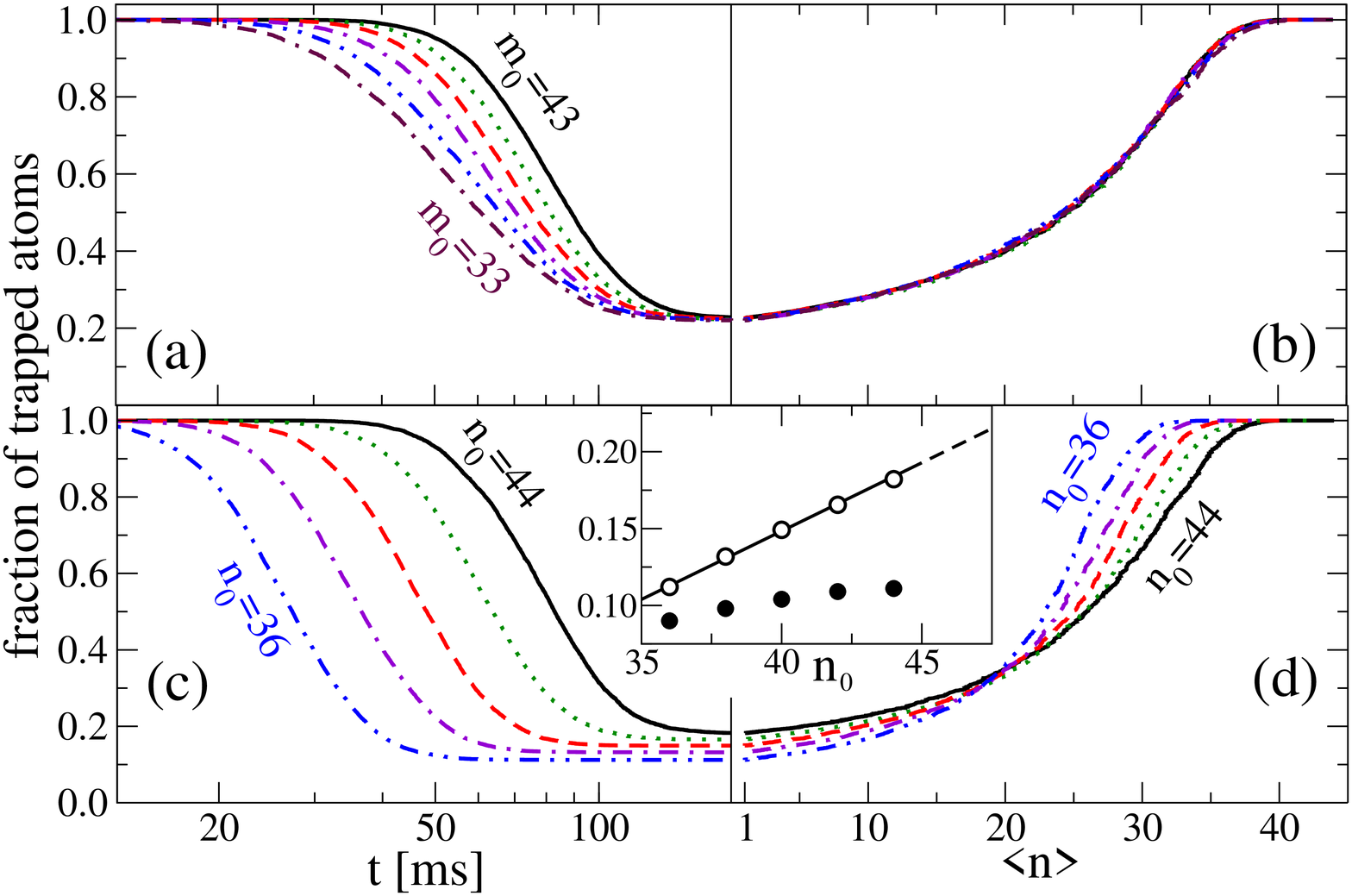}
\end{center}
\caption{Fraction of trapped atoms as functions of time [(a), (c)] and average principal 
quantum number [(b), (d)], for $n_0=44$ and different initial $33\le m_0\le 43$ [in (a) and (b) 
at $T(0)=10$ K] and circular Rydberg atoms with $36\le n_0\le 44$
[in (c) and (d) at $T(0)=12$ K]. Different curves are steps of two in $m_0$ and $n_0$. 
The inset shows the linear $n_0$-dependence of the fraction of trapped atoms (open circles), 
which disappears when strong magnetic field
effects are neglected (filled circles). The trap parameters are identical to those in Fig.~\ref{fig3_4_1}. From \cite{pohl2}.}\label{fig3_4_2}
\end{figure}

The cooling efficiency is generally expected to depend on the initial distribution of bound Rydberg states.  However, since these are generally unknown in experiments, calculations based on particular initial state distributions might be of limited practical applicability. 
Instead, Fig. \ref{fig3_4_2} shows the trapping efficiency for particular initial energies ($n_0$)\footnote{In the presence of a strong magnetic 
field, the field-free quantum numbers $n$ and $l$ are no longer good quantum 
numbers, but may, nevertheless, be used to label the atomic state, that
connects to its field-free counterpart as $B$ is decreased to zero.} and angular momenta ($m_0$). For fixed $n_0$ and varying $m_0$ atom loss accelerates with decreasing $m_0$, while the final fraction of trapped atoms is practically independent of $m_0$. When plotted against the 
average principal quantum number $\left<n\right>$, the results collapse on a single curve.
This insensitivity to $m_0$ is due to the fact, that the excited \Hbar is quickly driven to a 
circular Rydberg state during the radiative cascade \cite{flannery}. The following discussion can, thus, safely be restricted to circular Rydberg states. On the other hand, the initial value of $n$ has a clear impact on the trapping efficiency, as shown in 
Fig.\ref{fig3_4_2}(c) and \ref{fig3_4_2}(d). 
The fraction of trapped atoms increases linearly with $n_0$, largely a consequence of diamagnetic field effects on the atomic state.

Further insights into the cooling process can be gained by considering a simplified spherically symmetric field $B=\beta(r/\lambda)^{\gamma}$. Here the parameter $\lambda$ controls the gradient length and $\gamma$ the multipole order of the field, with $\gamma=1$ 
corresponding to the cusp trap case, and  
larger values of $\gamma$ reproducing the central trap shape of higher order multipole fields.
Furthermore, we neglect the diamagnetic term in the Hamiltonian, such that the external 
potential of circular states is given by $U_n=\mu_{\rm B}nB$. An atom in state 
$n$ with a total center-of-mass energy $E_n=K_n+U_n$, and which decays at position ${\bf r}_0$ 
to a lower lying state $n^{\prime}$, therefore loses energy of $\mu_{\rm B}(n^{\prime}-n)B$. 
According to the virial theorem, the final average kinetic energy after redistribution is given by
\begin{equation} \label{temp1}
\left<K_{\rm n^{\prime}}\right>=\left<K_n\right>-\frac{\gamma}{2+\gamma}\left(\frac{n-n^{\prime}}{n}\right)U_n(r_0)\;.
\end{equation}
From this relation, we can derive the average change in kinetic energy by averaging over ${\bf r}_0$ according to the distribution of radiative decay events. 

\begin{figure}[tb]
\begin{center}
\includegraphics[width=8.0cm]{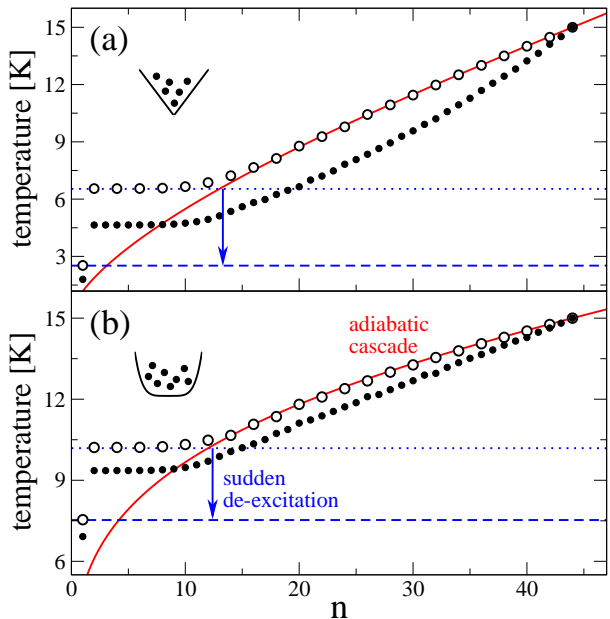}
\end{center}
\caption{$n$-dependence of the atomic temperature of circular Rydberg states for $n_0=44$, 
$T(0)=15$K, $\gamma=1$ (a) and $\gamma=5$ (b). 
The open circles show the numerical results neglecting the diamagnetic term in the atomic 
Hamiltonian, and the filled circles result from the full calculations. 
The solid line is obtained for the adiabatic cascade from Eq.~(\ref{temp4}), while the dotted 
and dashed lines indicate the temperature decrease due to the sudden de-excitation according 
to Eq.~(\ref{temp2}). From \cite{pohl2}.}\label{fig3_4_3}
\end{figure}

Let us first consider the case of $n$ being sufficiently low such that radiative decay to the ground state is 
much faster than the timescale of the atomic motion. In this case, the $B$-dependence 
of the transition rate is unimportant. We can, consequently, once more employ the virial theorem 
$\left<U(r_0)\right>=2\left<K_n\right>/\gamma$ and obtain for the temperature $T_n$
\begin{equation} \label{temp2}
T_{\rm 1}=\frac{n\gamma+2}{n(\gamma+2)}T_n\;.
\end{equation}

In the opposite limit of radiative decay proceeding slowly compared to the timescale of 
the atomic motion, the temperature decreases in consecutive steps according to Eq.~(\ref{temp1}) 
and one has to account for the position dependence of the transition rate according to 
Eq.~(\ref{decay_B2}).
Performing the ensemble average and considering the limit $(n-n^{\prime})/n\rightarrow0$, yields the following differential equation for the temperature, up to first order in $T_n/\varepsilon_n$
\begin{equation} \label{temp3}
\frac{dT}{dn}=\left(2+\frac{9\left<\delta U^2\right>}{\gamma}\frac{T}{\varepsilon_n}\right)\frac{T}{(2+\gamma)n}\;,
\end{equation}
where
$\left<\delta U^2\right>=\frac{\left<U^2\right>-\left<U\right>^2}{\left<U\right>^2}=\frac{\gamma2}{5\gamma+6}$. 
Eq.~(\ref{temp3}) permits a simple solution
\begin{equation} \label{temp4}
T_n=\frac{T_{n_0}\left(\frac{n}{n_0}\right)^{\frac{2}{2+\gamma}}}{1+\kappa\left[1-\left(\frac{n}{n_0}\right)^{\frac{2\gamma+6}{\gamma+2}}\right]}\;,
\end{equation}
with $\kappa=\frac{9\gamma}{2(5\gamma+6)(\gamma+3)}\frac{k_{\rm
    B}T_{n_0}}{\varepsilon_{n_0}}$. This expression reveals an approximate power-law cooling $\propto n^{2/(2+\gamma)}$, 
which is enhanced by magnetic field effects since $\kappa>0$. 
Fig.~\ref{fig3_4_3} shows the numerically calculated temperature 
$k_{\rm B}T_n=\frac{M}{3}\int v^2f(r,v,n,t)d{\bf r}d{\bf v}dt$. 
The two cooling regimes, i.e. the adiabatic cascade followed by a sudden 
de-excitation, can clearly be identified and 
nicely follow the analytical predictions. Both cooling mechanisms become less 
efficient with increasing multipole order, and are expected to yield optimal trapping efficiencies in quadrupole (cusp) traps.

We finally point out, that the above discussion exclusively includes radiative decay and assumes that positive-$m$ states have been formed in the trap. In a plasma, however, charged particle collisions, such as electron(positron)-impact induced (de)excitation, collisional ionization, and recombination crucially affect the formation and subsequent dynamics of Rydberg atoms in cold magnetized plasmas. The formation of Rydberg atoms due to such collisions will be discussed in the next section.

\section{Rydberg atom formation in cold, magnetized plasmas}\label{sec:ryatffp}
The production of highly-excited anti-hydrogen (\Hbar) atoms in Penning traps \cite{athena,atrap1,atrap2,athena2,athena_temp,alpha07,atrap07,atrap08} 
has focussed attention on the mechanisms of Rydberg atom formation in strong magnetic fields.
Generally, atom formation in plasmas largely proceeds through three-body recombination (TBR) or radiative recombination. In the 
former process, two electrons collide in the vicinity of an ion, whereby one electron forms a bound state and the secondary one absorbs 
the corresponding excess energy. At the low temperatures considered here, TBR is expected to yield the dominant contribution. 
This expectation originates from the strong $\sim T^{-9/2}$ temperature scaling of the TBR rate, known from non-magnetized plasmas.
However, in magnetized plasmas, the electron dynamics is drastically modified due to its strong transversal confinement by the 
magnetic field.

Below we review some consequences of these modifications for the collisional recombination dynamics. Special emphasis is placed on implications for the production of antihydrogen atoms and on the particular effects of the employed trapping configuration.
\subsection{Three-body recombination and capture}
Early numerical calculations of TBR rates were performed by Mansbach and Keck \cite{MK69}, based on a variational 
approach combined with classical trajectory Monte Carlo (CTMC) calculations. This study confirmed an intriguingly simple 
expression for the recombination rate
\begin{equation}
\nu_{\rm TBR} \propto \rho^2\frac{1}{T^{9/2}}
\end{equation}
in terms of the electron density $\rho$ and temperature $T$, which is in accord with measurements in hot and cold plasmas \cite{MK69,VS80} as well as ultracold plasmas \cite{fzr07,unp12}. This temperature and density scaling can be qualitatively understood from simple dimensional scaling arguments.
The classical distance of closest approach $b=e^2/k_BT$ also corresponds to the typical initial Rydberg size and yields the cross section $\sim\pi b^2$ of the corresponding ion electron collision. With the electron density $\rho_{\rm e}$ and the thermal electron velocity $v_{\rm th}=\sqrt{k_{\rm B}T/m_{\rm e}}$, the collision rate can thus be estimated as $\sim \rho v_{\rm th}b^2$. Multiplication of this rate by the probability $\sim \rho_{\rm e}b^3$ to find a second electron in the collision region gives for the TBR rate 
\begin{equation}\label{sect41_eq1}
\nu_{\rm TBR} = C_0 \rho^2 v_{\rm th}b^5\sim\rho^2\frac{1}{T^{9/2}}\;.
\end{equation}
The variational Monte Carlo simulations of \cite{MK69} yield for the proportionality constant $C_0=0.76$.

In an early work, Glinsky and O'Neil \cite{GCA2} studied the effect of a very strong magnetic field on the recombination dynamics,
using a similar Monte Carlo approach, formulated in the $B\rightarrow\infty$ limit, where electron trajectories are assumed to be 
pinned onto the magnetic field lines. Remarkably, it was found that, despite reducing the dimensionality of the electron dynamics,
an infinitely strong magnetic field does not affect the field-free scaling behavior Eq.(\ref{sect41_eq1}). The calculations recovered the strong 
$\sim T^{-9/2}$ scaling but with a reduced pre-coefficient $C_{\infty}=0.07$.

As for the field-free case \cite{MK69}, this work also pointed out the importance of a kinetic bottleneck in the electron's bound state
phase space. The corresponding bottleneck energy of about $E_{\rm bn}\approx4k_{\rm B}T$ distinguishes between truly recombined 
and temporarily bound electrons. Above  $E_{\rm bn}$, frequent electron-atom collisions establish a steady state binding energy distribution 
in equilibrium with the free plasma electrons, while below the bottleneck energy re-ionization becomes ineffective resulting in a collisional 
cascade to deeper binding. Within this framework, the recombination rate is then obtained as the one-way phase space flux through the energy 
surface defined by $E_{\rm bn}$.
In the next section this concept will be applied to determine the distribution of low-lying bound states produced in antihydrogen 
experiments.

Robicheaux and Hanson \cite{RobiHanson02} later performed CTMC calculations that relaxed several of the 
above assumptions. The simulations partially account for the center of mass coupling (see section \ref{comc}) by following the exact motion 
of the protons, and include the electronic ${\bf E}\times{\bf B}$ drift motion within the guiding center drift approximation (see section \ref{gca}).
For finite magnetic field strengths of $B=3{\rm T}$ and $B=5.4{\rm T}$, used in the first antihydrogen experiments \cite{athena,atrap1}, an 
increased recombination rate by about $60\%$ at typical temperatures in the range of $T=4$K to $T=16$K was found.

While the above considerations yield a comparably large rate for forming atoms below the bottleneck energy, their subsequent collisional cascade 
to deeper binding may, however, be much slower. Since de-excitation is mainly caused by replacement collisions \cite{GCA2}, where the projectile 
electron traverses the bound state orbit, the corresponding collisional cross section quickly drops due to the ever decreasing size of the atoms during 
the cascade. In \cite{fedichev97} it was pointed out that another class of electron trajectories, large impact parameter collisions, also affect the Rydberg 
atom's binding energy evolution. Here, the collisional drag between a Rydberg electron and the background plasma leads to a diffusive relaxation
of the bound state orbit to deeper binding. Based on simple scaling arguments, it was concluded that the rate of this relaxation increases monotonically 
with binding energy, which, consequently, would potentially provide an efficient mechanism to form deeply bound antihydrogen atoms. 
However, Bass and Dubin \cite{bassdubin04} later incorporated the effect of the bound electron's ${\bf E}\times{\bf B}$ drift motion on the diffusion 
coefficient and identified an adiabatic cutoff at large binding energies, that is determined by the corresponding drift velocity 
$v_{\rm D}= r_{\rm D}\omega_{\rm D}$, where $r_{\rm D}$ and $\omega_{\rm D}$ are the radius and frequency of the magnetron orbit (see Fig.\ref{fig2_6_1}).
If $v_{\rm D}$ is much less than the thermal electron velocity $v_{\rm th}$ ($v_{\rm D}/v_{\rm th}=\xi\ll1$), the atom's energy loss rate was found to scale as 
$\xi^{3/2}\log^2\xi$, i.e. to increase with increasing binding energy of the atom. However, for $\xi\gg1$, the adiabatic cutoff was shown to take over and 
the rate drops exponentially as $\xi^{7/6}\exp(-3(2\xi)^{2/3}/2)$. Consequently, de-excitation drastically slows down during the collisional cascade 
to deeper binding, posing a major limitation to the efficiency of low-lying state production in antihydrogen experiments.

Another problem, specific to the antihydrogen production schemes, is that the antiprotons only spend a limited amount
of time in the reaction region, i.e. around the central saddle of the nested Penning trap (see Fig. \ref{NestedPenning}). This was already 
pointed out in \cite{GCA2} to violate the steady state condition, required to arrive at the recombination rate, Eq.(\ref{sect41_eq1}). Indeed, recent 
 measurements of the recombination rate in a nested Penning trap \cite{athena_temp,athena_temp2} have found a significant departure from the 
temperature scaling predicted by Eq.(\ref{sect41_eq1}), which may arise from the limited antiproton-positron interaction times.
A more detailed investigation of this effect for antihydrogen experiments has been performed in \cite{robicheaux}, which also included the 
antiproton dynamics outside the reaction region in the side-wells of the nested Penning trap. The calculations use a Monte Carlo procedure 
to trace the recombination dynamics of a single antiproton, traversing the central positron cloud for a finite amount of time. To make such 
calculations feasible, only a small box centered around the antiproton is considered explicitly, which is continuously filled with positrons 
according to their equilibrium phase-space distribution \cite{robicheaux}. The simulations qualitatively confirmed observations of preceding 
experiments. Subsequent measurements by the ATRAP collaboration, however, found surprisingly large numbers of deeply bound atomic 
states, inconsistent with previous calculations \cite{GabRev05}. In the following section we will provide a more detailed interpretation  
of these measurements, and show that discrepancies can be traced back to a break-down of the guiding center approximation.

\subsection{Two-step capture and formation of deeply-bound states} \label{sect_twostep}
\begin{figure} [b!]
\includegraphics[width=3.0in]
{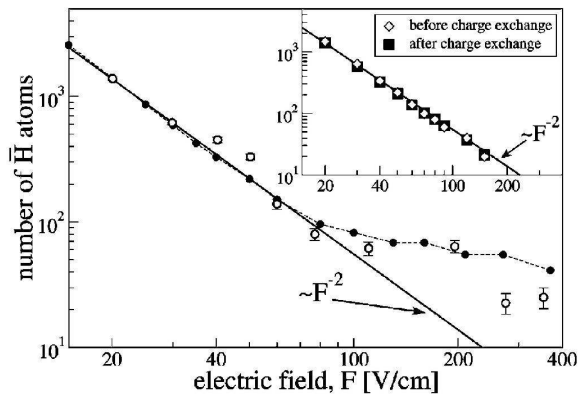}
\caption{Field-ionization spectrum of \Hbar atoms that survive a given 
electric field $F$. The experimental measurement \cite{GabRev05} (open circles) are compared to
the exact Monte Carlo calculations (filled circles). The inset demonstrates that charge exchange (see section \ref{sect_chx})
does not alter the field-ionization spectrum. Here GCA simulations before  and after charge exchange are compared. The
theoretical curves have been rescaled to match the measured
atom number at $20$ V/cm. From \cite{pohl1} \label{FI}}
\end{figure}

In Refs.\cite{atrap2,GabRev05} the ATRAP collaboration performed state selective field ionization measurements
to obtain additional information about the internal states of the produced antihydrogen atoms. Here a variable electric field
was applied in front of the detector, which ionizes weakly bound atoms. By monitoring the number of detected atoms as a function
of the electric field amplitude, the resulting field ionization spectrum yields the number of atoms that are 
sufficiently deeply bound to survive a given electric field $F$. The measured spectrum shown in Fig.\ref{FI} has two distinct features. 
At small fields, corresponding to comparably weakly bound states, it follows a simple power-law decay $\sim F^{-2}$. At larger fields, 
implying potentially deeply bound states, one finds a significant departure from the $\sim F^{-2}$ behavior, and much larger numbers of 
atoms are observed.

\begin{figure}[b!]
\includegraphics[width=2.8in]{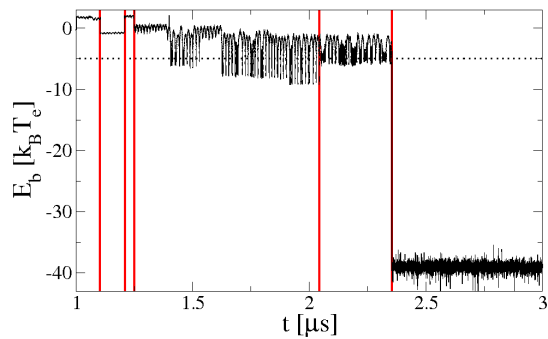}\hskip10pt
\caption{
Time evolution of the atomic binding energy, illustrating the two-step formation process.
The electron is initially captured near the kinetic bottleneck as indicated by the horizontal
dotted line. During this phase, the atom undergoes a number of replacement collisions
(vertical lines) until its bound electron is eventually driven down to very deep binding
($\sim -40 k_B T_{\rm {e^+}}$) by a single collision near $t \sim 2.3\:\mu$s. From Ref. \cite{pohl3}.
}\label{orbitt}
\end{figure}

Fig.\ref{FI} also shows the result of GCA Monte-Carlo simulations \cite{pohl1}, demonstrating good agreement with the observed 
low-field power-law. The individual binding energy evolutions obtained from such simulations show that the majority of atoms with 
sufficiently large binding energies are formed within a two-step process. A typical example is shown in Fig.\ref{orbitt}. First, an 
electron is quickly captured via three-body collisions into a highly excited state. This bound state undergoes frequent collisions, 
establishing a steady state equilibrium with the surrounding plasma electrons. However, there is a small but finite probability for a 
strong single replacement collision that drives the atom to a very deeply bound state below the bottleneck energy. As discussed 
above, such deeply bound states are stable against re-ionizing collisions. This process appears to be characteristic to the antihydrogen 
production schemes, since the short interaction time of the antiproton with the positron plasma ensures that de-excitation below the 
bottleneck largely occurs via a single collision.

\begin{figure}[tb!]
\includegraphics*[width=2.6in]{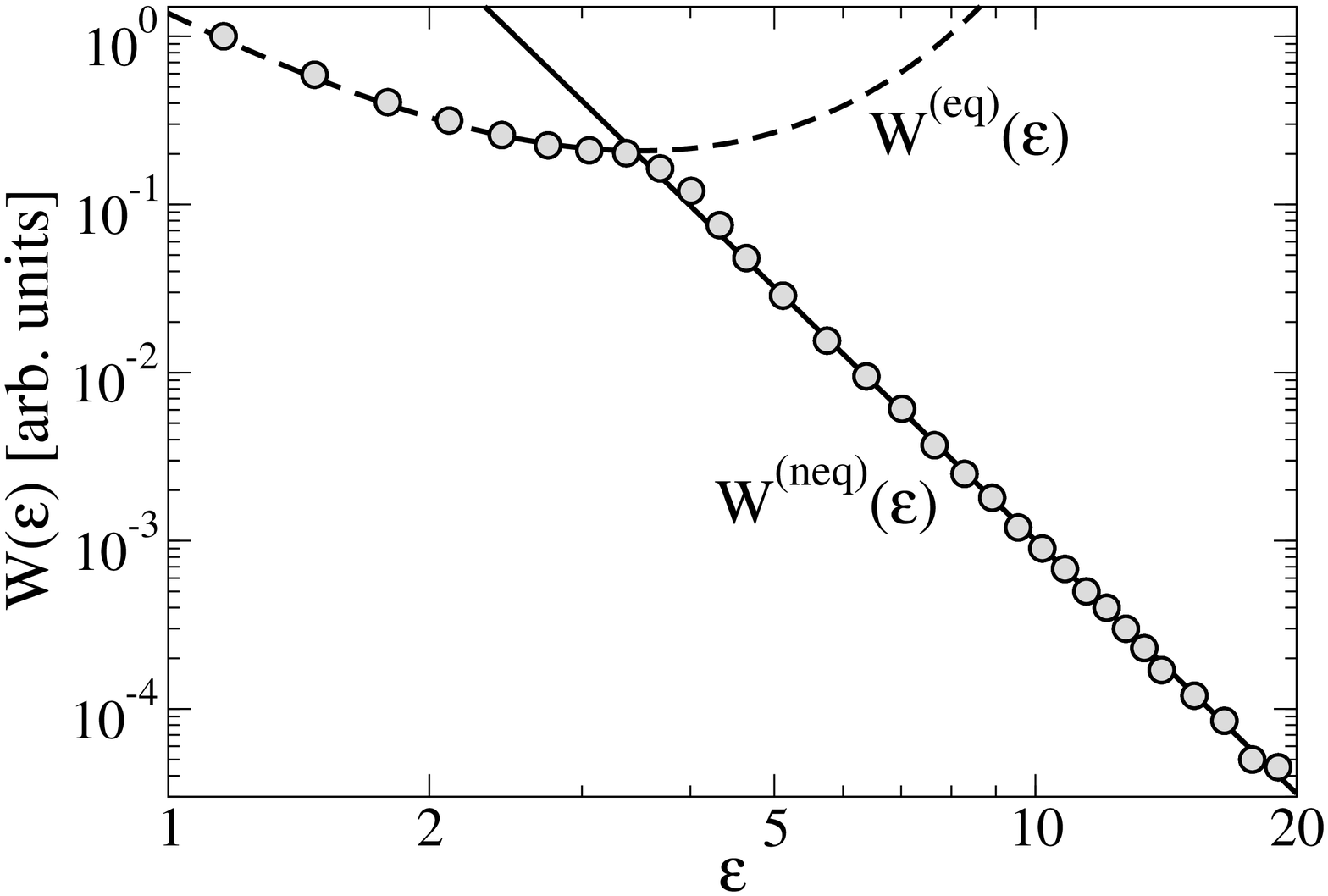}
\caption{Numerically calculated binding energy distribution (dots) compared with the equilibrium  distribution $W_{
\rm eq}$ according to
Eq.(\ref{dist_eq}) (dashed line) and the power-law dependence
Eq.(\ref{pl}) with $\epsilon_{\rm b}=4$ (solid line) arising from the two-step capture. From Ref. \cite{pohl3}}
\label{e_dist}
\end{figure}

The simplicity of this two-step process permits to develop a explanation of the observed $\sim F^{-2}$ power-law decay of the field ionization spectrum. After a short time, binding energies above the kinetic bottleneck establish an equilibrium distribution \cite{GCA2,driscoll}
\begin{equation} \label{dist_eq}
W_{\rm (eq)}\propto E^{-7/2}\exp\left(\frac{E}{k_{\rm B}T_{\rm e^+}}\right)\;.
\end{equation}
Fig.\ \ref{e_dist} shows the binding energy distribution obtained from the described GCA Monte-Carlo calculations \cite{pohl1} for parameters of the ATRAP experiment \cite{GabRev05}, and demonstrates good agreement with Eq.\ (\ref{dist_eq}) for binding energies $E \lesssim 4k_{\rm B}T_{rm e}$ above the bottleneck. Below $E_{\rm bn}$, however, the distribution rapidly departs from its equilibrium form.
This behavior can be understood by assuming that those energy states are populated by a single de-exciting collision, which drives the atom through the kinetic bottleneck.
Under this assumption, the distribution of final energies $E=E_{\rm f}$ is proportional to the corresponding de-excitation rate $K(E_{\rm i},E_{\rm f})$,
which can be obtained from \cite{MK69}
\begin{equation}
\label{de-exc_rate}
K(E_{\rm i},E_{\rm f})=R(E_{\rm f},E_{\rm i})W_{\rm eq}(E_{\rm i})/\rho_{\rm e^{+}}\;,
\end{equation}
where $R(E_{\rm f},E_{\rm i})$ is the collisional kernel for transitions between the
binding energies $E_{\rm i}$ and $E_{\rm f}$, and $\rho_{\rm e}$ is the electron density.
In the limits of $B=0$ and $B\rightarrow\infty$, the energy dependence of $R(E_{\rm f},E_{\rm i})$
is well described by $R(E_{\rm f},E_{\rm i})\propto e^{-\epsilon_i} (-E_f)^{-(\epsilon_{\rm b}+1)}$ \cite{MK69,GCA2},
where $\epsilon_i$ and $\epsilon_{\rm bn}$ denote the initial and bottleneck energies in units of
$k_{\rm B}T_{e}$. Hence, the non-equilibrium binding energy distribution $W_{\rm neq}$ below the bottleneck satisfies the relation
\begin{equation} \label{pl}
W_{\rm neq}(E)dE\sim K(E_{\rm i},E)dE\propto E^{-(\epsilon_{\rm bn}+1)}dE\;.
\end{equation}
Substitution of the known value $\epsilon_{\rm bn}=4$ for the bottleneck energy yields a good fit to the numerically calculated deep-binding energy tail of Fig.\ref{e_dist}.

In order to connect this result to the measured field ionization spectrum of Fig.\ref{FI}, Eq. (\ref{pl}) is transformed
into the corresponding distribution of critical ionization fields $F$. As we have discussed in section \ref{sect_efield}, 
the GCA allows relate the binding energy $E$ and the ionization field $F$ to the respective 
atomic radius $\rho$ and axial bounce amplitude $z_{\rm m}$. We recall that $F\sim \rho/(\rho^2+z_{\rm m}^2)^{3/2}$
and $E_{\rm b}\propto1/\sqrt{\rho^2+z_{\rm m}^2}$. The ionization field distribution can then be obtained by first 
transforming Eq. (\ref{pl}) to spherical coordinates with radial coordinate $r=1/E$. Subsequently, the result is transformed to cylindrical
coordinates ($\rho,z_{\rm m}$), with $r^2=\rho^2+z_{\rm m}^2$, and finally to $F=\rho/r^3$, which yields
\begin{equation} \label{pl2}
P(F)dF \sim \int_{0}^{1}\frac{x^{\frac{\epsilon_{\rm bn}+2}{3}}}{\sqrt{x^{2/3}-x^2}}dx\: F^{-\frac{\epsilon_{\rm bn}+2}{2}}dF\;.
\end{equation}
By integrating $P(F)$ over the ionization field $F$, we obtain the total number of atoms surviving a given electric field \cite{pohl1}
\begin{equation} \label{NF}
N(F) \propto \int_F^{\infty} \tilde{F}^{-\frac{\epsilon+2}{2}} d\tilde{F} \propto F^{-\epsilon_{\rm bn} /2}\;.
\end{equation}
which for $\epsilon_{\rm bn}=4$ gives $N(F) \propto F^{-2}$, as observed in the experiment \cite{GabRev05}
and in agreement with numerical GCA simulations (see Fig.\ \ref{e_dist}). Very recently, Bass and Dubin \cite{bd09} performed extensive calculations
of various Rydberg state-changing in a magnetized plasmas, also including the afore described diffusive de-excitation as well as radiative decay. The
obtained power-laws for collisional de-excitation was also found to be consistent with the experimentally observed field ionization spectrum.

While the above considerations provide a simple picture for the observed power-law at low ionization fields, they fail to explain 
the high-field departure from it. An estimate of the magneton radii corresponding to such ionization fields, however, suggests a break-down of 
the GCA \cite{GabRev05} at the high-field tails of the measured field ionization spectrum. Relaxing the GCA within the described Monte-Carlo
calculations is challenging as it requires to follow the fast electronic cyclotron motion.
This problem can be partly solved by using a symplectic integrator \cite{vrinceanuLA}, based on a splitting of the Hamiltonian into two parts: the total Coulomb potential energy, and the total kinetic energy, which includes all magnetic field terms. Because this integrator consequently describes the cyclotron motion exactly,
it yields high accuracy even though timesteps are chosen considerably larger than $\omega_{\rm c}^{-1}$.
The occurrence of close collisions, which lead to the formation of very deeply-bound atoms, necessitates that progressively smaller
timesteps are chosen. Adjusting the timestep accordingly, however, destroys the
symplectic nature of the integrator, resulting in a drift of the total energy.
This problem can be overcome \cite{pohl1,pohl3} within an adapted version of the time-transformed leap-frog scheme
introduced in \cite{leap-frog}, which introduces a distance-dependent scaled time through an auxiliary phase-space variable.
The combined integrator tracts the exact electron dynamics in both limiting cases, i.e.
for strong magnetic as well as dominating Coulomb fields, ensuring an efficient and, at the same time,  
accurate solution of the electron dynamics. 

\begin{figure}[htb!]
\includegraphics*[width=2.6in]{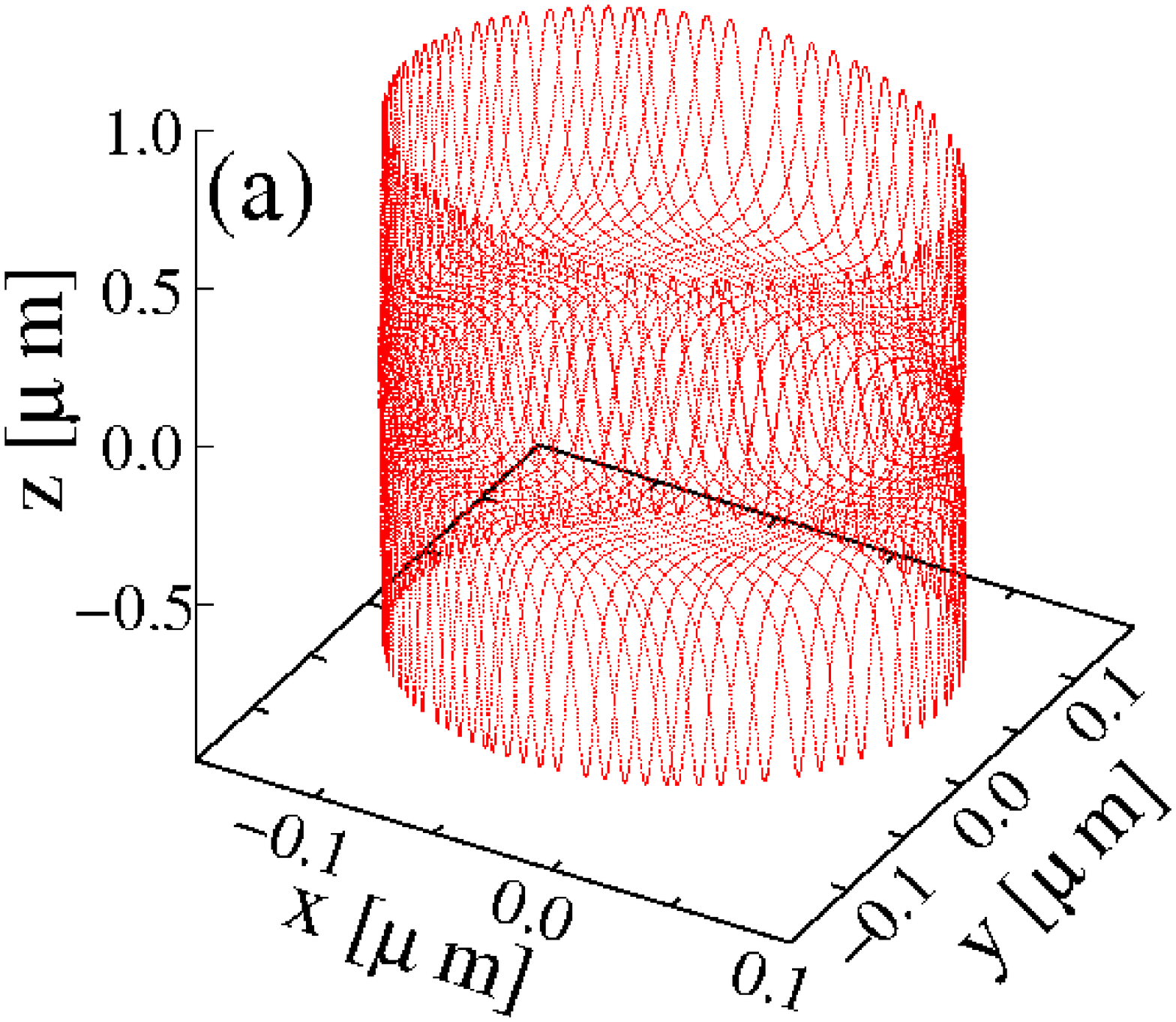}
\includegraphics*[width=2.6in]{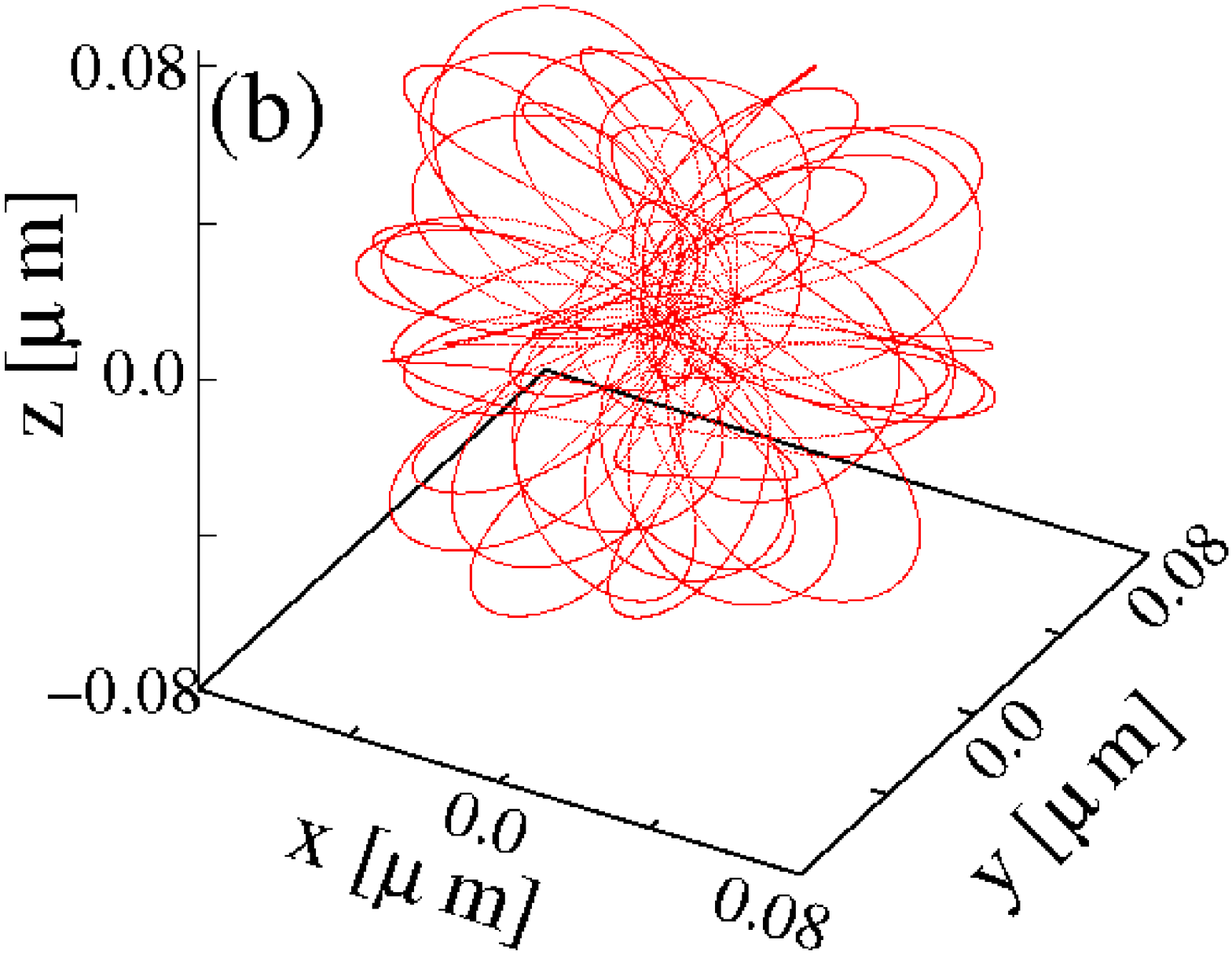}
\caption{Typical example of a small GCA atom trajectory (a), and a small nGC chaotic trajectory (b). From Ref. \cite{pohl3}.\label{trajectory}}
\end{figure}

The three-body capture into deeply-bound \Hbar atoms in magnetic fields can be reliably monitored with this
numerical regularization technique. The comparison between the result of the corresponding Monte-Carlo simulations and the experimental field ionization spectrum, shown in Fig.\ \ref{e_dist}, yields good agreement and gives strong arguments for the formation of non-guiding center (nGC) atoms. 
In Fig. \ref{trajectory} we show typical examples of a formed small GCA atom (a) and a nGC atom (b), highlighting the differences in how these
atoms would ionize in an electric field. While the recombined GCA atom have tightly confined transverse trajectories, they are  highly elongated in the axial direction, which makes them susceptible to field ionization. The corresponding nGCA atoms, on the other hand, have tight axial and transverse confinements, and chaotically fill a nearly spherical spatial volume. 
The three degrees of freedom (cyclotron, magnetron, and axial bounce motions) are intricately coupled and the energy 
stored in the cyclotron motion is shared among all three degrees of freedom. 
These much more compact atoms, thus, survive large ionizing fields, resulting in a departure from the F$^{-2}$ power-law behavior as shown in Fig. \ref{FI}.

\subsection{Heavy-particle collisions in strong magnetic fields} \label{sect_chx}
The requirement of producing {\emph{cold}} antihydrogen atoms is an important prerequisite for their use in subsequent cooling, trapping,  
and spectroscopy experiments. In an effort to meet this condition, the ATRAP collaboration used a "driven production" method, 
in which cold antiprotons were gently heated out of the side-wells of the nested Penning trap \cite{atrap2}. By driving the antiprotons 
just above the central well into the positron plasma, a production of low-energy ($\sim1$meV) antihydrogen atoms has been expected.
This expectation was dispelled in a later experiment \cite{AtrapVelocity}, that measured the speed of \Hbar atoms by adding a
time-varying electric field in front of the detector, acting as a velocity selective barrier.
By monitoring the fraction of atoms that passes through this oscillating field, the experiment allowed to deduce the characteristic 
velocity of formed atoms. To some surprise, the extracted kinetic energy of $\sim200$meV was much larger than the otherwise expected
thermal positron energy of 0.3 meV (4.2 K).

Below we discuss charge-exchange collisions with fast antiprotons in the side wells of the nested
Penning trap as a possible explanation for the observed fast \Hbar atoms \cite{pohl1}.

\subsubsection{Rydberg charge exchange cross section}
Bradenbrink \etal  \cite{sidky,sidky2} performed a semi-classical study of electron transfer from Rydberg atoms in collision with singly-charged ions in 
the presence of laboratory magnetic fields. In this work, the initial state was constructed from a microcanonical phase-space distribution.

For the present problem, the initial phase space distribution of Rydberg atoms is determined by the preceding three-body recombination dynamics, 
and can be obtained from the final states of the Monte-Carlo calculations, described in the previous section \cite{pohl1}.
Within a CTMC approach, i.e. by firing a large number of antiprotons with
a given incident energy onto these classical atomic states, one obtains the cross section for charge exchange for each microscopic internal and translational initial state of the formed atoms. 
Instead of characterizing the corresponding bound states by their respective energies, it is more appropriate to label the initial and final atomic states of the \Hbar atoms by their critical ionization fields, $F_{\rm i}$ and $F_{\rm f}$. In this way, one obtains the state-selective cross section
$\sigma_{\rm cx}(F_{\rm i},F_{\rm f};v_{\bar{H}},v_{\bar{p}})$ for exchange between an incident
antiproton with velocity $v_{\bar{p}}$ and an \Hbar with velocity $v_{\bar{H}}$ ionizing at a minimum field  
$F_{\rm i}$. The resulting \Hbar atom, moving at $v_{\bar{p}}$, will ionize at a maximum field $F_{\rm f}$. 

\begin{figure}[htb!]
\includegraphics[width=2.8in]{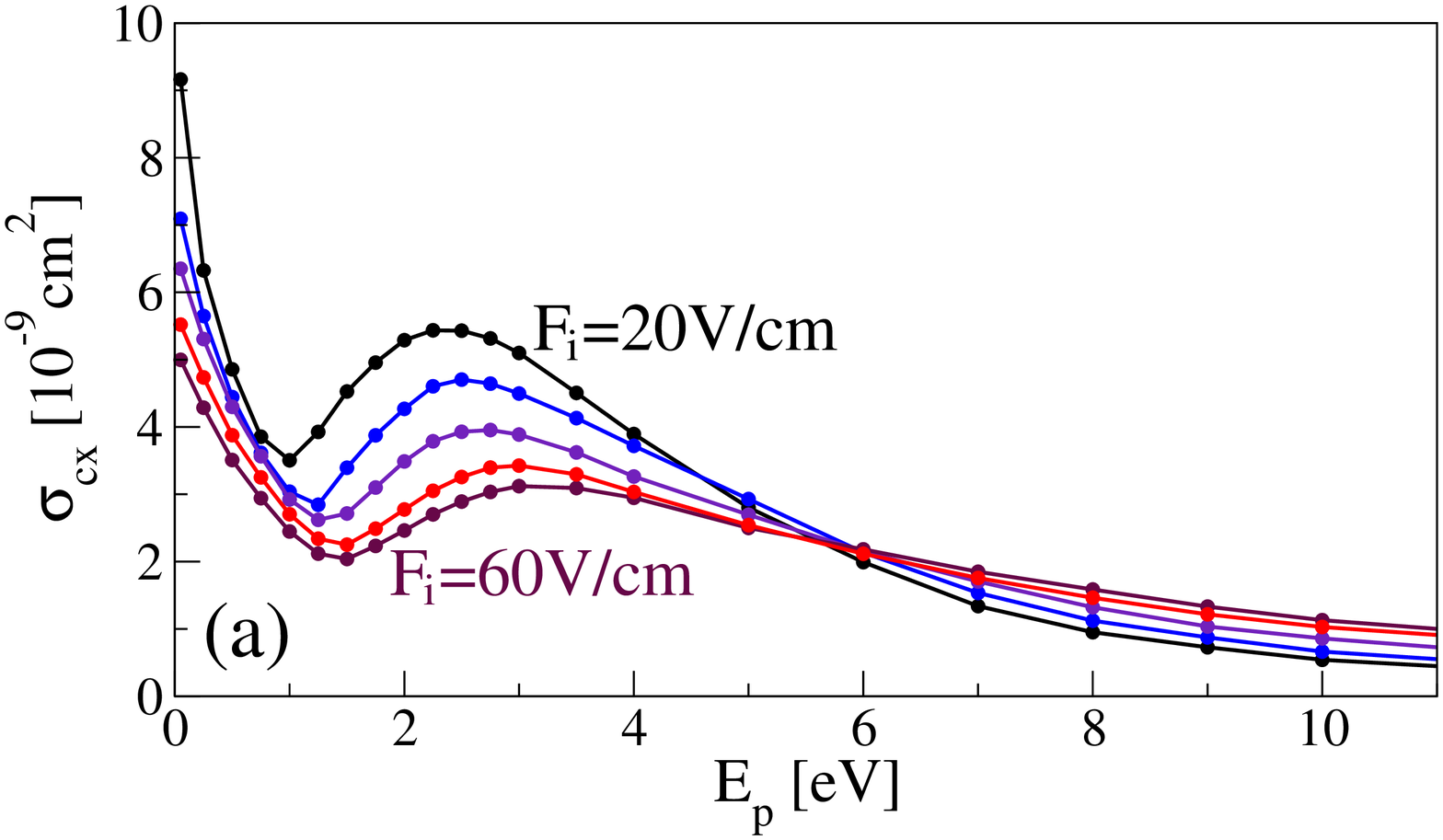}\hskip10pt
\includegraphics[width=2.95in]{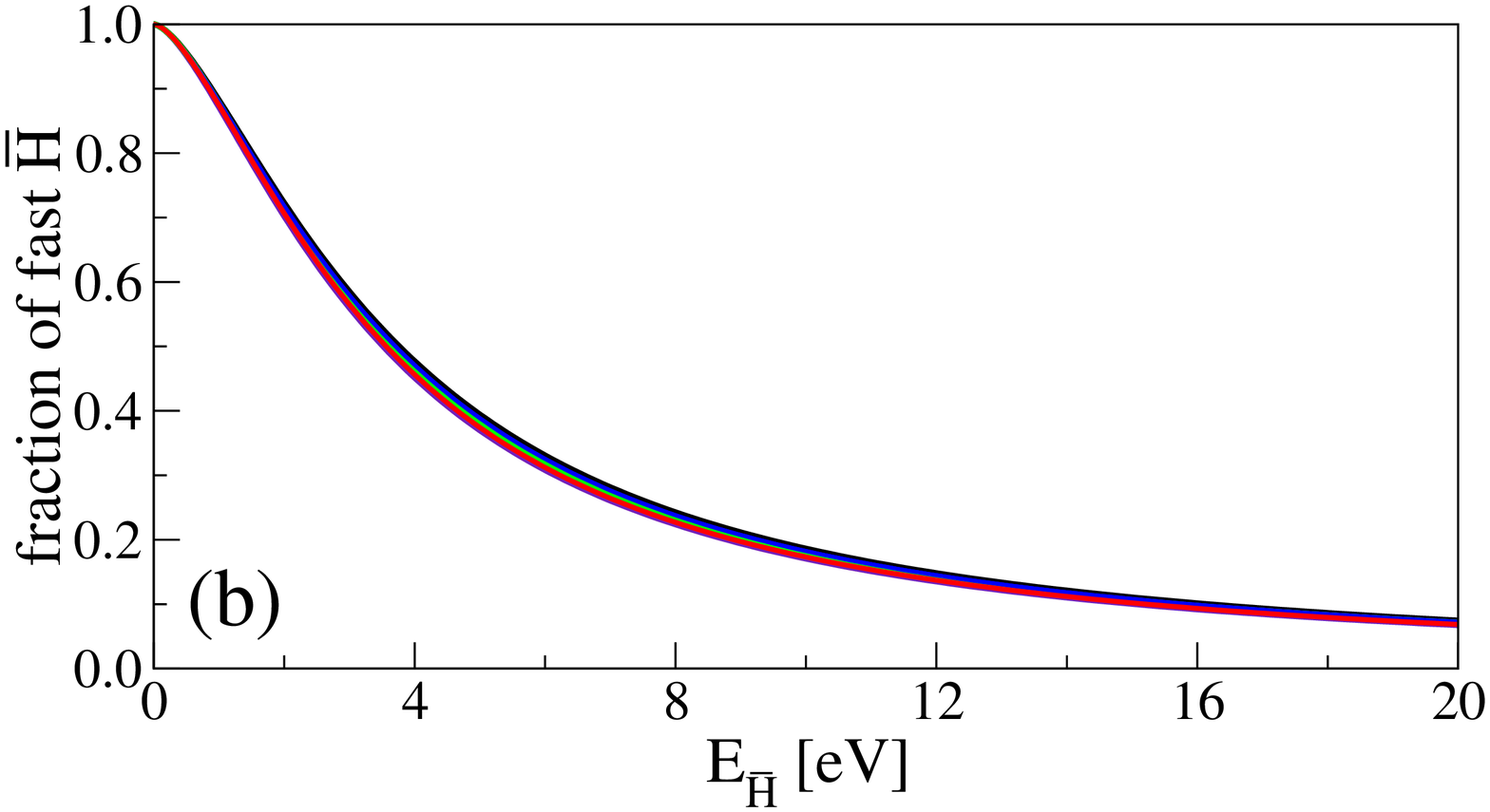}
\caption{(a) Charge exchange cross section from atoms with a kinetic energy of $1$meV as a
function of antiproton axial energy, for different ionization fields in the range, $20-60$  V/cm. (b) The fraction of charge
exchanged atoms is independent of the electric field. From Ref. \cite{pohl1}}
\label{cx}
\end{figure}

In Fig. \ref{cx}, we show the average total capture cross section
\begin{equation}
\bar{\sigma}_{\rm cx}(F_{\rm i};v_{\bar{H}},v_{\bar{p}})
=\int \sigma_{\rm cx}(F_{\rm i},F_{\rm f};v_{\bar{H}},v_{\bar{p}})dF_{\rm f}
\end{equation}
as a function of $v_{\bar{p}}$ for different values of $F_{\rm i}$. The calculations yield large charge exchange cross sections, due to the large spatial extent of Rydberg atoms, that result in large geometric cross sections. Moreover, there are pronounced maxima, arising from a 
matching of the Rydberg positron and the projectile antiproton velocities, known from field-free charge exchange processes \cite{XC1,XC2}.

Using the calculated cross sections, the probability that the initially formed \Hbar atoms undergo charge exchange in the side wells can be written as
\begin{equation}\label{prob}
P_{\rm cx}(F_{\rm i},F_{\rm f};v_{\bar{H}},v_{\bar{p}}) = 1 - \exp\left(-\sigma_{cx} n_{\bar{p}} d \frac{v_{\bar{p}}}{v_{\bar{H}}}\right)\;,
\end{equation}
where $\rho_{\bar{p}}$ and $d$ denote the density and the length of the \pbar plasma in the side wells of
the nested Penning trap.
The fraction of detected slow \Hbar atoms with velocity $v$ that {\it did not} undergo charge exchange after traversing the side well
is given by
\footnotesize
\begin{equation}
f_{\rm ncx}(F,v)=N(F)\kappa(v) \phi_{\bar{H}}(v)\int\phi_{\bar{p}}(v_{\bar{p}})\left[1-
P_{\rm cx}\left(F,F_{\rm f};v,v_{\bar{p}}\right)\right]dF_{\rm f}d^3v_{\bar{p}}\;,
\label{prob1}
\end{equation}
\normalsize

where $N(F)$ is the number of initially formed \Hbar that survive the field $F$ and $\kappa(v)$ is the velocity dependent 
detection efficiency \cite{pohl1}. The functions $\phi_{\bar{H}}(v)$ and $\phi_{\bar{p}}(v)$ denote the initial velocity distributions
of the \Hbar atoms and antiprotons, and are assumed to be Gaussian with respective temperatures $T_{\bar{H}}$ and $T_{\bar{p}}$.
In analogy, the fraction of fast \Hbar produced by charge exchange collisions is obtained from
\begin{equation}
f_{\rm cx}(F,v)=\phi_{\bar{p}}(v)\int N(F_{\rm i}) \phi_{\bar{H}}(v_{\bar{H}})
P_{\rm cx}(F_{\rm i},F;v_{\bar{H}},v)dF_{\rm i}d^3v_{\bar{H}}\;.
\label{prob2}
\end{equation}

\begin{figure}[htb!]
\includegraphics[width=2.6in]{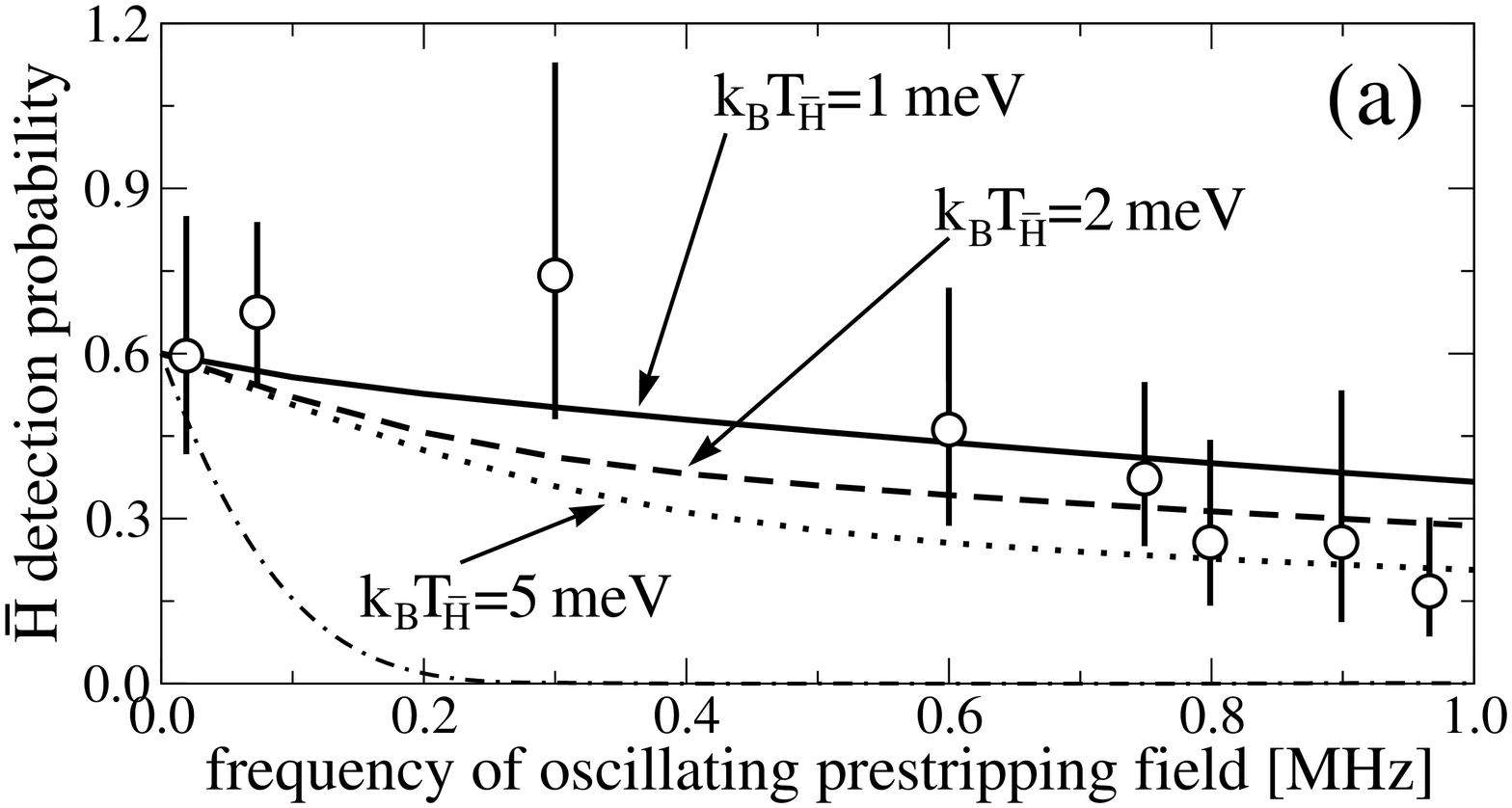}\hskip10pt
\includegraphics[width=2.85in]{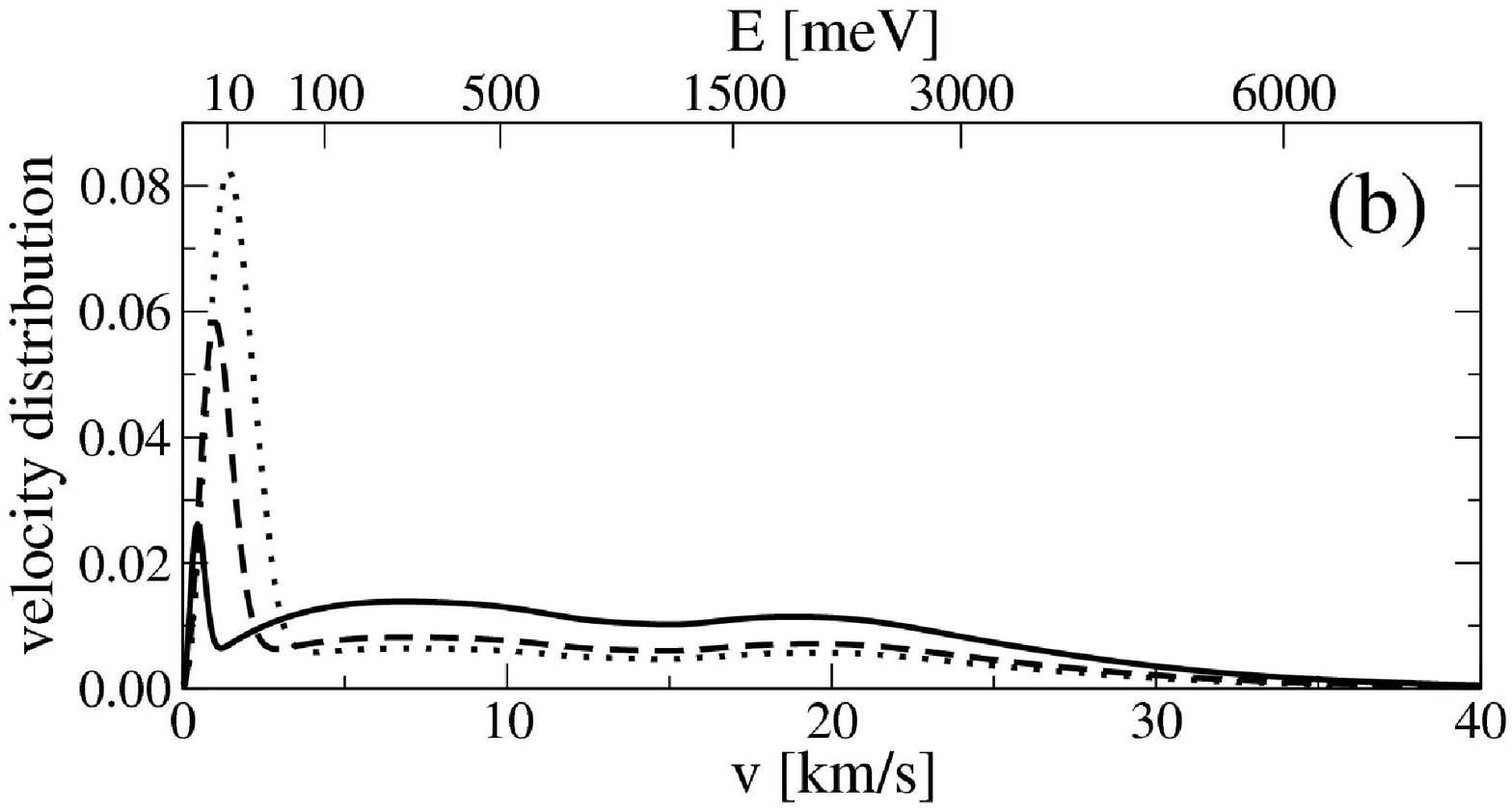}
\caption{(a) Proposed
charge exchange mechanism produces a frequency dependent field ionization spectrum consistent
with experimental data \cite{AtrapVelocity}
(circles).
The spectra have been calculated for $k_{\rm B}T_{\bar{p}}=8$eV and $k_{\rm B}T_{\bar{H}}=1$meV (solid line),
$k_{\rm B}T_{\bar{H}}=2$ meV
(dashed line) and $k_{\rm B}T_{\bar{H}}=5$meV (dotted line). The dot-dashed line shows the spectrum that results by neglecting \pos
charge transfer
for $kT_{\bar{H}}= 2$meV ($T_{\bar{H}}= 30$K). (b)
Corresponding velocity distributions after charge exchange. From Ref. \cite{pohl1}} \label{Velocity}
\end{figure}

Note that the fraction of fast atoms is almost independent of the ionization field (see Fig. \ref{cx}(b)), such that the field ionization spectrum, shown in Fig. \ref{FI}) remains  unaffected by charge exchange processes \cite{pohl1}.
The velocity distribution of atoms that reach the detector is then obtained by adding the contributions from the slow [Eqs. (\ref{prob1})] and the fast [Eqs. (\ref{prob2})] atoms, and integrating over the ionization field
\begin{equation}
\phi(v)=\int [f_{\rm ncx}(F,v)+f_{\rm cx}(F,v)]dF
\end{equation}
As shown in Fig. \ref{Velocity}(b), the velocity distribution contains two prominent features.
There is a low-velocity peak due to the initially recombined \Hbar atoms that did not undergo charge exchange in the side-wells. 
On the other hand, the extended high velocity tails arise from charge exchange with the fast \pbar atoms in the side wells.
The latter produces the high-frequency tail in the observed field spectrum.
Fig.\ \ref{Velocity}(a) demonstrates good agreement with the experimentally observed field ionization spectrum for assumed \Hbar
temperatures ranging from $1$meV to $5$meV. Without the inclusion of charge exchange reactions, a much higher velocity
would be obtained from a fit to the experimental data.

The discussions of this section have shown that a multitude of Rydberg atom collision processes in strong magnetic fields are at work in 
antihydrogen experiments. They critically determine the internal states as well as the translational energies of the produced atoms, 
both being important for their subsequent use for trapping and spectroscopy experiments.

\section{Outlook and Perspectives} \label{outlook}
Giant dipole or decentered states of atoms in crossed electric and magnetic fields, as they have been reviewed in the present work, 
possess very different properties from conventional atomic Rydberg states. The latter are centered around the Coulomb singularity
and represent in this sense compact quantum objects. On the contrary, giant dipole states have one or several of their
electrons located far from the nucleus, and their states behave more like a displaced oscillator rather than a centered
Coulomb or magnetized Coulomb state. Giant dipole states inherently possess huge oriented electric dipole moments and are of fragile
character, representing a rather unique state of highly excited quantum matter. This naturally raises the question for the existence
of giant dipole systems beyond atoms, i.e. beyond single electron bound giant dipole states or multi-electron
giant dipole resonances. A next step would certainly be to explore the existence of giant dipole molecules.
At this point one can imagine several scenarios. An apparent possibility would be the simple analogue
of single electron atomic giant dipole states, where a decentered single electron is bound to a remaining positively
charged molecular core. A second type would be in the spirit of the molecular trilobite states \cite{gds,hgs}:
One of the atoms of, e.g., a diatomic molecule, is excited to a single electron decentered state in crossed fields, while the remaining
 ground state atom probes, as a perturber, the extended wave function of the Rydberg electron.
In this way, a weakly bound decentered Rydberg molecule could emerge. Finally, a third type would be due to cations and
anions forming decentered molecules. Particularly the latter case raises the intriguing question whether bound compounds of such ions
 could be sufficiently long lived, and if yes, what their properties would be. Definitely its characteristics would be vastly
different from that of 'ordinary' quantum matter. Theoretical and experimental efforts in this
direction would be necessary, possibly starting with the scheme outlined in this review: Ground state atoms or
molecules are excited to Rydberg states and then via a sequence of electric fields switches transported to 
the decentered configuration. 

The reverse process, i.e. the formation of Rydberg atoms, including giant dipole states, in collisions of charged particles, constantly takes 
place in cold antimatter plasmas realized in antihydrogen experiments. While the experiments primarily target the 
production of ultracold ground state atoms, the physics of magnetized Rydberg atoms is important for understanding and 
ultimately optimizing collisional \Hbar formation. Currently, there appear to be several obstacles to be overcome for the 
production of ultracold trappable ground state atoms. One of such issues is that the recombination rate typically turns 
out to be too large, such that \Hbar atoms typically form before the antiprotons have cooled-down sufficiently through 
interaction with the positron cloud. The problem may be solved if external field control could be realized. For example 
ionization of high lying Rydberg states via microwave radiation may place an energy barrier above the bottleneck and 
thus limit recombination. Several open questions concerning the coupling to the different magnetized degrees of 
freedom and the effects on the plasma dynamics will have to be clarified here. Microwave ionization of Rydberg atoms 
proves to be a complex, separate research field itself, and its extension to strongly magnetized Rydberg atoms would be 
inherently interesting. Moreover, one major motivation for the discussed creation of strongly magnetized rubidium 
plasmas \cite{ckz05} is to exploit the strong-field suppression of recombination for a realization of a strongly coupled
neutral plasma, for which a further suppression would certainly be highly beneficial. Finding possible routes to external field
control of Rydberg atom formation in ultracold environments would, hence, be an important achievement for several reasons. 
 
The development of trapping techniques for ultracold Rydberg atoms, as for example in inhomogeneous magnetic field 
configurations, is currently of high interest.
Our emphasis in this review has been to report on the fundamental properties of such
atoms including the spin-spatial symmetries and structural properties. Due to the inherent coupling of the external (center of mass)
and internal (electronic) degrees of freedom one typically encounters hybrid states: The electronic configuration varies
within a center of mass quantum state with varying position of the center of mass. Such an entanglement of external
and internal degrees of freedom is absent in the field-free case or in homogeneous magnetic fields. Moreover, trapping of 
Rydberg atoms is in general an intricate task since the variation of the energy levels with varying field strength leads generically 
to (avoided) level crossings and consequently an uncontrollable nonadiabatic quantum dynamics. Having shown here that stable 
trapping is possible in magnetic traps, it remains an open challenge, both theoretically and
experimentally, to demonstrate trapping and work out flexible trapping configurations, e.g. through coupling to optical fields. 
The controlled coherent processing of many-body Rydberg systems, which represents an important perspective
of the field, depends critically on the possibility to individually trap and prepare Rydberg atoms. This supports the hope that 
fundamentally new quantum coherent systems, such as controllably interacting many-body Rydberg systems, 
are conceivable and might enrich the diversity of ultracold quantum matter in general.

\section{Acknowledgment}
The work of TP and HRS were partly supported by NSF through a grant to
the Institute for Theoretical Atomic, Molecular and Optical Physics (ITAMP).
PS acknowledges support from the Deutsche Forschungsgemeinschaft in the
framework of the Excellence Initiative through the Heidelberg Graduate School
of Fundamental Physics (GSC 129/1) and through the grant SCHM 885/10.
PS acknowledges fruitful collaborations with L.S. Cederbaum, I. Lesanovsky,
J. Shertzer and J. Ackermann. TP and HRS are grateful to G. Gabrielse and Y. Yamazaki 
for fruitful collaborations.


\begin{appendix}
\end{appendix}

\bibliography{literature}
\end{document}